\pdfoutput=1
\documentclass[12pt,a4paper]{article}

\usepackage{ifthen} 
\newboolean{pdflatex}
\setboolean{pdflatex}{true} 

\newboolean{articletitles}
\setboolean{articletitles}{true} 

\newboolean{uprightparticles}
\setboolean{uprightparticles}{false} 

\newboolean{inbibliography}
\setboolean{inbibliography}{false} 

\usepackage{microtype}
\usepackage{lineno}  
\usepackage{xspace} 
\usepackage{caption} 

\usepackage{graphicx}  
\usepackage{color}
\usepackage{colortbl}
\graphicspath{{./figs/}} 

\usepackage{amsmath} 
\usepackage{amssymb}
\usepackage{amsfonts}
\usepackage{upgreek} 

\newcommand*\patchAmsMathEnvironmentForLineno[1]{%
\expandafter\let\csname old#1\expandafter\endcsname\csname #1\endcsname
\expandafter\let\csname oldend#1\expandafter\endcsname\csname
end#1\endcsname
 \renewenvironment{#1}%
   {\linenomath\csname old#1\endcsname}%
   {\csname oldend#1\endcsname\endlinenomath}%
}
\newcommand*\patchBothAmsMathEnvironmentsForLineno[1]{%
  \patchAmsMathEnvironmentForLineno{#1}%
  \patchAmsMathEnvironmentForLineno{#1*}%
}
\AtBeginDocument{%
\patchBothAmsMathEnvironmentsForLineno{equation}%
\patchBothAmsMathEnvironmentsForLineno{align}%
\patchBothAmsMathEnvironmentsForLineno{flalign}%
\patchBothAmsMathEnvironmentsForLineno{alignat}%
\patchBothAmsMathEnvironmentsForLineno{gather}%
\patchBothAmsMathEnvironmentsForLineno{multline}%
\patchBothAmsMathEnvironmentsForLineno{eqnarray}%
}

\allowdisplaybreaks

\usepackage{hyperref}    
\usepackage[all]{hypcap} 

\def\lhcb {\mbox{LHCb}\xspace}

\ifthenelse{\boolean{uprightparticles}}%
{

 \def\Pmu         {\ensuremath{\upmu}\xspace}

 \def\Ppi         {\ensuremath{\uppi}\xspace}

 \def\Ppsi        {\ensuremath{\uppsi}\xspace}

 \def\PDelta      {\ensuremath{\Delta}\xspace}                 
 \def\PXi      {\ensuremath{\Xi}\xspace}                 
 \def\PLambda      {\ensuremath{\Lambda}\xspace}                 
 \def\PSigma      {\ensuremath{\Sigma}\xspace}                 
 \def\POmega      {\ensuremath{\Omega}\xspace}                 
 \def\PUpsilon      {\ensuremath{\Upsilon}\xspace}                 
 
 \def\PB      {\ensuremath{\mathrm{B}}\xspace}                 
                  
 \def\PD      {\ensuremath{\mathrm{D}}\xspace}

 \def\PJ      {\ensuremath{\mathrm{J}}\xspace}                 
 \def\PK      {\ensuremath{\mathrm{K}}\xspace}

 \def\Pb      {\ensuremath{\mathrm{b}}\xspace}                 
 \def\Pc      {\ensuremath{\mathrm{c}}\xspace}

 \def\Pi      {\ensuremath{\mathrm{i}}\xspace}

 \def\Pp      {\ensuremath{\mathrm{p}}\xspace}

 \def\Ps      {\ensuremath{\mathrm{s}}\xspace}

}
{

 \def\Pmu         {\ensuremath{\mu}\xspace}

 \def\Ppi         {\ensuremath{\pi}\xspace}

 \def\Ppsi        {\ensuremath{\psi}\xspace}                 
                  
 \mathchardef\PDelta="7101
 \mathchardef\PXi="7104
 \mathchardef\PLambda="7103
 \mathchardef\PSigma="7106
 \mathchardef\POmega="710A
 \mathchardef\PUpsilon="7107
                  
 \def\PB      {\ensuremath{B}\xspace}                 
                  
 \def\PD      {\ensuremath{D}\xspace}

 \def\PJ      {\ensuremath{J}\xspace}                 
 \def\PK      {\ensuremath{K}\xspace}

 \def\Pb      {\ensuremath{b}\xspace}                 
 \def\Pc      {\ensuremath{c}\xspace}

 \def\Pi      {\ensuremath{i}\xspace}

 \def\Pp      {\ensuremath{p}\xspace}

 \def\Ps      {\ensuremath{s}\xspace}

}

\def\mup        {{\ensuremath{\Pmu^+}}\xspace}
\def\mun        {{\ensuremath{\Pmu^-}}\xspace} 
\def\mumu       {{\ensuremath{\Pmu^+\Pmu^-}}\xspace}

\def\squark    {{\ensuremath{\Ps}}\xspace}

\def\cquark    {{\ensuremath{\Pc}}\xspace}

\def\bquark    {{\ensuremath{\Pb}}\xspace}

\def\pion   {{\ensuremath{\Ppi}}\xspace}

\def\pip    {{\ensuremath{\pion^+}}\xspace}
\def\pim    {{\ensuremath{\pion^-}}\xspace}
\def\pipm   {{\ensuremath{\pion^\pm}}\xspace}
\def\pimp   {{\ensuremath{\pion^\mp}}\xspace}

\def\kaon    {{\ensuremath{\PK}}\xspace}
  \def\Kbar    {{\kern 0.2em\overline{\kern -0.2em \PK}{}}\xspace}

\def\Kz      {{\ensuremath{\kaon^0}}\xspace}
\def\Kzb     {{\ensuremath{\Kbar^0}}\xspace}
\def\Kp      {{\ensuremath{\kaon^+}}\xspace}
\def\Km      {{\ensuremath{\kaon^-}}\xspace}
\def\Kpm     {{\ensuremath{\kaon^\pm}}\xspace}

\def\KS      {{\ensuremath{\kaon^0_{\rm\scriptscriptstyle S}}}\xspace}

  \def\Dbar    {{\kern 0.2em\overline{\kern -0.2em \PD}{}}\xspace}
\def\D       {{\ensuremath{\PD}}\xspace}

\def\Dz      {{\ensuremath{\D^0}}\xspace}

\def\B       {{\ensuremath{\PB}}\xspace}
\def\Bbar    {{\ensuremath{\kern 0.18em\overline{\kern -0.18em \PB}{}}}\xspace}

\def\Bu      {{\ensuremath{\B^+}}\xspace}

\def\Bp      {{\ensuremath{\Bu}}\xspace}

\def\Bd      {{\ensuremath{\B^0}}\xspace}
\def\Bs      {{\ensuremath{\B^0_\squark}}\xspace}
\def\Bsb     {{\ensuremath{\Bbar^0_\squark}}\xspace}

\def\Bds     {{\ensuremath{\B^0_{(\squark)}}}\xspace}

\def\jpsi     {{\ensuremath{{\PJ\mskip -3mu/\mskip -2mu\Ppsi\mskip 2mu}}}\xspace}
\def\psitwos  {{\ensuremath{\Ppsi{(2S)}}}\xspace}

  \def\Y#1S{\ensuremath{\PUpsilon{(#1S)}}\xspace}

\def\proton      {{\ensuremath{\Pp}}\xspace}

\def\Lz          {{\ensuremath{\PLambda}}\xspace}
\def\Lbar        {{\ensuremath{\kern 0.1em\overline{\kern -0.1em\PLambda}}}\xspace}

\def\Lb      {{\ensuremath{\Lz^0_\bquark}}\xspace}

\def\to                 {\ensuremath{\rightarrow}\xspace}

\def\CP                {{\ensuremath{C\!P}}\xspace}

\def\AT#1     {\ensuremath{A_{\mathrm{T}}^{#1}}\xspace}           

\def\C#1      {\ensuremath{\mathcal{C}_{#1}}\xspace}                       
\def\Cp#1     {\ensuremath{\mathcal{C}_{#1}^{'}}\xspace}                    
\def\Ceff#1   {\ensuremath{\mathcal{C}_{#1}^{\mathrm{(eff)}}}\xspace}        
\def\Cpeff#1  {\ensuremath{\mathcal{C}_{#1}^{'\mathrm{(eff)}}}\xspace}       
\def\Ope#1    {\ensuremath{\mathcal{O}_{#1}}\xspace}                       
\def\Opep#1   {\ensuremath{\mathcal{O}_{#1}^{'}}\xspace}                    

\newcommand{\tev}{\ifthenelse{\boolean{inbibliography}}{\ensuremath{~T\kern -0.05em eV}\xspace}{\ensuremath{\mathrm{\,Te\kern -0.1em V}}}\xspace}
\newcommand{\gev}{\ensuremath{\mathrm{\,Ge\kern -0.1em V}}\xspace}
\newcommand{\mev}{\ensuremath{\mathrm{\,Me\kern -0.1em V}}\xspace}
\newcommand{\kev}{\ensuremath{\mathrm{\,ke\kern -0.1em V}}\xspace}
\newcommand{\ev}{\ensuremath{\mathrm{\,e\kern -0.1em V}}\xspace}
\newcommand{\gevc}{\ensuremath{{\mathrm{\,Ge\kern -0.1em V\!/}c}}\xspace}
\newcommand{\mevc}{\ensuremath{{\mathrm{\,Me\kern -0.1em V\!/}c}}\xspace}
\newcommand{\gevcc}{\ensuremath{{\mathrm{\,Ge\kern -0.1em V\!/}c^2}}\xspace}
\newcommand{\gevgevcccc}{\ensuremath{{\mathrm{\,Ge\kern -0.1em V^2\!/}c^4}}\xspace}
\newcommand{\mevcc}{\ensuremath{{\mathrm{\,Me\kern -0.1em V\!/}c^2}}\xspace}

\def\mum  {\ensuremath{{\,\upmu\rm m}}\xspace}

\def\invfb   {\ensuremath{\mbox{\,fb}^{-1}}\xspace}

\newcommand{\stat}{\scalebox{0.75}{\ensuremath{\mathrm{\,(stat)}}}\xspace}
\newcommand{\syst}{\scalebox{0.75}{\ensuremath{\mathrm{\,(syst)}}}\xspace}
\newcommand{\norm}{\scalebox{0.75}{\ensuremath{\mathrm{\,(norm)}}}\xspace}
\newcommand{\pdg} {\scalebox{0.75}{\ensuremath{\mathrm{\,(PDG)}}}\xspace}
\newcommand{\fsfd}{\scalebox{0.75}{\ensuremath{\,(f_s/f_d)}}\xspace}

\newcommand{\chisq}{\ensuremath{\chi^2}\xspace}

\newcommand{\chisqip}{\ensuremath{\chi^2_{\rm IP}}\xspace}

\def\gsim{{~\raise.15em\hbox{$>$}\kern-.85em
          \lower.35em\hbox{$\sim$}~}\xspace}
\def\lsim{{~\raise.15em\hbox{$<$}\kern-.85em
          \lower.35em\hbox{$\sim$}~}\xspace}

\def\sPlot{\mbox{\em sPlot}}

\def\pt         {\mbox{$p_{\rm T}$}\xspace}

\def\evtgen     {\mbox{\textsc{EvtGen}}\xspace}

\def\geant      {\mbox{\textsc{Geant4}}\xspace}

\def\photos     {\mbox{\textsc{Photos}}\xspace}

\def\pythia     {\mbox{\textsc{Pythia}}\xspace}

\def\tell1  {TELL1\xspace}
\def\ukl1   {UKL1\xspace}

\def\phani{\phantom{1}}
\def\phanii{\phantom{11}}

\usepackage{lmodern}

\makeatletter
\ifcase \@ptsize \relax
  \newcommand{\miniscule}{\@setfontsize\miniscule{4}{5}}
\or
  \newcommand{\miniscule}{\@setfontsize\miniscule{5}{6}}
\or
  \newcommand{\miniscule}{\@setfontsize\miniscule{5}{6}}
\fi
\makeatother

\DeclareRobustCommand{\optbar}[1]{\shortstack{{\miniscule (\rule[.5ex]{1.25em}{.18mm})}
  \\ [-.7ex] $#1$}}
\def\BorBbar    {\kern 0.18em\optbar{\kern -0.18em B}{}\xspace}
\def\DorDbar    {\kern 0.18em\optbar{\kern -0.18em D}{}\xspace}
\def\KorKbar    {\kern 0.18em\optbar{\kern -0.18em K}{}\xspace}

\def\BsToJpsiKShhp {\ensuremath{\Bs\to\jpsi\KS h^+ h^{\left(\prime\right) -}}\xspace}
\def\BxToJpsiKShhp {\ensuremath{\Bds\to\jpsi\KS h^+ h^{\left(\prime\right) -}}\xspace}

\def\BdToPsitwoSKS {\ensuremath{\Bd \to \psitwos \KS }\xspace}

\def\BxToPsitwoSKS {\ensuremath{\Bds \to \psitwos \KS }\xspace}

\def\BdToJpsiKS {\ensuremath{\Bd \to \jpsi \KS }\xspace}

\def\BsToJpsiKS {\ensuremath{\Bs \to \jpsi \KS }\xspace}

\def\BxToJpsiKS {\ensuremath{\Bds \to \jpsi \KS }\xspace}

\def\BdToJpsiKSpipi {\ensuremath{\Bd \to \jpsi \KS \pip\pim}\xspace}

\def\BsToJpsiKSpipi {\ensuremath{\Bs \to \jpsi \KS \pip\pim}\xspace}

\def\BxToJpsiKSpipi {\ensuremath{\Bds \to \jpsi \KS \pip\pim}\xspace}

\def\BdToJpsiKSKpi {\ensuremath{\Bd \to \jpsi \KS \Kpm\pimp}\xspace}

\def\BsToJpsiKSKpi {\ensuremath{\Bs \to \jpsi \KS \Kpm\pimp}\xspace}

\def\BxToJpsiKSKpi {\ensuremath{\Bds \to \jpsi \KS \Kpm\pimp}\xspace}

\def\BdToJpsiKSKK {\ensuremath{\Bd \to \jpsi \KS \Kp\Km}\xspace}

\def\BsToJpsiKSKK {\ensuremath{\Bs \to \jpsi \KS \Kp\Km}\xspace}

\def\BxToJpsiKSKK {\ensuremath{\Bds \to \jpsi \KS \Kp\Km}\xspace}

\def\BdToJpsiKzpipi {\ensuremath{\Bd \to \jpsi \Kz \pip\pim}\xspace}
\def\BsToJpsiKzbpipi {\ensuremath{\Bs \to \jpsi \Kzb \pip\pim}\xspace}
\def\BdToJpsiKzKpi {\ensuremath{\Bd \to \jpsi \KorKbar^0 \Kpm \pimp}\xspace}
\def\BsToJpsiKzKpi {\ensuremath{\Bs \to \jpsi \KorKbar^0 \Kpm \pimp}\xspace}
\def\BxToJpsiKzKpi {\ensuremath{\Bds \to \jpsi \KorKbar^0 \Kpm \pimp}\xspace}
\def\BdToJpsiKzKK {\ensuremath{\Bd \to \jpsi \Kz \Kp\Km}\xspace}
\def\BsToJpsiKzbKK {\ensuremath{\Bs \to \jpsi \Kzb \Kp\Km}\xspace}

\usepackage{multirow}

\usepackage{cite} 
\usepackage{mciteplus}

\DeclareGraphicsExtensions{.pdf,.png,.eps,.mps}

\begin{document}

\renewcommand{\thefootnote}{\fnsymbol{footnote}}
\setcounter{footnote}{1}

\begin{titlepage}
\pagenumbering{roman}

\vspace*{-1.5cm}
\centerline{\large EUROPEAN ORGANIZATION FOR NUCLEAR RESEARCH (CERN)}
\vspace*{1.5cm}
\hspace*{-0.5cm}
\begin{tabular*}{\linewidth}{lc@{\extracolsep{\fill}}r}
\ifthenelse{\boolean{pdflatex}}
{\vspace*{-2.7cm}\mbox{\!\!\!\includegraphics[width=.14\textwidth]{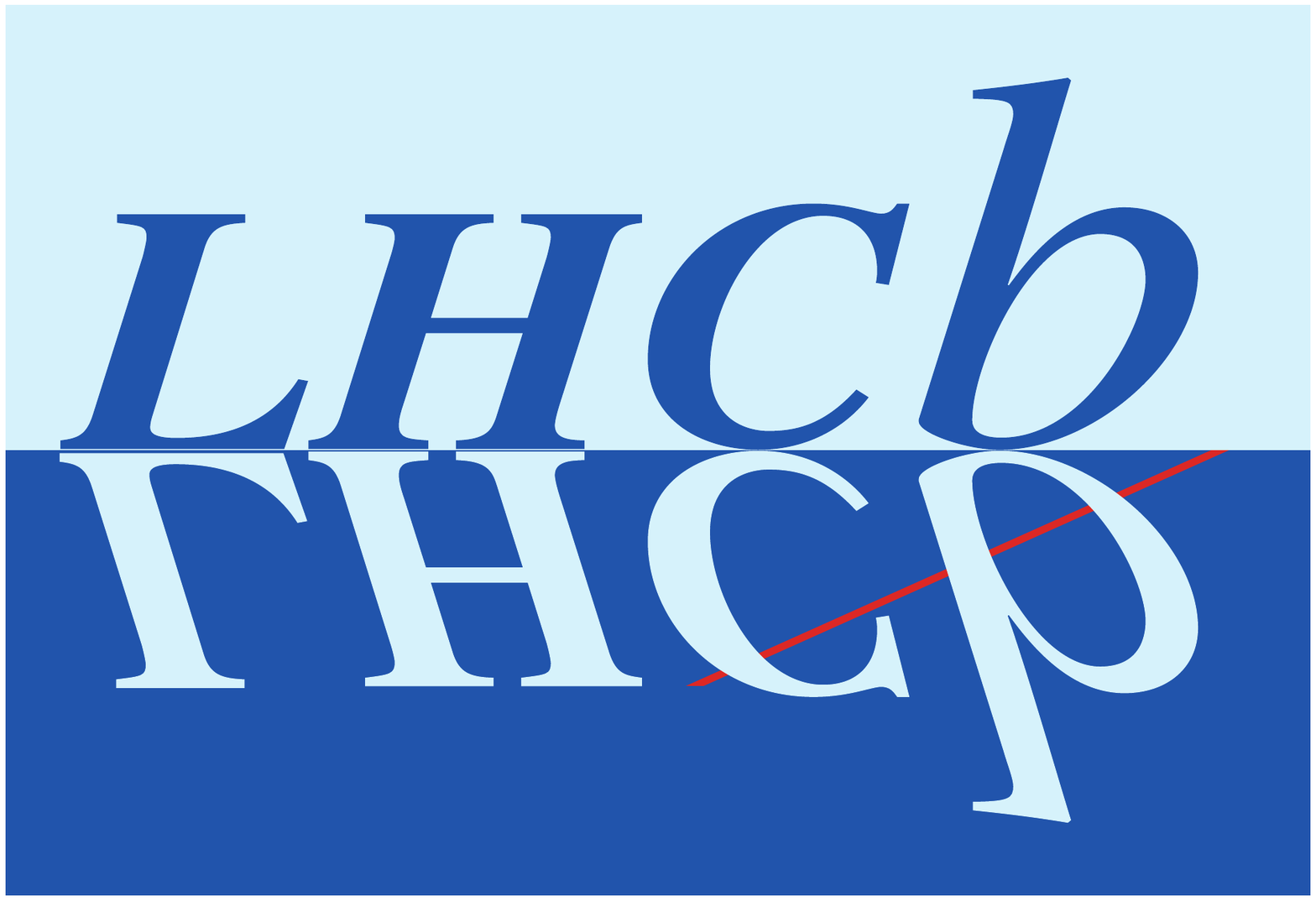}} & &}%
{\vspace*{-1.2cm}\mbox{\!\!\!\includegraphics[width=.12\textwidth]{lhcb-logo.eps}} & &}%
\\
 & & CERN-PH-EP-2014-085 \\  
 & & LHCb-PAPER-2014-016 \\  
 & & \today \\ 
 & & \\
\end{tabular*}

\vspace*{3.0cm}

{\bf\boldmath\huge
\begin{center}
  Observation of the \BsToJpsiKSKpi decay
\end{center}
}

\vspace*{1.5cm}

\begin{center}
  The LHCb collaboration\footnote{Authors are listed on the following pages.}
\end{center}

\vspace{\fill}

\begin{abstract}
  \noindent
  Decays of the form \BxToJpsiKShhp ($h^{(\prime)} = K, \pi$) are searched for in proton-proton collision data corresponding to an integrated luminosity of $1.0 \invfb$ recorded with the \lhcb detector.
  The first observation of the \BsToJpsiKSKpi decay is reported, with significance in excess of 10 standard deviations.
  The \BdToJpsiKSKK decay is also observed for the first time.
  The branching fraction of \BdToJpsiKSpipi is determined, to significantly better precision than previous measurements, using \BdToJpsiKS as a normalisation channel.
  Branching fractions and upper limits of the other \BxToJpsiKShhp modes are determined relative to that of the \BdToJpsiKSpipi decay.
\end{abstract}

\vspace*{1.5cm}

\begin{center}
  Submitted to JHEP
\end{center}

\vspace{\fill}

{\footnotesize 
\centerline{\copyright~CERN on behalf of the \lhcb collaboration, license \href{http://creativecommons.org/licenses/by/3.0/}{CC-BY-3.0}.}}
\vspace*{2mm}

\end{titlepage}


\newpage
\setcounter{page}{2}
\mbox{~}
\newpage

\centerline{\large\bf LHCb collaboration}
\begin{flushleft}
\small
R.~Aaij$^{41}$, 
B.~Adeva$^{37}$, 
M.~Adinolfi$^{46}$, 
A.~Affolder$^{52}$, 
Z.~Ajaltouni$^{5}$, 
J.~Albrecht$^{9}$, 
F.~Alessio$^{38}$, 
M.~Alexander$^{51}$, 
S.~Ali$^{41}$, 
G.~Alkhazov$^{30}$
% ...

{\footnotesize \it
% ...
$ ^{5}$Clermont Universit\'{e}, Universit\'{e} Blaise Pascal, CNRS/IN2P3, LPC, Clermont-Ferrand, France\\
% ...
$ ^{9}$Fakult\"{a}t Physik, Technische Universit\"{a}t Dortmund, Dortmund, Germany\\
% ...
$ ^{30}$Petersburg Nuclear Physics Institute (PNPI), Gatchina, Russia\\
% ...
$ ^{37}$Universidad de Santiago de Compostela, Santiago de Compostela, Spain\\
$ ^{38}$European Organization for Nuclear Research (CERN), Geneva, Switzerland\\
% ...
$ ^{41}$Nikhef National Institute for Subatomic Physics, Amsterdam, The Netherlands\\
% ...
$ ^{46}$H.H. Wills Physics Laboratory, University of Bristol, Bristol, United Kingdom\\
% ...
$ ^{51}$School of Physics and Astronomy, University of Glasgow, Glasgow, United Kingdom\\
$ ^{52}$Oliver Lodge Laboratory, University of Liverpool, Liverpool, United Kingdom\\
% ...
}
\end{flushleft}

\cleardoublepage


\renewcommand{\thefootnote}{\arabic{footnote}}
\setcounter{footnote}{0}


\pagestyle{plain} 
\setcounter{page}{1}
\pagenumbering{arabic}


\section{Introduction}
\label{sec:Introduction}

All current experimental measurements of \CP violation in the quark sector are well described by the Cabibbo-Kobayashi-Maskawa mechanism~\cite{PhysRevLett.10.531, PTP.49.652}, which is embedded in the framework of the Standard Model (SM).
However, it is known that the size of \CP violation in the SM is not sufficient to account for the asymmetry between matter and antimatter observed in the Universe; hence, additional sources of \CP violation are being searched for as manifestations of non-SM physics.

The measurement of the phase $\phi_s \equiv -2\,{\rm arg}\left( -V_{ts} V_{tb}^* / V_{cs} V_{cb}^* \right)$ associated with $\Bs$--$\Bsb$ mixing is of fundamental interest (see, {\it e.g.}, Ref.~\cite{LHCb-PAPER-2012-031} and references therein). 
To date, only the decays $\Bs \to \jpsi \phi$~\cite{Abazov:2011ry,Aaltonen:2012ie,Aad:2012kba,LHCb-PAPER-2011-021,LHCb-PAPER-2013-002}, $\Bs \to \jpsi \pip\pim$~\cite{LHCb-PAPER-2011-031,LHCb-PAPER-2012-006} and $\Bs \to \phi\phi$~\cite{LHCb-PAPER-2013-007} have been used to measure $\phi_s$.
To maximise the sensitivity to all possible effects of non-SM physics, which might affect preferentially states with certain quantum numbers, it would be useful to study more decay processes.
Decay channels involving \jpsi mesons are well-suited for such studies since
the $\jpsi \to \mumu$ decay provides a distinctive experimental signature and
allows a good measurement of the secondary vertex position.
Observation of the decay $\Bs \to \jpsi \pip\pim\pip\pim$, with a significant contribution from the $\jpsi f_1(1285)$ component, has recently been reported by LHCb~\cite{LHCb-PAPER-2013-055}.
There are several unflavoured mesons, including $a_1(1260)$, $f_1(1285)$, $\eta(1405)$, $f_1(1420)$ and $\eta(1475)$, that are known to decay to $\KS K^\pm \pi^\mp$~\cite{PDG2012}, and that could in principle be produced in \Bs decays together with a \jpsi meson.
If such decays are observed, they could be used in future analyses to search for \CP violation.

No measurements exist of the branching fractions of \BxToJpsiKSKpi decays.
The decays \BdToJpsiKSpipi~\cite{Affolder:2001qi,Abe:2001wa,Choi:2011fc} and \BdToJpsiKSKK~\cite{Aubert:2003ii,Jessop:1999cr} have been previously studied, though the measurements of their branching fractions have large uncertainties.
In addition to being potential sources of ``feed-across'' background to \BxToJpsiKSKpi, these decays allow studies of potential exotic charmonia states.
For example, the decay chain $\Bp \to X(3872)\Kp$ with $X(3872) \to \jpsi \pip\pim$ has been observed by several experiments~\cite{Choi:2003ue,Aubert:2004ns,LHCb-PAPER-2013-001}, and it is of interest to investigate if production of the $X(3872)$ state in $\Bd$ decays follows the expectation from isospin symmetry. 
Another reported state, dubbed the $X(4140)$, has been seen in the decay chain $\Bp \to X(4140) \Kp$, $X(4140) \to \jpsi \phi$ by some experiments~\cite{Aaltonen:2009tz,Abazov:2013xda,Chatrchyan:2013dma} but not by others~\cite{LHCb-PAPER-2011-033}, and further experimental studies are needed to understand if the structures in the $\jpsi \phi$ system in $\Bp \to \jpsi \phi \Kp$ decays are of resonant nature.
In addition, the relative production of an isoscalar meson in association with a $\jpsi$ particle in $\Bd$ and $\Bs$ decays can provide a measurement of the mixing angle between the $\frac{1}{\sqrt{2}}\left| u\bar{u}+d\bar{d} \rangle \right.$ and $\left| s\bar{s} \rangle \right.$ components of the meson's wavefunction~\cite{Fleischer:2011au,Fleischer:2011ib,Stone:2013eaa}.
Therefore studies of $\BxToJpsiKSKpi$ decays may provide further insights into light meson spectroscopy.

In this paper, the first measurements of \Bd and \Bs meson decays to $\jpsi \KS \Kpm\pimp$ final states are reported.
All $\jpsi \KS h^+ h^{(\prime) -}$ final states are included in the analysis, where $h^{(\prime)} = K, \pi$. 
The inclusion of charge-conjugate processes is implied throughout the paper.
The \jpsi and \KS mesons are reconstructed through decays to $\mup\mun$ and
$\pip\pim$ final states, respectively.
The analysis strategy is to reconstruct the \B meson decays with minimal bias on their phase-space to retain all possible resonant contributions in the relevant invariant mass distributions.
If contributions from broad resonances overlap, an amplitude analysis will be necessary to resolve them.  
Such a study would require a dedicated analysis to follow the exploratory work reported here.

This paper is organised as follows.
An introduction to the LHCb detector and the data sample used in the analysis is given in Sec.~\ref{sec:detector}, followed by an overview of the analysis procedure in Sec.~\ref{sec:overview}.
The selection algorithms and fit procedure are described in Secs.~\ref{sec:selection} and~\ref{sec:fit}, respectively.
In Sec.~\ref{sec:phase-space} the phase-space distributions of the decay modes with significant signals are shown.
Sources of systematic uncertainty are discussed in Sec.~\ref{sec:systematics} and the results are presented together with a summary in Sec.~\ref{sec:results}.

\section{The LHCb detector}
\label{sec:detector}

The analysis is based on a data sample corresponding to an integrated luminosity of $1.0 \invfb$ of $pp$ collisions at centre-of-mass energy $\sqrt{s} = 7 \tev$ recorded with the \lhcb detector at CERN.
The \lhcb detector~\cite{Alves:2008zz} is a single-arm forward
spectrometer covering the \mbox{pseudorapidity} range $2<\eta <5$,
designed for the study of particles containing \bquark or \cquark
quarks. The detector includes a high-precision tracking system
consisting of a silicon-strip vertex detector (VELO) surrounding the $pp$
interaction region, a large-area silicon-strip detector located
upstream of a dipole magnet with a bending power of about
$4{\rm\,Tm}$, and three stations of silicon-strip detectors and straw
drift tubes~\cite{LHCb-DP-2013-003} placed downstream of the magnet.
The combined tracking system provides a momentum measurement with
relative uncertainty that varies from 0.4\% at low momentum to 0.6\% at 100\gevc,
and impact parameter resolution of 20\mum for
tracks with large transverse momentum, \pt. 
Different types of charged hadrons are distinguished by information
from two ring-imaging Cherenkov detectors~\cite{LHCb-DP-2012-003}. 
Photon, electron and
hadron candidates are identified by a calorimeter system consisting of
scintillating-pad and preshower detectors, an electromagnetic
calorimeter and a hadronic calorimeter. Muons are identified by a
system composed of alternating layers of iron and multiwire
proportional chambers~\cite{LHCb-DP-2012-002}.

The trigger~\cite{LHCb-DP-2012-004} consists of hardware and software stages.
The events selected for this analysis are triggered at the hardware stage by a single muon candidate with $\pt>1.48\gevc$ or a pair of muon candidates with \pt product greater than $(1.296 \gevc)^2$.
In the software trigger, events are initially required to have either two oppositely charged muon candidates with combined mass above $2.7 \gevcc$, or at least one muon candidate or one track with $\pt > 1.8 \gevc$ with impact parameter greater than $100\mum$ with respect to all $pp$ interaction vertices (PVs). 
In the subsequent stage, only events containing $\jpsi \to \mumu$ decays that are significantly displaced from the PVs are retained.
 
Simulated events are used to study the detector response to signal decays and to investigate potential sources of background.
In the simulation, $pp$ collisions are generated using \pythia~\cite{Sjostrand:2006za} with a specific \lhcb configuration~\cite{LHCb-PROC-2010-056}.  
Decays of hadronic particles are described by \evtgen~\cite{Lange:2001uf}, in which final state radiation is generated using \photos~\cite{Golonka:2005pn}. 
The interaction of the generated particles with the detector and its
response are implemented using the \geant toolkit~\cite{Allison:2006ve, *Agostinelli:2002hh} as described in Ref.~\cite{LHCb-PROC-2011-006}.

\section{Analysis overview}
\label{sec:overview}

The main objective of the analysis is to measure the relative branching fractions of the \BxToJpsiKShhp decays.
Since the most precise previous measurement of any of these branching fractions is ${\cal B}(\BdToJpsiKzpipi) = (10.3 \pm 3.3 \pm 1.5) \times 10^{-4}$~\cite{Affolder:2001qi}, where the first uncertainty is statistical and the second is systematic, conversion of relative to absolute branching fractions would introduce large uncertainties.
To alleviate this, a measurement of the branching fraction of \BdToJpsiKSpipi relative to that of \BdToJpsiKS is also performed.
For this measurement the optimisation of the selection criteria is performed based on simulation, whereas for the \BxToJpsiKShhp relative branching fraction measurements, the optimisation procedure uses data.
The two sets of requirements are referred to as ``simulation-based'' and ``data-based'' throughout the paper.
The use of two sets of requirements is to avoid bias in the measurements, since the selection requirements for the yield of the numerator in each branching fraction ratio are optimised on independent samples.
Furthermore, the regions of the invariant mass distributions potentially containing previously unobserved decays were not inspected until after all analysis procedures were established.

The relative branching fractions are determined from
\begin{eqnarray}
  \label{eq:standardBF1}
  \frac{{\cal B}(\BdToJpsiKSpipi)}{{\cal B}(\BdToJpsiKS)} & = & 
  \frac{\epsilon_{\BdToJpsiKS}}{\epsilon_{\BdToJpsiKSpipi}}
  \frac{N_{\BdToJpsiKSpipi}}{N_{\BdToJpsiKS}} \, ,\\
  \label{eq:standardBF2}
  \frac{{\cal B}(\BxToJpsiKShhp)}{{\cal B}(\BdToJpsiKSpipi)} & = &
  \frac{\epsilon_{\BdToJpsiKSpipi}}{\epsilon_{\BxToJpsiKShhp}}
  \left( \frac{f_d}{f_q} \right)
  \frac{N_{\BxToJpsiKShhp}}{N_{\BdToJpsiKSpipi}} \, ,
\end{eqnarray}
where $\epsilon$ represents the total efficiency, including effects from acceptance, trigger, reconstruction, and selection and particle identification requirements.
The relative efficiencies are determined from samples of simulated events, generated with either a phase-space distribution for previously unobserved decay modes, or including known contributions from resonant structures.
The relevant ratio of fragmentation fractions, denoted $f_d/f_q$, is either trivially equal to unity or is taken from previous measurements, $f_s/f_d = 0.259 \pm 0.015$~\cite{LHCb-PAPER-2011-018,LHCb-PAPER-2012-037,LHCb-CONF-2013-011}.
The measured numbers $N$ of decays for each channel are determined from fits to the appropriate invariant mass spectra.
To determine the ratios in Eq.~(\ref{eq:standardBF2}), a simultaneous fit to all $\jpsi\KS h^+h^{(\prime)-}$ final states is used to account for possible feed-across coming from kaon--pion misidentification.
The contribution from $\psitwos$ decays to the $\jpsi\KS\pip\pim$ final state is vetoed, and the veto is inverted to determine the relative branching fraction for \BdToPsitwoSKS using a relation similar to that of Eq.~(\ref{eq:standardBF1}).
In Eq.~(\ref{eq:standardBF2}) effects due to the width difference between mass eigenstates in the \Bs system~\cite{DeBruyn:2012wj} are neglected, since the final states in \BsToJpsiKShhp decays are expected to be \CP mixtures.
(The quantity determined using Eq.~(\ref{eq:standardBF2}) is the time-integrated branching fraction.)

The long lifetime of \KS mesons and the large boost of particles produced in LHC $pp$ collisions causes a significant fraction of \KS decays to occur outside the VELO detector.
Following Refs.~\cite{LHCb-PAPER-2012-035,LHCb-PAPER-2013-015,LHCb-PAPER-2013-042,LHCb-PAPER-2013-061,LHCb-PAPER-2014-006}, two categories are considered: ``long'', where both tracks from the $\KS\to\pip\pim$ decay products contain hits in the VELO, and ``downstream'', where neither does. 
The long candidates have better momentum and vertex resolution, so different selection requirements are imposed for candidates in the two \KS decay categories, and the ratios given in Eqs.~(\ref{eq:standardBF1}) and~(\ref{eq:standardBF2}) are determined independently for each.
These are then combined and the absolute branching fractions obtained by multiplying by the relevant normalisation factor.
Upper limits are set for modes where no significant signal is observed

In addition, the phase-space is inspected for resonant contributions from either exotic or conventional states in channels where significant signals are seen.
The presence or absence of resonances could guide future analyses.
However, no attempt is made to determine the relative production rates of the different possible contributions. 

\section{Selection requirements}
\label{sec:selection}

After a set of preselection requirements to allow \B candidates to be formed, additional criteria are imposed based on the output of a recursive algorithm designed to optimise the signal significance for \BxToJpsiKShhp decays.
For the measurement of the ratio of \BdToJpsiKSpipi and \BdToJpsiKS branching fractions, the same requirements are also applied to \BdToJpsiKS candidates, with the exception of those on variables that are related to the two extra pions in the numerator final state.

To optimise the simulation-based selection, used only for the determination of the relative branching fraction of \BdToJpsiKSpipi and \BdToJpsiKS decays, the algorithm is applied to simulated signal events and to background events in the data.
These background events are taken from invariant mass sideband regions that are not otherwise used in the analysis.
For the tuning of the data-based selection, used for the relative branching
fraction measurements of \BxToJpsiKShhp decays, the properties of the
\BdToJpsiKSpipi decays in data are used instead of simulation, since a highly pure signal can be isolated with loose requirements.
Since the amount of background varies depending on whether each of $h$ and $h^{\prime}$ is a pion or kaon, different requirements are imposed for each final state. 
For both simulation- and data-based selections, different sets of requirements are obtained for long and downstream categories.

In the preselection, the $\jpsi \to \mumu$ decay is reconstructed from two oppositely charged tracks with hits in the VELO, the tracking stations and the muon chambers.
The tracks are required to have $\pt > 500 \mevc$, to be positively identified as muons~\cite{LHCb-DP-2013-001}, to form a common vertex with $\chisq < 16$, and to have an invariant mass within $\pm 80 \mevcc$ of the known \jpsi mass~\cite{PDG2012}. 

The $\KS\to\pip\pim$ decay is reconstructed from pairs of tracks with opposite charge, each with momentum greater than $2 \gevc$, that form a common vertex with $\chisq < 20$.
The mass of the pion pair must be within $\pm30 \mevcc$ of the known \KS mass~\cite{PDG2012}.
When considering the pair under the hypothesis that one of the tracks is a misidentified proton, the invariant mass for candidates in the long (downstream) \KS category must differ by more than $10 \mevcc$ ($25 \mevcc$) from the known \Lz baryon mass~\cite{PDG2012}.

Candidates for the pions and kaons coming directly from the \B decay (referred to as ``bachelor'' tracks) are selected if they have impact parameter \chisqip, defined as the difference in \chisq of the primary $pp$ interaction vertex reconstructed with and without the considered particle, greater than $4$ and $\pt> 250 \mevc$.
They must have momentum less than $100 \gevc$ to allow reliable particle identification, and must not be identified as muons.
Kaons, pions and protons are distinguished using the difference in the natural logarithm of the likelihoods (DLL) obtained from the particle identification subdetectors under the different mass hypotheses for each track~\cite{LHCb-DP-2012-003}.
Bachelor pions are selected with the requirements ${\rm DLL}_{K\pi}<0$ and ${\rm DLL}_{p\pi}<10$, while kaons must satisfy ${\rm DLL}_{K\pi}>2$ and ${\rm DLL}_{pK}<10$.
The particle identification efficiencies, determined from control samples of $D^0 \to \Km\pip$ decays reweighted to match the kinematic properties of the signal, are found to range from around 73\% for \BxToJpsiKSpipi to around 93\% for \BxToJpsiKSKK decays.
The bachelor candidates are required to form a vertex with $\chisq < 10$.

The \B candidates are reconstructed using a kinematic fit~\cite{Hulsbergen:2005pu} to their decay products, including the requirements that the \B meson is produced at a PV and that the \jpsi and \KS decay products combine to the known masses of those mesons~\cite{PDG2012}.
Candidates with invariant mass values between 5180 and $5500 \mevcc$ are retained for the fits to determine the signal yields described in Sec.~\ref{sec:fit}.

The recursive algorithm tunes requirements on a number of variables that are found to discriminate between signal and background and that are not strongly correlated.
The most powerful variables are found to be the significance of the separation of the \KS vertex from the PV for the long category and the \B candidate \chisqip.
The other variables are the following:
the \B, \jpsi and \KS candidates' vertex \chisq p-values;
the \jpsi and \KS candidates' and the bachelor tracks' \chisqip values;
the separation of the \jpsi vertex from the PV divided by its uncertainty;
the angle between the \B momentum vector and the vector from the PV to the \B decay vertex;
and the \B candidate \pt.
These variables are found to be only weakly correlated with the \B candidate mass or the position in the phase-space of the decay.
For the simulation-based selection, the efficiency of the requirements relative to those made during preselection is around 50\%.  
For the data-based selection the corresponding value is between around 40\% for \BxToJpsiKSpipi and around 55\% for \BxToJpsiKSKK decays, where the background is low due to the particle identification requirements and the narrow signal peak.
The efficiency of the requirement that the \B meson decay products lie within
the detector acceptance also depends on the final state, ranging from around 10\% for \BxToJpsiKSpipi to almost 15\% for \BxToJpsiKSKK decays.

Backgrounds may arise from decays of $b$ baryons.
In addition to decay modes where the \KS meson is replaced by a \Lz baryon, which are removed by the veto described above, there may be decays such as $\Lb \to \jpsi \KS p h^-$, which have the same final state as the signal under consideration except that a kaon or pion is replaced by a proton.
There is currently no measurement of such decays that could enable the level of potential background to be assessed, though the yields observed in the $\Lb\to\jpsi\proton\Km$ channel~\cite{LHCb-PAPER-2013-032,LHCb-PAPER-2014-003} suggest that it may not be negligible.
Therefore, this background is vetoed by recalculating the candidate mass under the appropriate mass hypothesis for the final state particles and removing candidates that lie within $\pm 25 \mevcc$ of the known \Lb mass~\cite{PDG2012}.

In the $\jpsi\KS\pip\pim$ final state, the $\pip\pim$ system could potentially arise from a \KS meson that decays close to the \B candidate vertex.  
This background is removed by requiring that the $\pip\pim$ invariant mass is more than $25 \mevcc$ from the known \KS mass~\cite{PDG2012}.
In addition, in \BdToJpsiKSpipi decays, there is a known contribution from the decay chain $\BdToPsitwoSKS$, $\psi(2S)\to\jpsi\pip\pim$.  
There could potentially be a similar contribution in the \Bs decay to the same final state.  
Such decays are removed from the sample by vetoing candidates with invariant masses of the $\jpsi\pip\pim$ system within $\pm 15 \mevcc$ of the known $\psi(2S)$ mass~\cite{PDG2012}.

In around 2\% of events retained after all criteria are applied, more than one candidate is selected.
A pseudorandom algorithm is used to select only a single candidate from these events.

\section{Determination of signal yields}
\label{sec:fit}

\begin{figure}[!t]
  \centering
  \includegraphics[width=.49\textwidth]{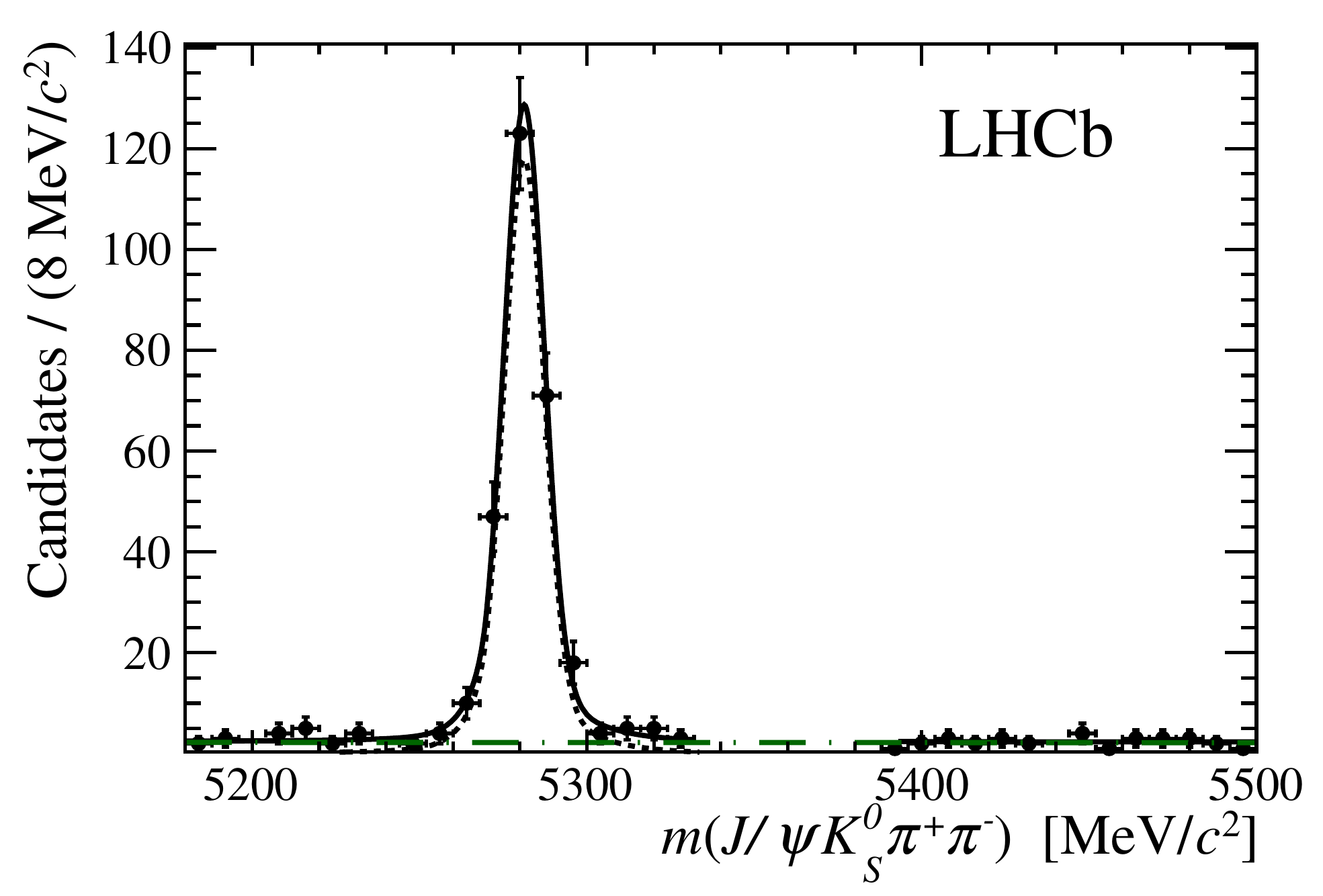}
  \includegraphics[width=.49\textwidth]{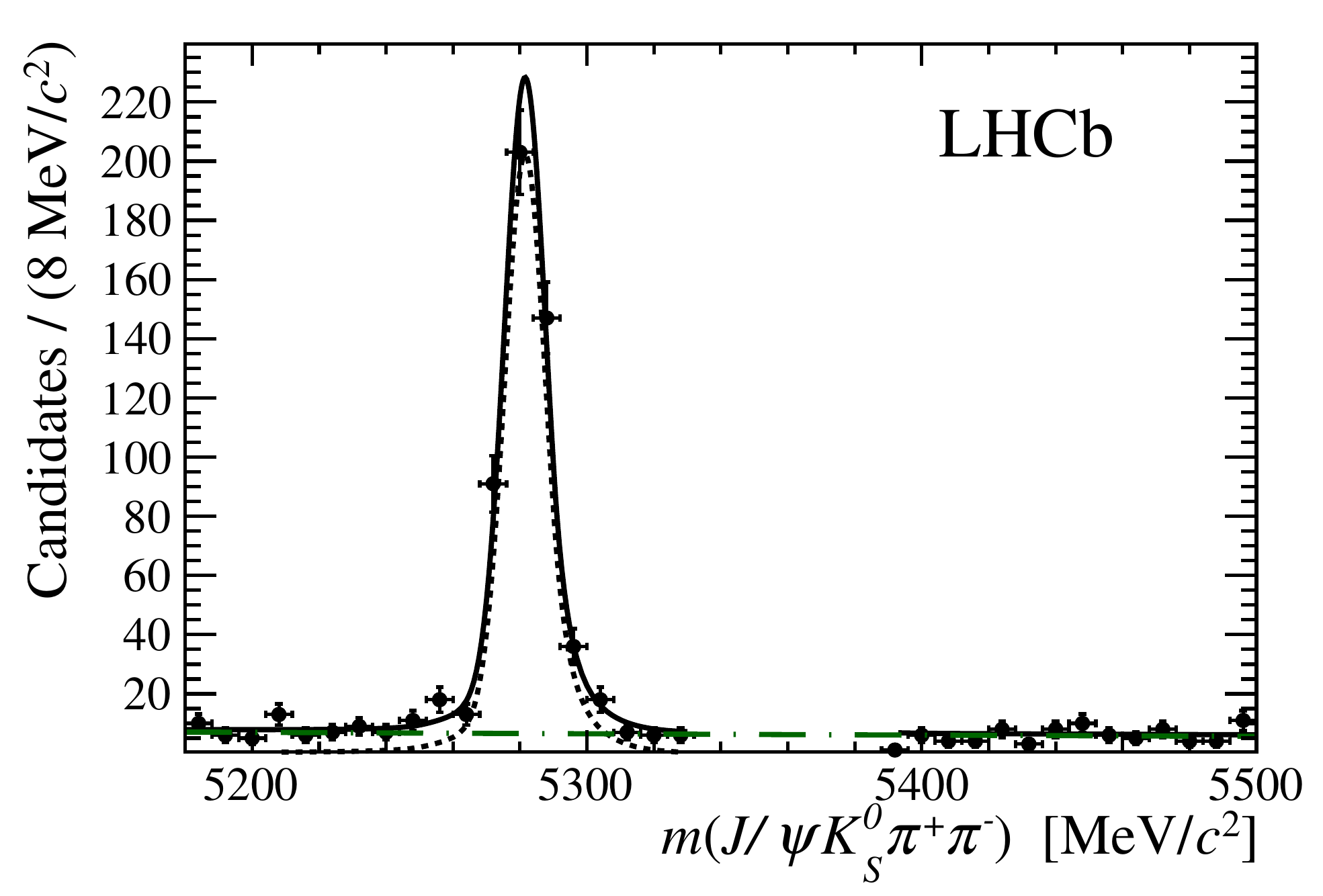}
  \caption{\small 
    Invariant mass distributions of (left) long and (right) downstream \BdToJpsiKSpipi candidates with simulation-based selection, with fit projections overlaid.
    The solid line shows the total fit result, while the dashed line shows the signal component and the dot-dashed line shows the combinatorial background.
    The \Bs region is not examined in these fits.
  }
  \label{fig:BToJpsiKSpipi-partI}
\end{figure}

After all selection requirements are applied, the only sources of candidates in the selected invariant mass ranges are expected to be signal decays, feed-across from \BxToJpsiKShhp decays with kaon--pion misidentification, and combinatorial background.  
The suppression to negligible levels of other potential sources of background, such as $b$ baryon decays, is confirmed with simulation.
For each mode, the ratios of yields under the correct particle identification hypothesis and as feed-across are found to be at the few percent level from the kaon and pion control samples from $\Dz \to \Km\pip$ decays reweighted to the appropriate kinematic distributions.
The feed-across contribution can therefore be neglected in the fit to the $\jpsi\KS\pip\pim$ final state, as is done in the fit to the candidates passing the simulation-based selection, shown in Fig.~\ref{fig:BToJpsiKSpipi-partI}.

\begin{figure}[!t]
  \centering
  \includegraphics[width=.49\textwidth]{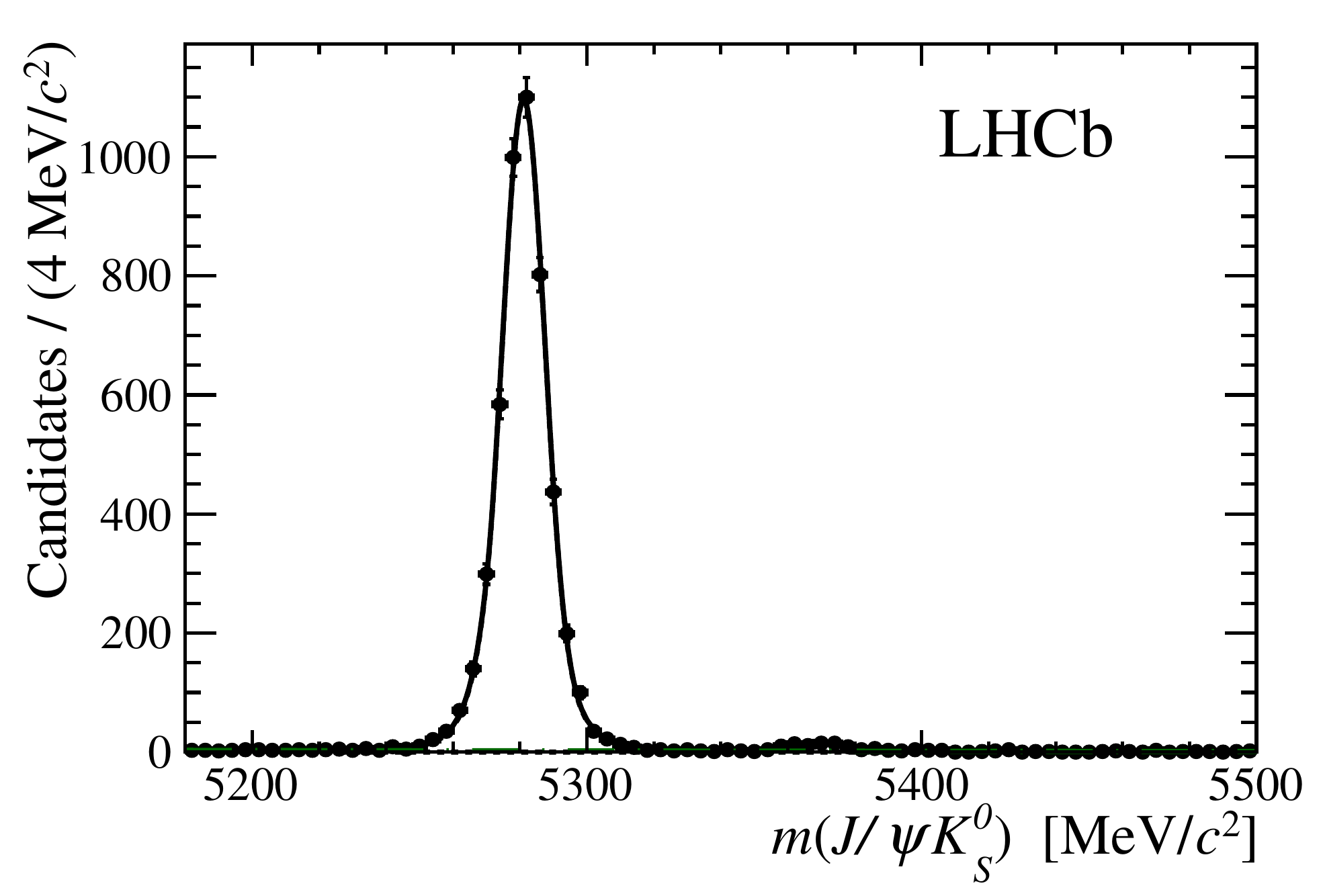}
  \includegraphics[width=.49\textwidth]{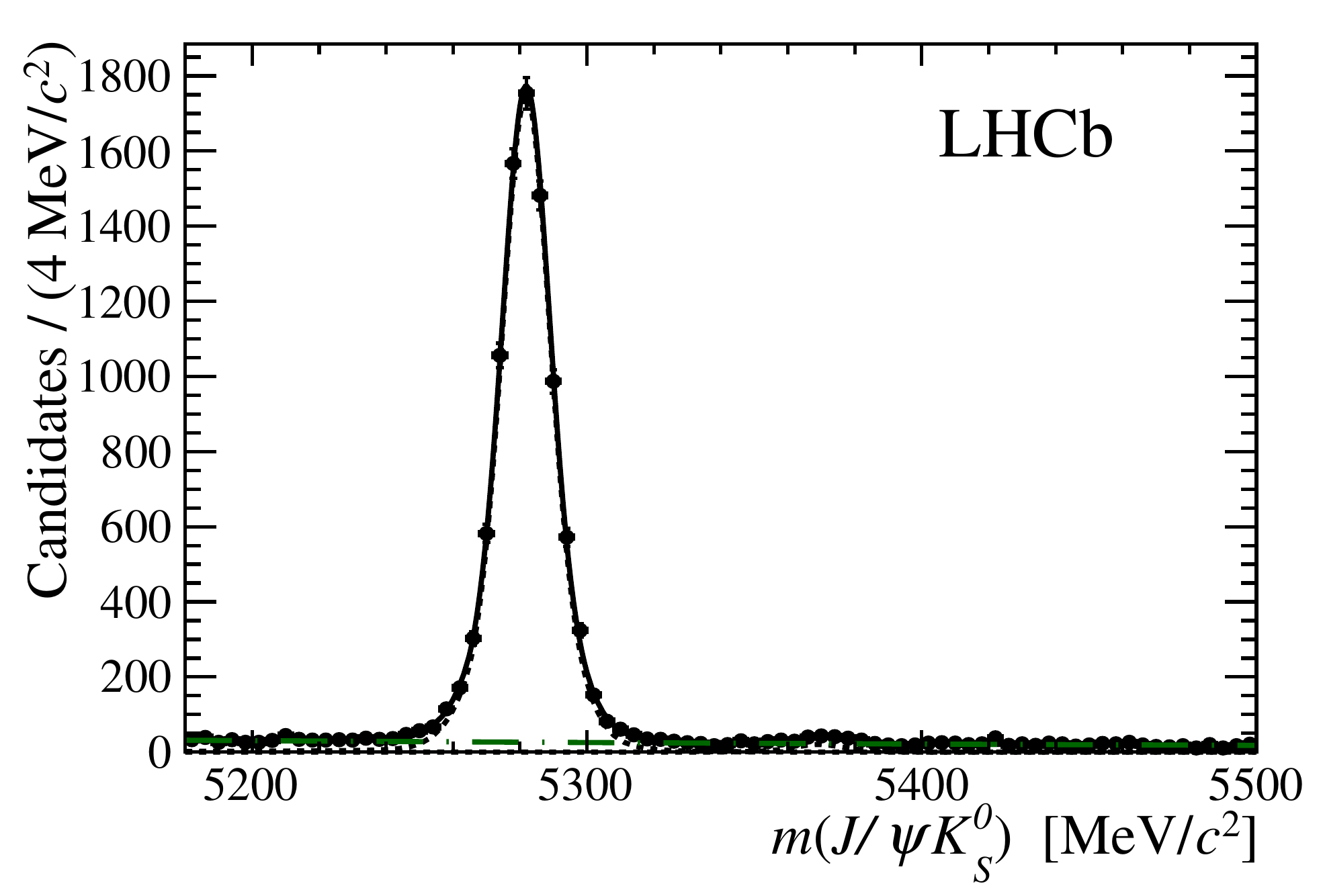}
  \caption{\small 
    Invariant mass distributions of (left) long and (right) downstream \BxToJpsiKS candidates with simulation-based selection, with fit projections overlaid.
    The solid line shows the total fit result, while the dashed line shows the signal component and the dot-dashed line shows the combinatorial background.
  }
  \label{fig:BToJpsiKS}
\end{figure}

The signal shape is parametrised in the same way for all \BxToJpsiKShhp and \BxToJpsiKS decays, and follows the approach used in Ref.~\cite{LHCb-PAPER-2013-015}.
Namely, the signal is described with the sum of two Crystal Ball functions~\cite{Skwarnicki:1986xj} with common mean and independent tails on opposite sides of the peak.
This shape is found to give an accurate description of simulated signal decays.
In the fit to data, the tail parameters are fixed according to  values determined from simulation.
The mean and the widths as well as the relative normalisation of the two Crystal Ball functions are allowed to vary freely in the fit to data. 
The \Bs region is excluded from the fit to candidates passing the simulation-based selection in the $\jpsi\KS\pip\pim$ final state.
In the fit to the $\jpsi\KS$ candidates, shown in Fig.~\ref{fig:BToJpsiKS}, a \Bs component is included with shape identical to that for the \Bd decays except with mean value shifted by the known value of the $\Bs$--$\Bd$ mass difference~\cite{PDG2012}. 

\begin{figure}[!t]
  \centering
  \includegraphics[width=.49\textwidth]{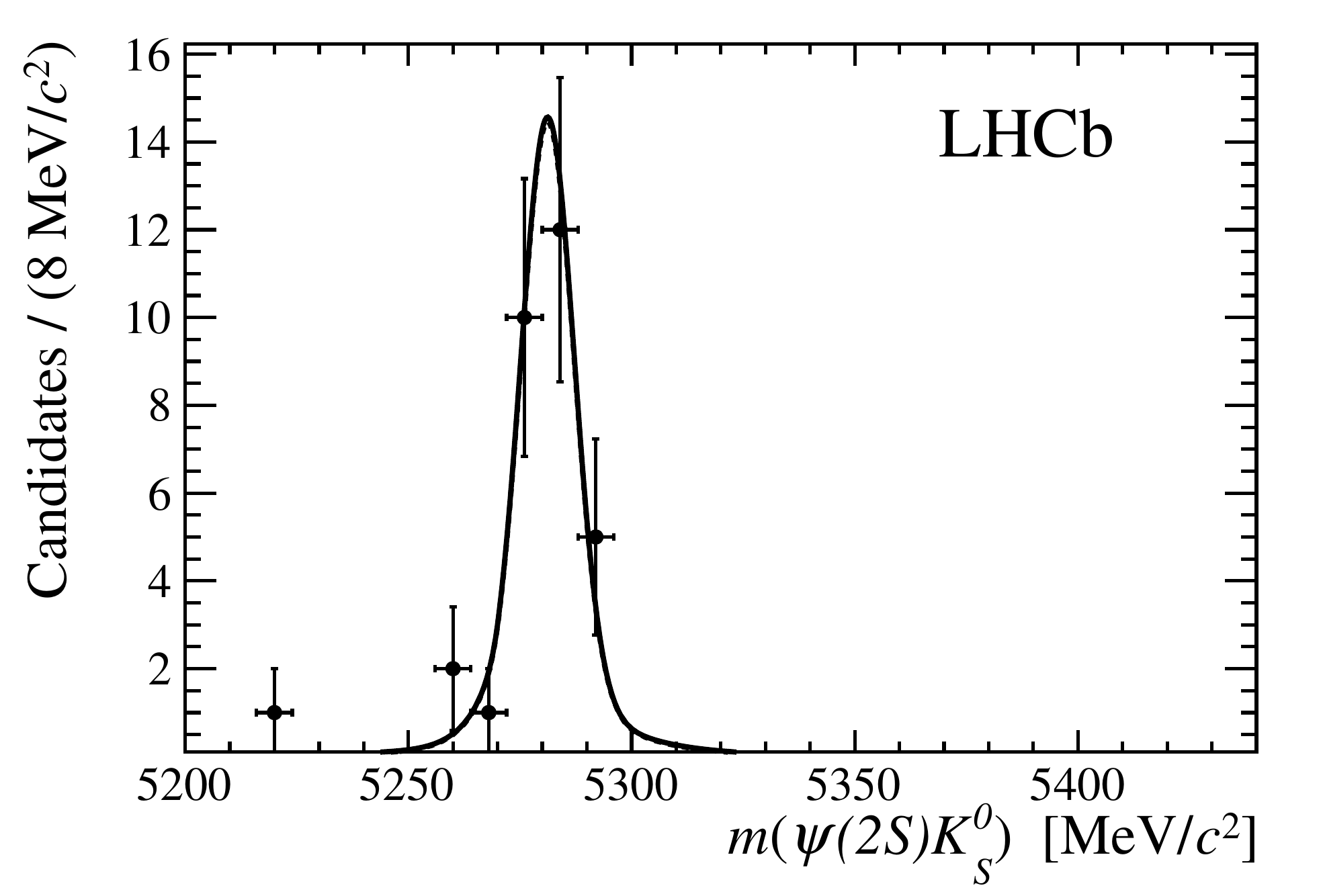}
  \includegraphics[width=.49\textwidth]{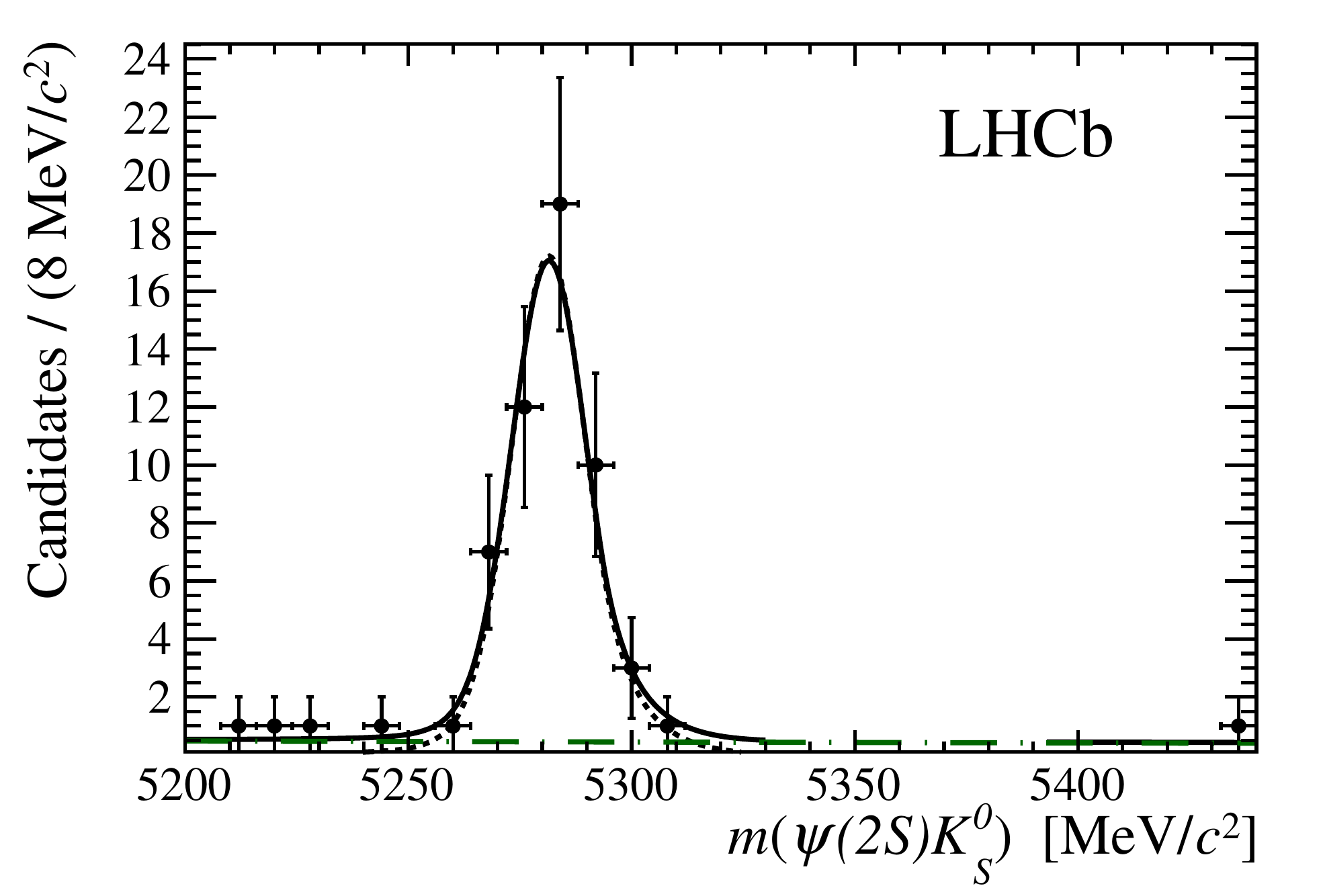}
  \caption{\small 
    Invariant mass distributions of (left) long and (right) downstream \BdToPsitwoSKS candidates with simulation-based selection, with fit projections overlaid.
    The solid line shows the total fit result, while the dashed line shows the signal component and the dot-dashed line shows the combinatorial background.
    The \Bs region is not examined in these fits.
  }
  \label{fig:BTopsi2SKS}
\end{figure}

The signal yields are obtained from extended unbinned maximum likelihood fits to the mass distributions of the reconstructed candidates.
Independent fits are carried out for candidates in the long and downstream categories.
In addition to the signal components, an exponential function is included to describe the combinatorial background with both yield and slope parameter allowed to vary freely.
The results of the fits to the $\jpsi\KS\pip\pim$ and $\jpsi\KS$ invariant mass distributions are summarised in Table~\ref{table:FitResultsLoose}.
The ratio of \BsToJpsiKS and \BdToJpsiKS yields is consistent with that found in a dedicated study of those channels~\cite{LHCb-PAPER-2013-015}.
Also included in Table~\ref{table:FitResultsLoose} are the results of fits to the \BxToJpsiKSpipi sample with the $\psi(2S)$ veto inverted to select candidates consistent with \BxToPsitwoSKS decays, shown in Fig.~\ref{fig:BTopsi2SKS}.
These fits provide a consistency check of the analysis procedures, since the measured ratio of the $\BdToPsitwoSKS$ and $\BdToJpsiKS$ branching fractions can be compared to its known value~\cite{PDG2012}.
For consistency with the fit to \BxToJpsiKSpipi candidates, the \Bs region is not examined in these fits.

\begin{table}[!tb]
  \begin{center}
    \caption{\small
      Yields determined from the fits to the \BxToJpsiKSpipi, \BxToJpsiKS and \BxToPsitwoSKS samples with simulation-based selection.
    }
    \label{table:FitResultsLoose}
    \begin{tabular}{l@{\hspace{5mm}}cc@{\hspace{5mm}}cc@{\hspace{5mm}}cc}
      \hline \\ [-2.5ex]
                 & \multicolumn{2}{c}{\BxToJpsiKSpipi} & \multicolumn{2}{c}{\BxToJpsiKS} & \multicolumn{2}{c}{\BxToPsitwoSKS} \\
                 & long         & downstream   & long          & downstream   & long          & downstream \\
    \hline \\ [-2.5ex]
    $N_{\Bd}$     & $269 \pm 18$ & $483 \pm 26$ & $4869 \pm 71$ & $9870 \pm 107$ & $25 \pm 6$ & $41 \pm 9$ \\ [0.3ex]
    $N_{\Bs}$     & ---          & ---          & $\phanii75 \pm 10$   & $115 \pm 20$   & ---        & ---        \\ [0.3ex]
    \hline
    \end{tabular}
  \end{center}
\end{table}

The fit to the sample selected with data-based criteria is similar to that for the sample selected with simulation-based criteria, but with some important differences.
Signal shapes are included for both \Bd and \Bs decays to each of the final states considered.
The signal components are described with the same sum of two Crystal Ball functions as used in the fits to the sample selected with simulation-based criteria, with tail parameters fixed according to values determined from simulation.
For each final state, the shape of the \Bs component is identical to that for the \Bd decays, with mean value shifted by the known value of the $\Bs$--$\Bd$ mass difference~\cite{PDG2012}.
To reduce the number of freely varying parameters in the fit, the relative widths of the signal shapes in the final states with long and downstream \KS candidates are constrained to be identical for all signal components.
The combinatorial background is modelled as a linear function, rather than the exponential model used in the fits to the samples obtained from the simulation-based selection.
The use of the linear shape is found to make the fit more stable in channels with low background yields, such as \BxToJpsiKSKK, and it is preferable to use the same shape for all channels in the simultaneous fit.
The linear function has independent parameters in each final state.
A single extended unbinned maximum likelihood fit is performed for the long and downstream categories, with all final states fitted simultaneously.
This procedure allows the amount of each feed-across contribution to be constrained according to the observed yields and known misidentification rates.
The shapes of the feed-across contributions are described with kernel functions~\cite{Cranmer:2000du} obtained from simulation.
All correlations between fitted yields are found to be less than 10\% and are neglected when determining the branching fraction ratios.

\begin{figure}[!t]
  \centering
  \includegraphics[width=.49\textwidth]{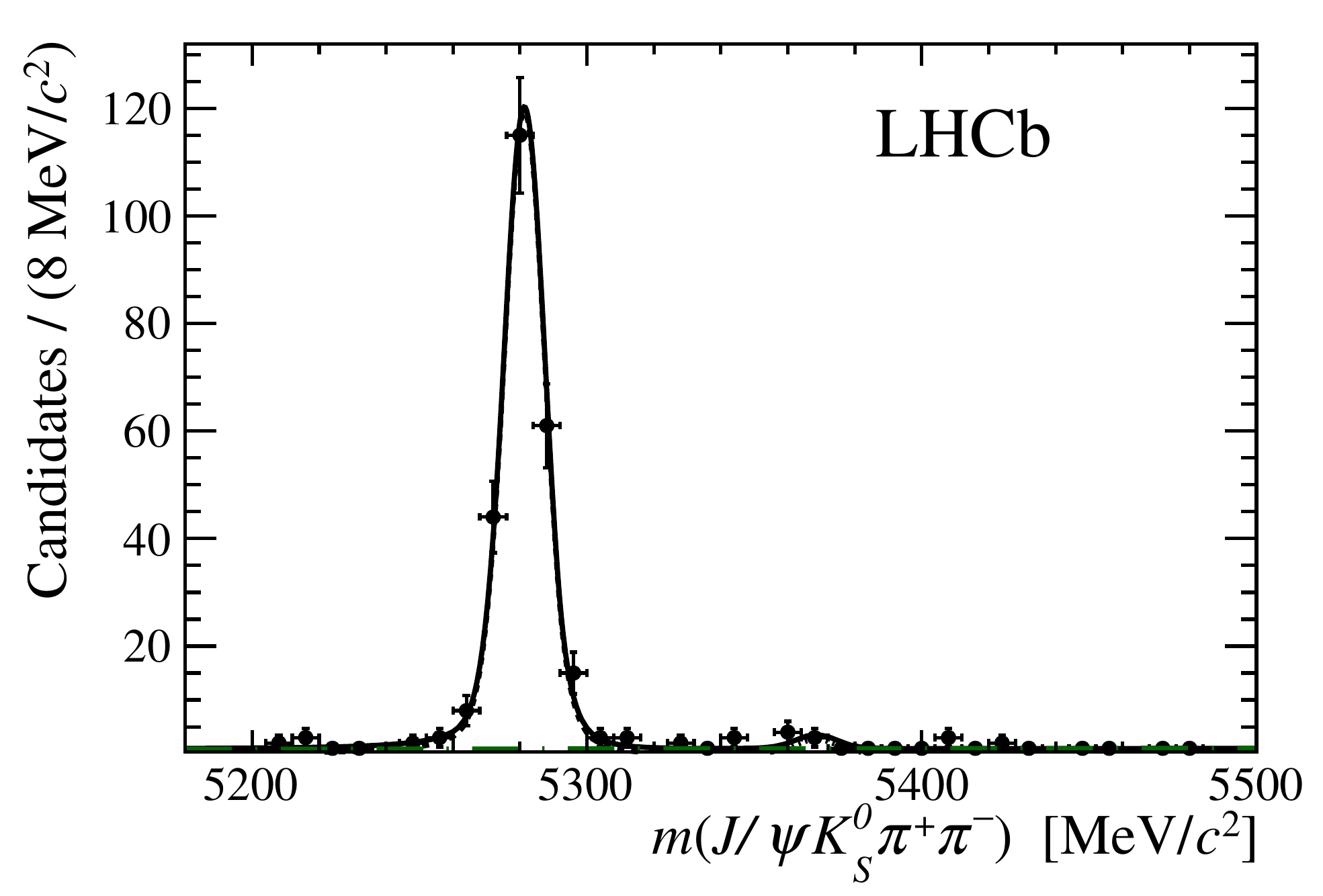}
  \includegraphics[width=.49\textwidth]{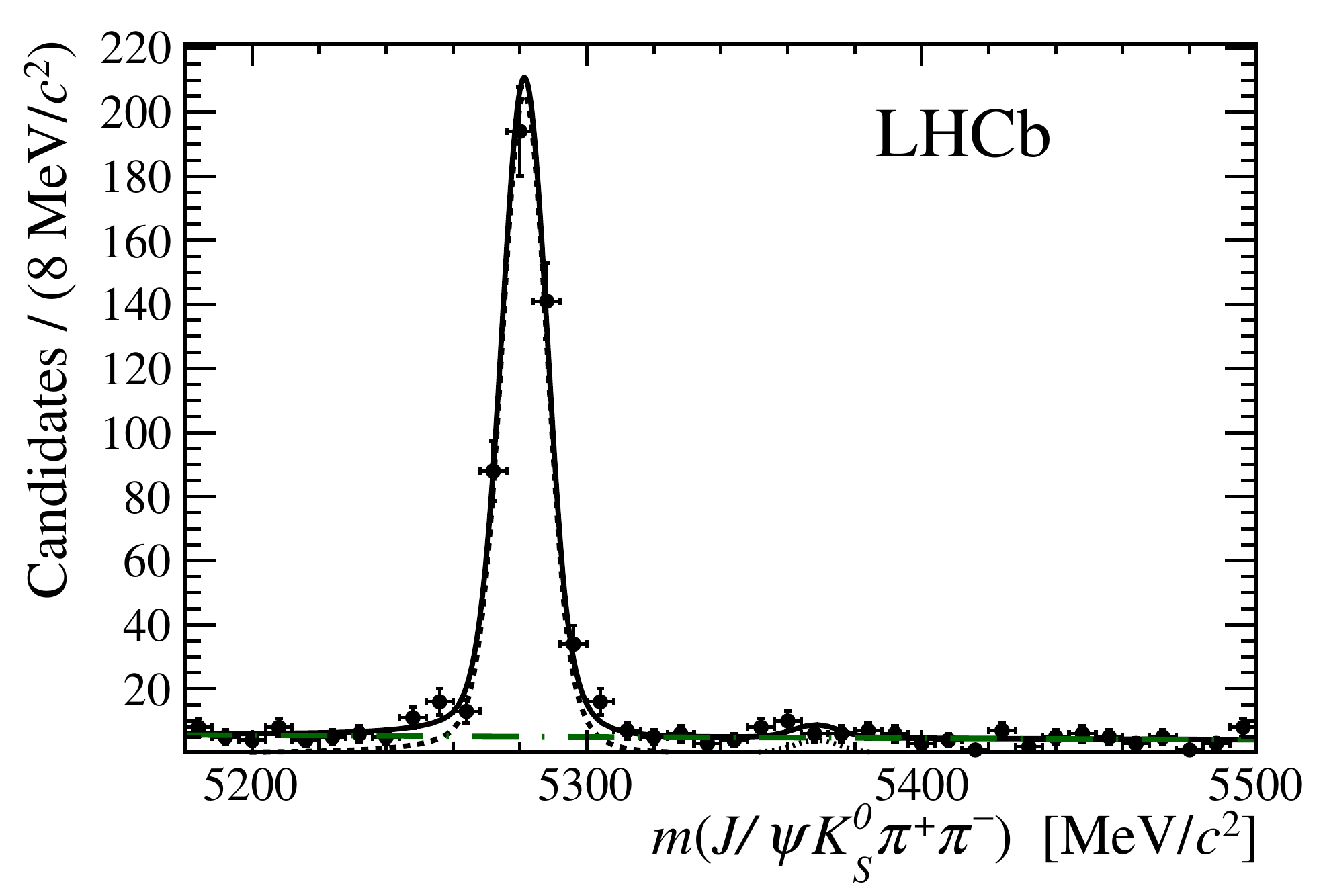}
\includegraphics[width=.49\textwidth]{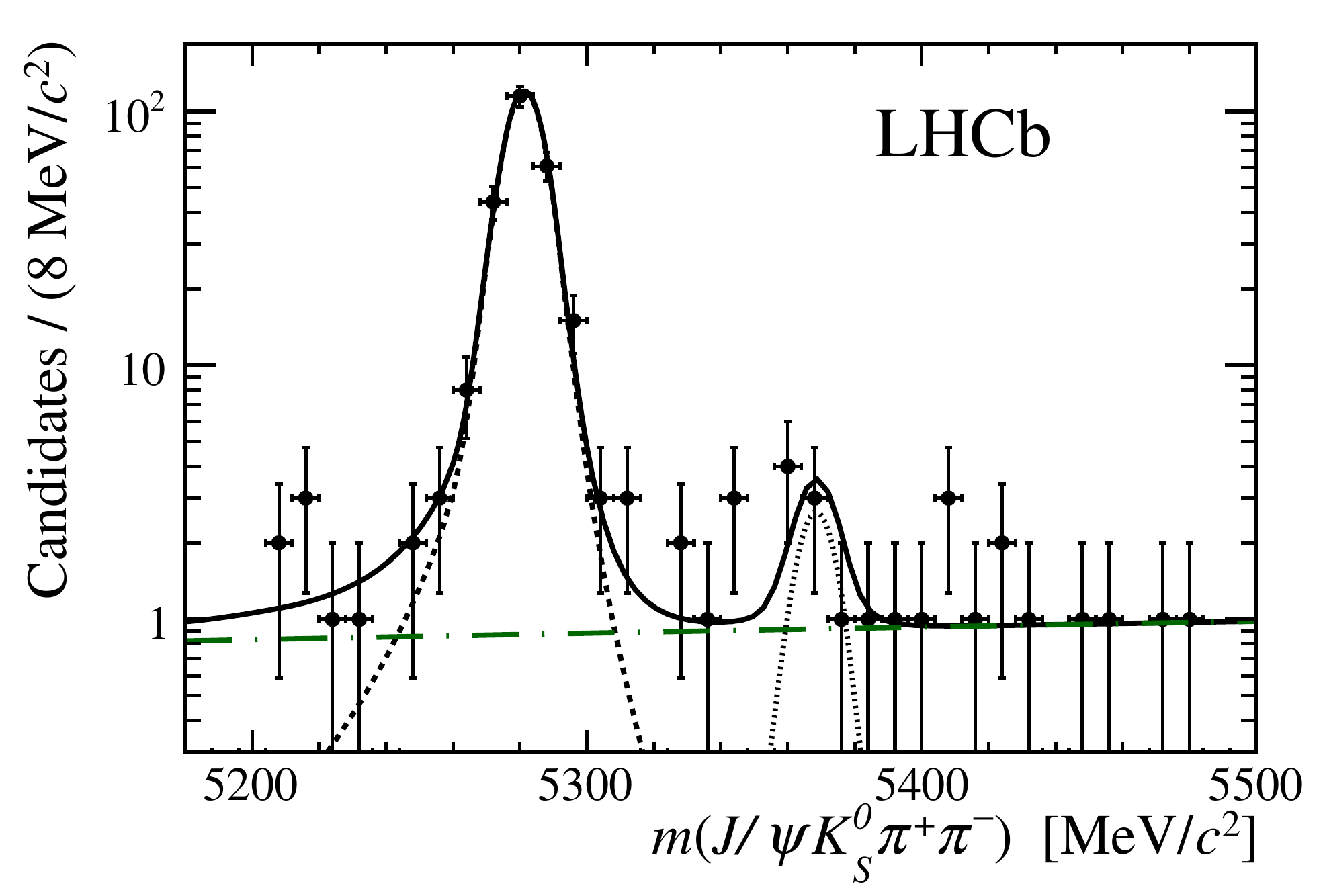}
\includegraphics[width=.49\textwidth]{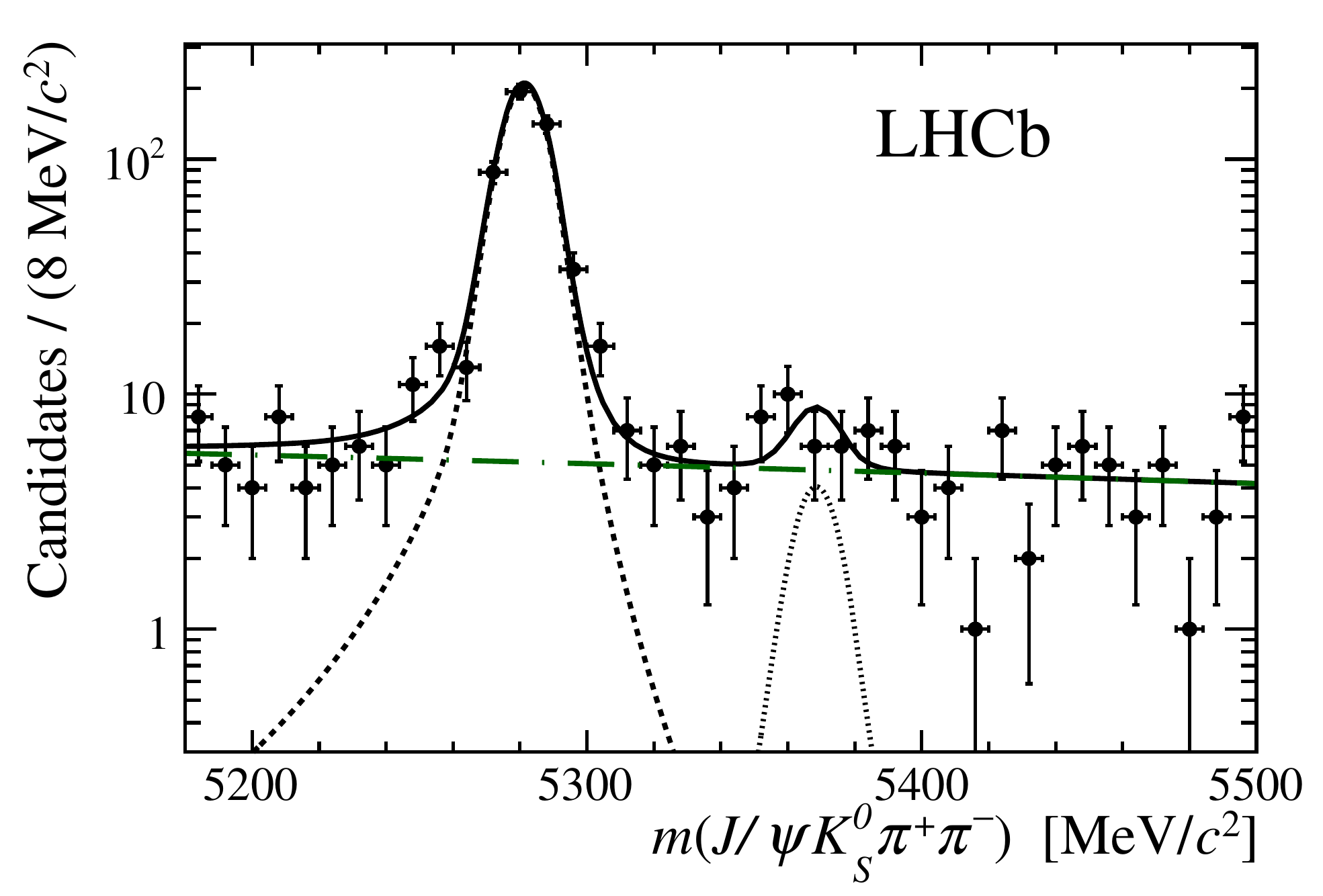}
  \caption{\small 
    Invariant mass distributions of (left) long and (right) downstream \BxToJpsiKSpipi candidates, with data-based selection, shown with (top) linear and (bottom) logarithmic $y$-axis scales, with fit projections overlaid.
    The solid line shows the total fit result, while the dashed and dotted
    lines show the \Bd and \Bs signal components, respectively, and the dot-dashed line shows the combinatorial background.
  }
  \label{fig:BToJpsiKSpipi}
\end{figure}

\begin{figure}[!t]
  \centering
  \includegraphics[width=.49\textwidth]{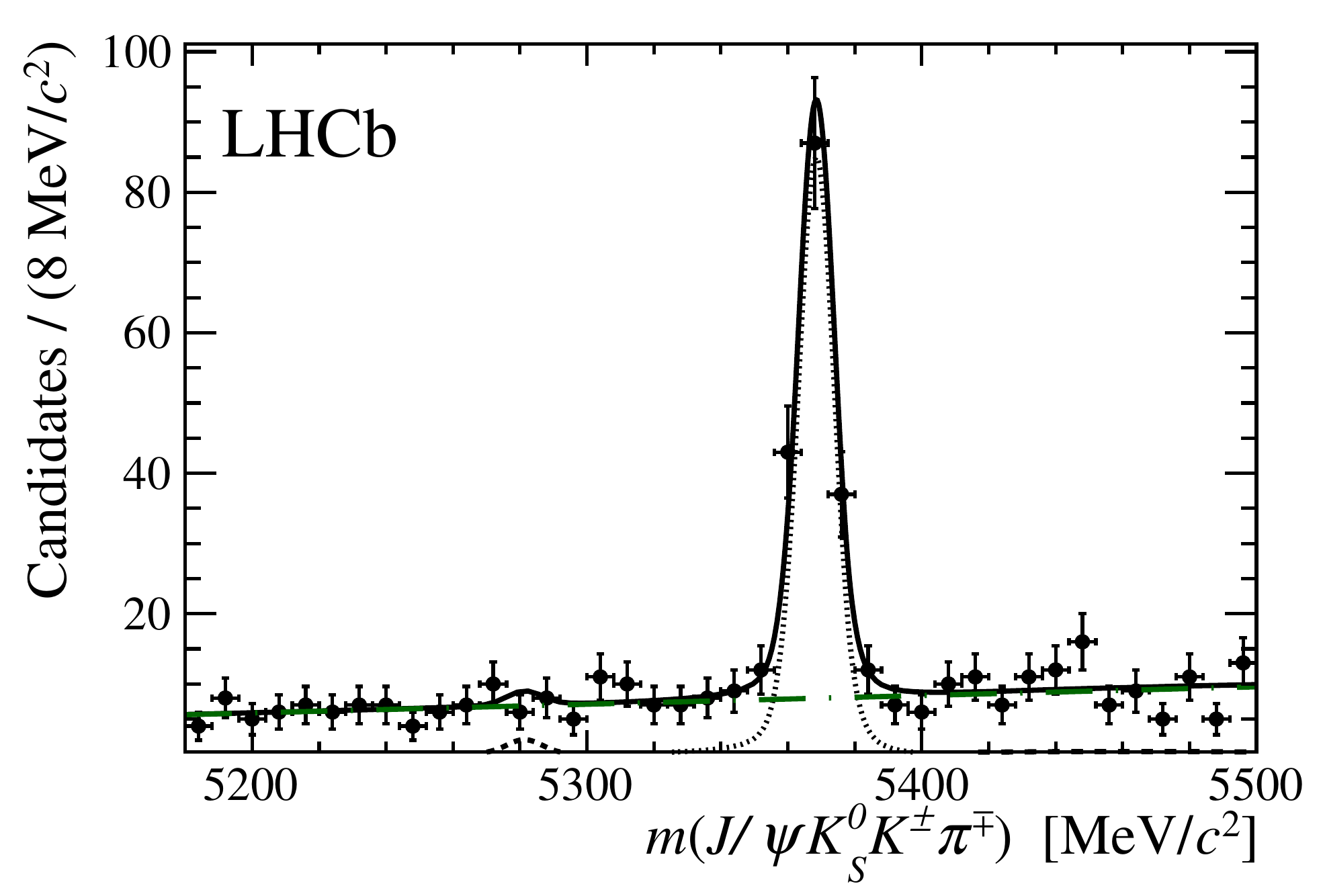}
  \includegraphics[width=.49\textwidth]{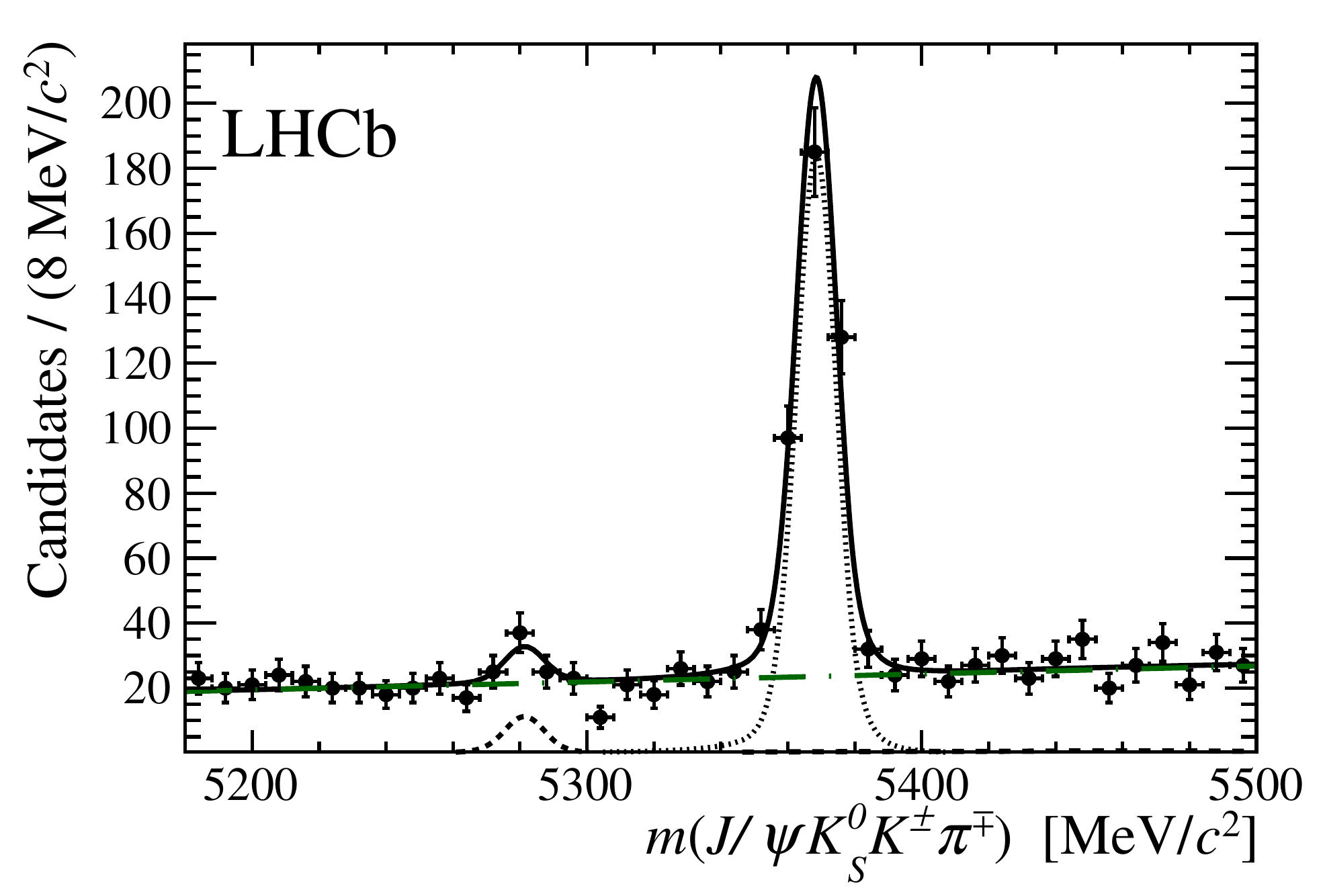}
\includegraphics[width=.49\textwidth]{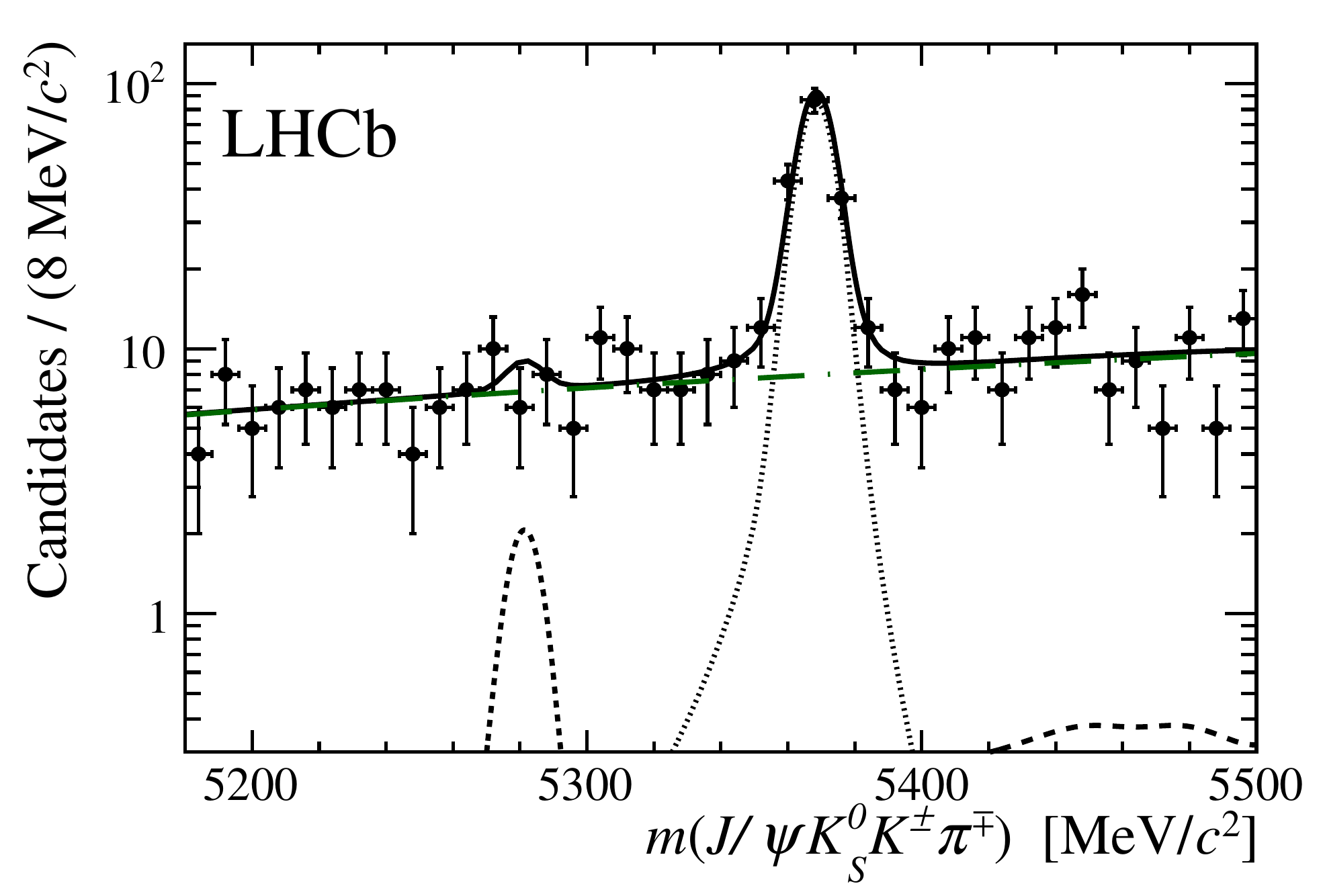}
\includegraphics[width=.49\textwidth]{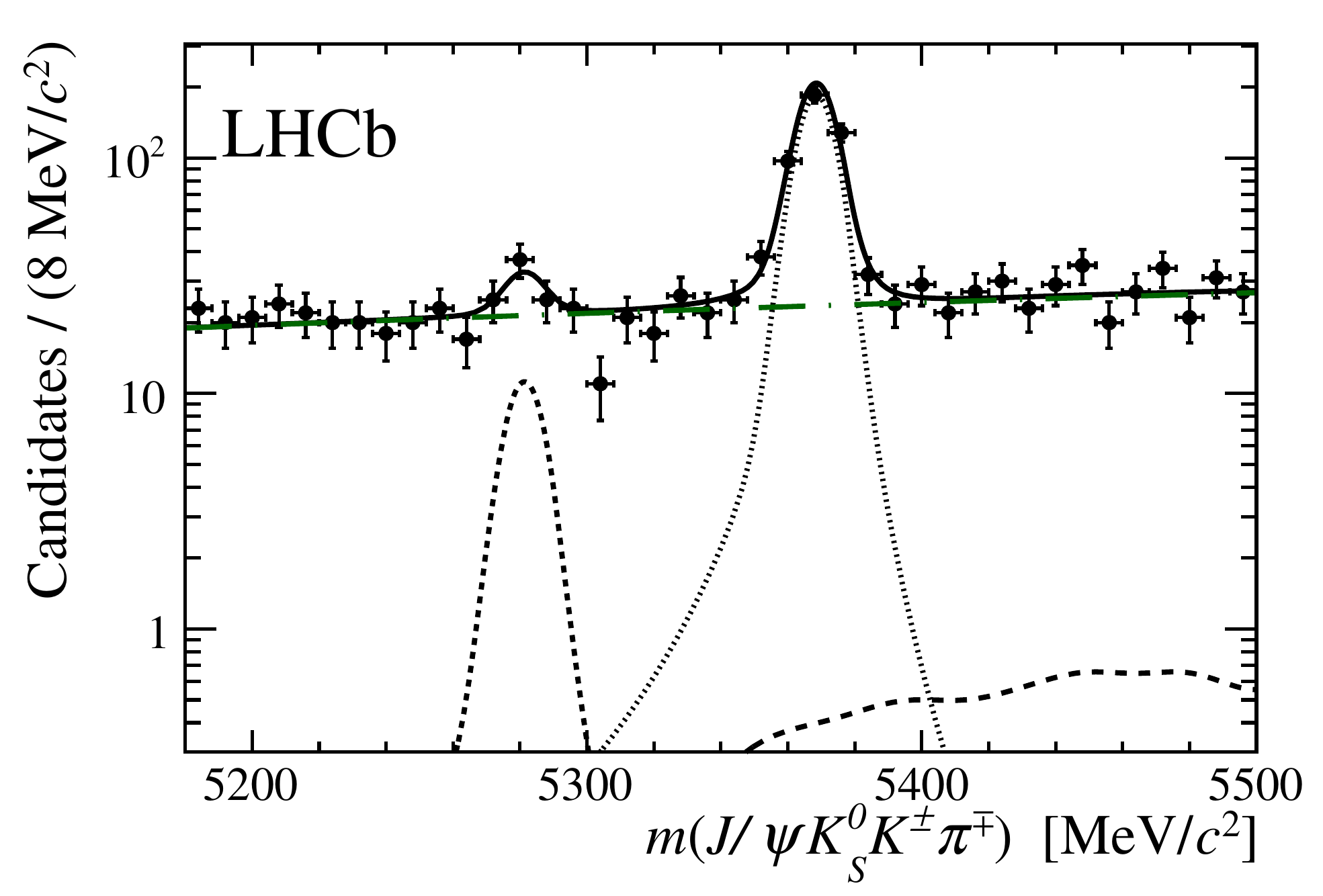}
  \caption{\small 
    Invariant mass distributions of (left) long and (right) downstream \BxToJpsiKSKpi candidates, with data-based selection, shown with (top) linear and (bottom) logarithmic $y$-axis scales, with fit projections overlaid.
    The solid line shows the total fit result, while the dashed and dotted
    lines show the \Bd and \Bs signal components, respectively, the long-dashed line shows the feed-across contribution and the dot-dashed line shows the combinatorial background.
  }
  \label{fig:BToJpsiKSKpi}
\end{figure}

\begin{figure}[!t]
  \centering
  \includegraphics[width=.49\textwidth]{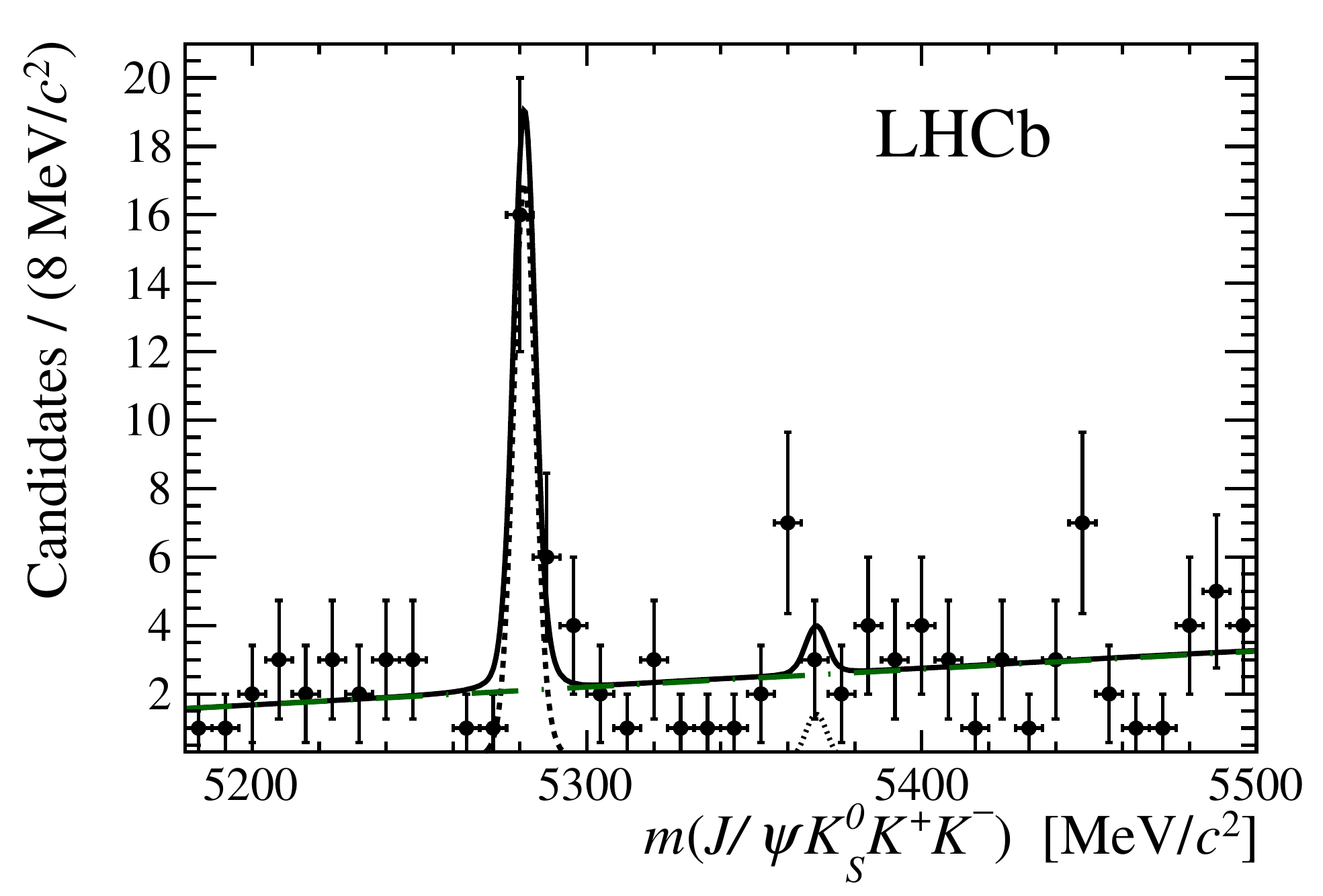}
  \includegraphics[width=.49\textwidth]{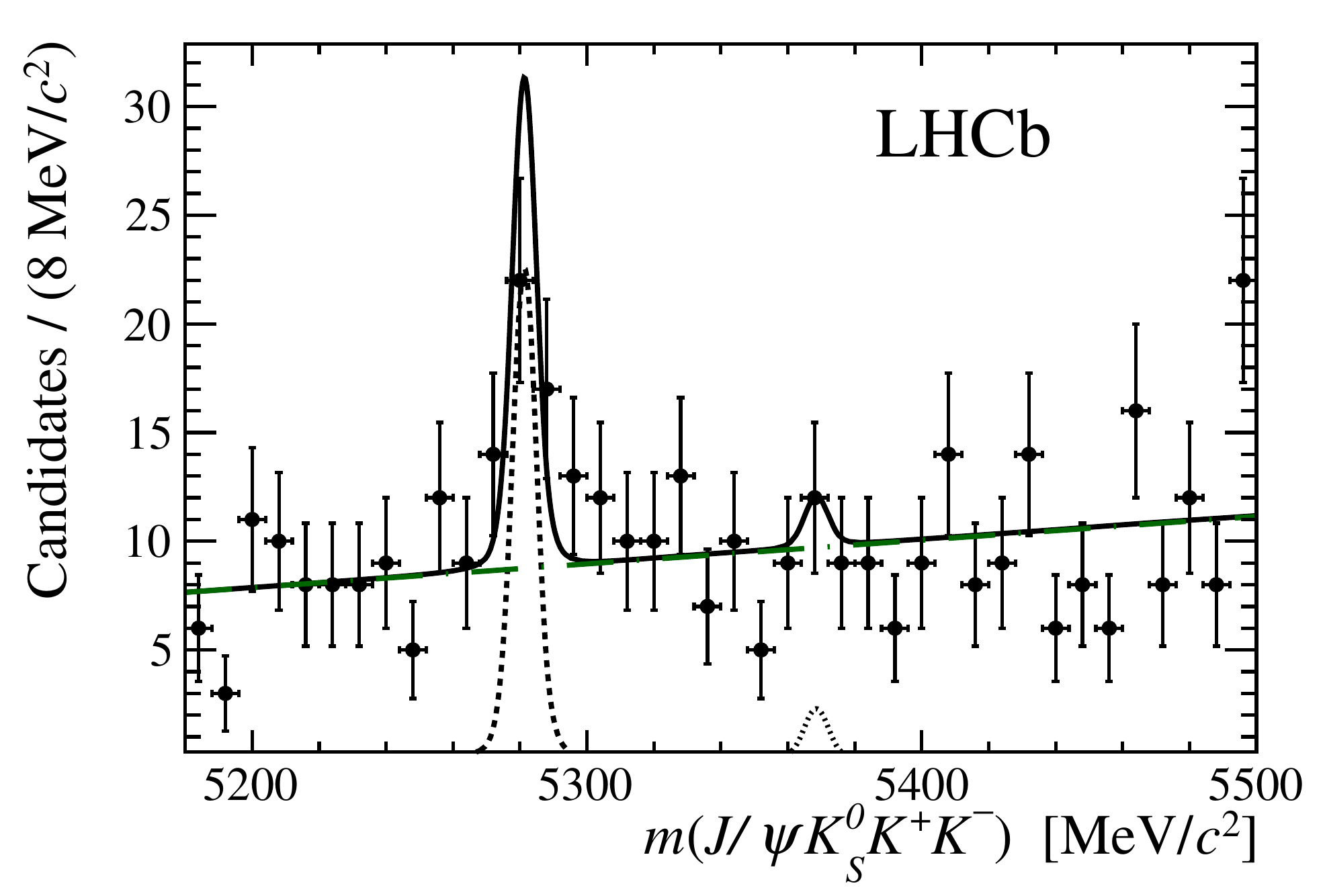}
\includegraphics[width=.49\textwidth]{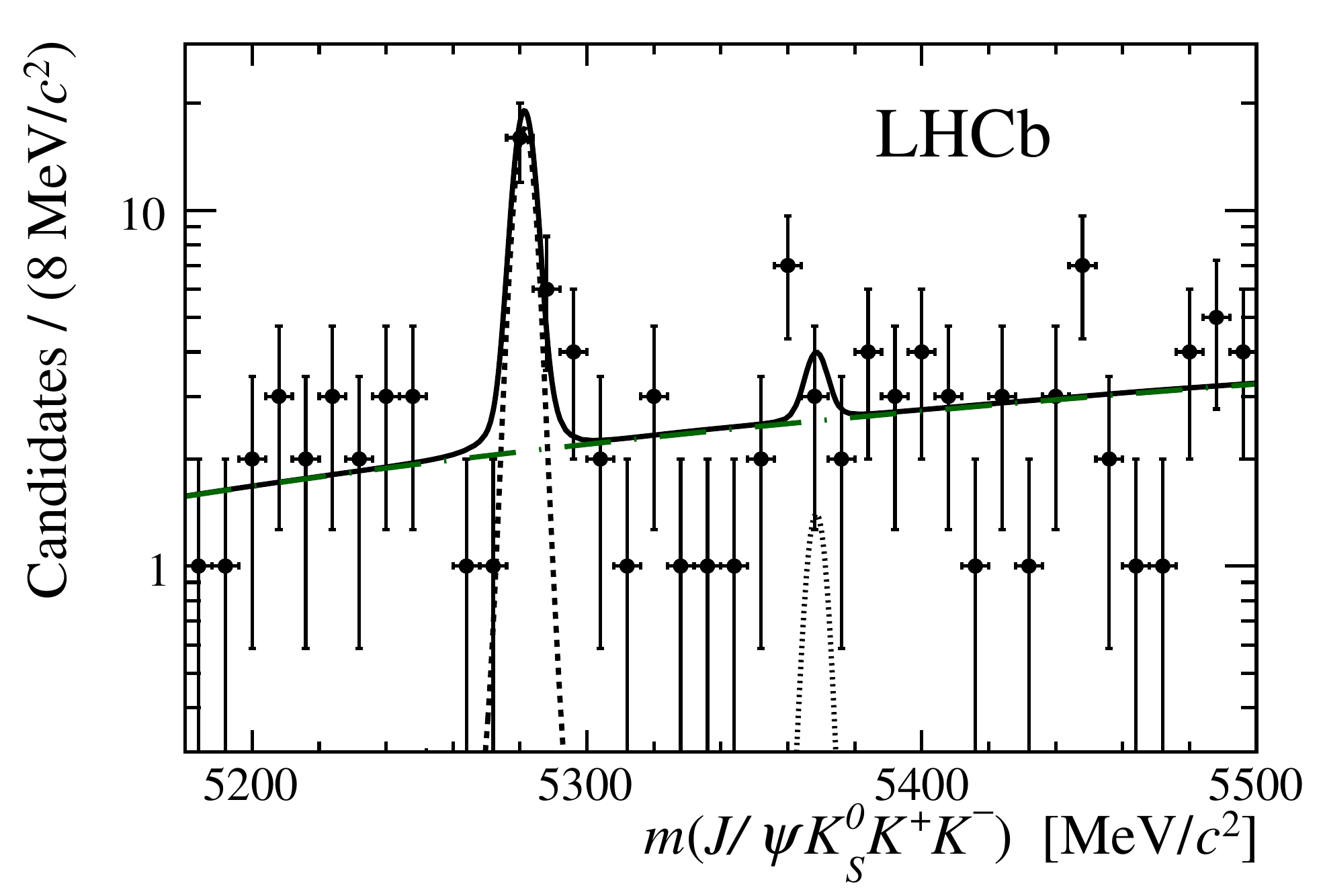}
\includegraphics[width=.49\textwidth]{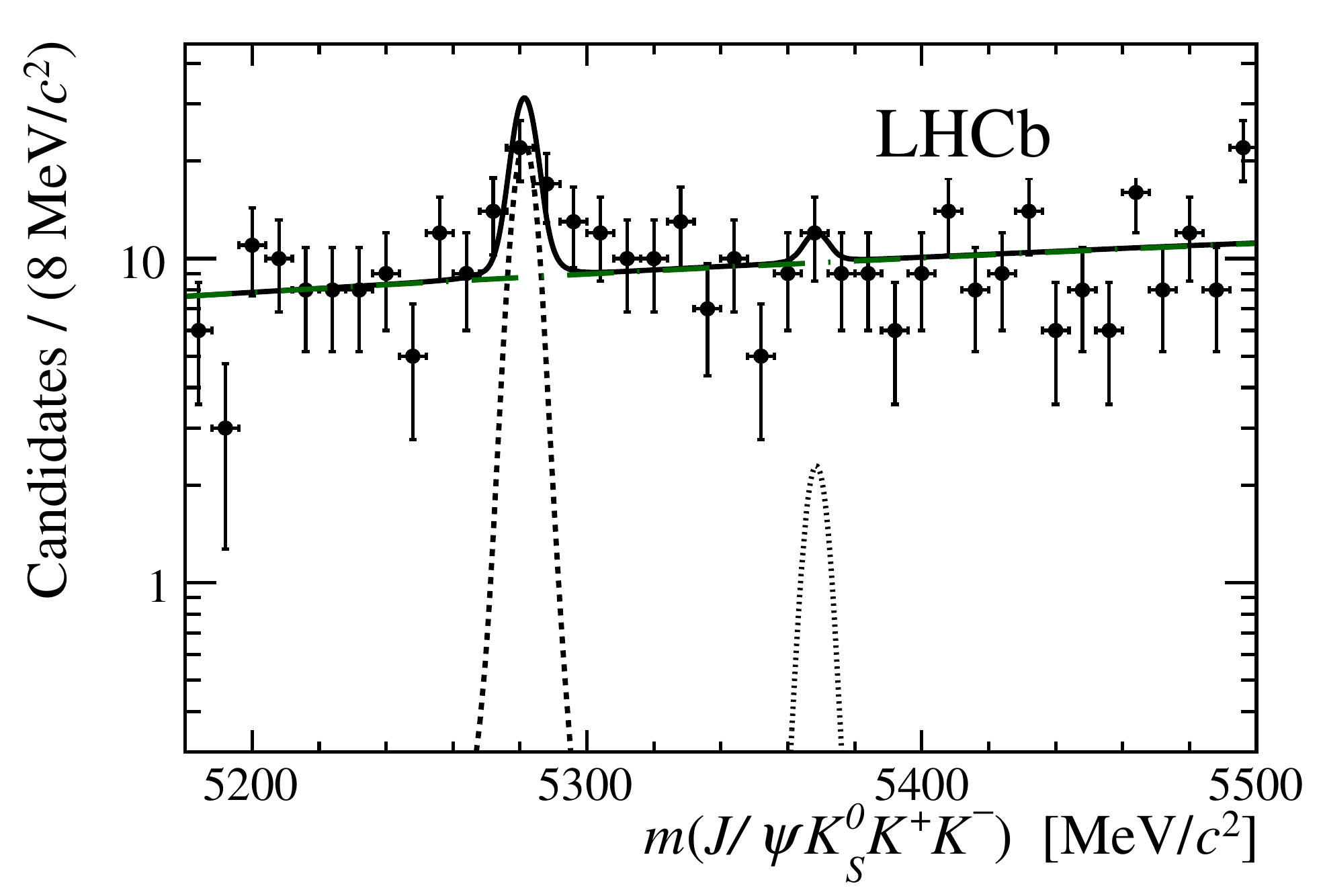}
  \caption{\small 
    Invariant mass distributions of (left) long and (right) downstream \BxToJpsiKSKK candidates, with data-based selection, shown with (top) linear and (bottom) logarithmic $y$-axis scales, with fit projections overlaid.
    The solid line shows the total fit result, while the dashed and dotted
    lines show the \Bd and \Bs signal components, respectively, and the dot-dashed line shows the combinatorial background.
  }
  \label{fig:BToJpsiKSKK}
\end{figure}

The results of the fit to the samples obtained with the data-based selection are shown in Fig.~\ref{fig:BToJpsiKSpipi} for the $\jpsi\KS\pip\pim$ hypothesis, in Fig.~\ref{fig:BToJpsiKSKpi} for the $\jpsi\KS\Kpm\pimp$ hypothesis and in Fig.~\ref{fig:BToJpsiKSKK} for the $\jpsi\KS\Kp\Km$ hypothesis.
A summary of the fitted yields is given in Table~\ref{table:FitResultsTight}.

\begin{table}[!tb]
  \begin{center}
    \caption{\small
      Yields determined from the simultaneous fit to the \BxToJpsiKSpipi, \BxToJpsiKSKpi and \BxToJpsiKSKK samples with data-based selection.
    }
    \label{table:FitResultsTight}
    \begin{tabular}{l@{\hspace{5mm}}cc@{\hspace{5mm}}cc@{\hspace{5mm}}cc}
      \hline \\ [-2.5ex]
                 & \multicolumn{2}{c}{\BxToJpsiKSpipi} & \multicolumn{2}{c}{\BxToJpsiKSKpi} & \multicolumn{2}{c}{\BxToJpsiKSKK} \\
                 & long         & downstream   & long          & downstream   & long          & downstream \\
    \hline \\ [-2.5ex]
    $N_{\Bd}$     & $246\,^{+17}_{-16}$ & $471\,^{+24}_{-23}$ & $\phanii4\,^{+6\phani}_{-5}$ & $\phani23 \pm 10$ & $18\,^{+5}_{-4}$ & $27\,^{+8}_{-7}$ \\ [0.3ex]
    $N_{\Bs}$     & $\phanii5\,^{+4\phani}_{-3}$ & $\phanii9\,^{+6\phani}_{-5}$ & $154\,^{+15}_{-14}$ & $371 \pm 23$ & $\phani2\,^{+3}_{-2}$ & $\phani3\,^{+5}_{-4}$ \\ [0.3ex]
    \hline
    \end{tabular}
  \end{center}
\end{table}

\section{Phase-space distributions of signal decays}
\label{sec:phase-space}

Clear signals are seen for \BdToJpsiKSpipi, \BsToJpsiKSKpi and \BdToJpsiKSKK decays. 
The significance of each of the signals is discussed in Sec.~\ref{sec:results}.
The distributions of the two- and three-body invariant mass combinations of the signal decay products are examined using the \sPlot\ technique~\cite{Pivk:2004ty} with the \B candidate invariant mass as the discriminating variable.

\begin{figure}[!t]
  \centering
  \includegraphics[width=0.49\textwidth]{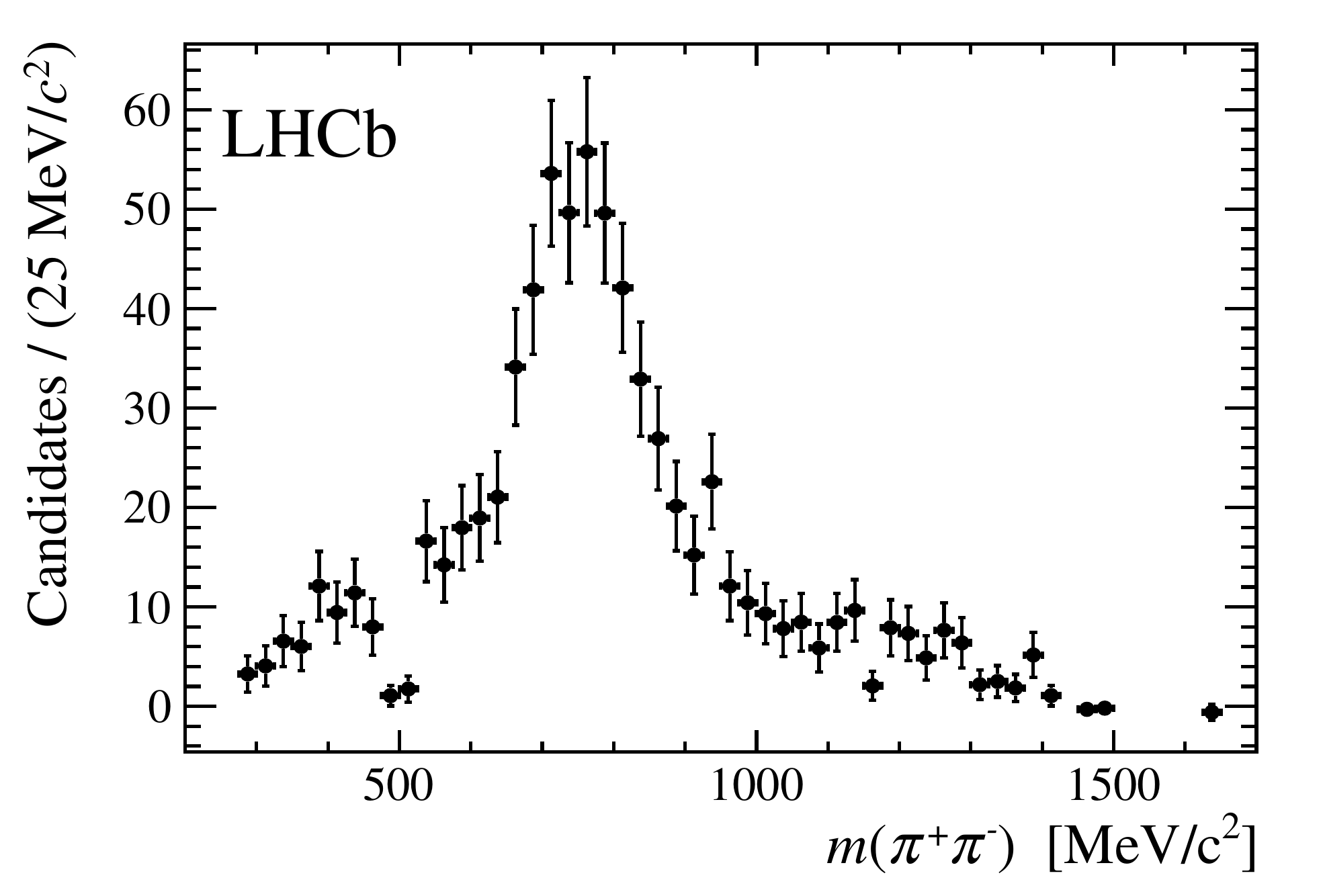}
  \includegraphics[width=0.49\textwidth]{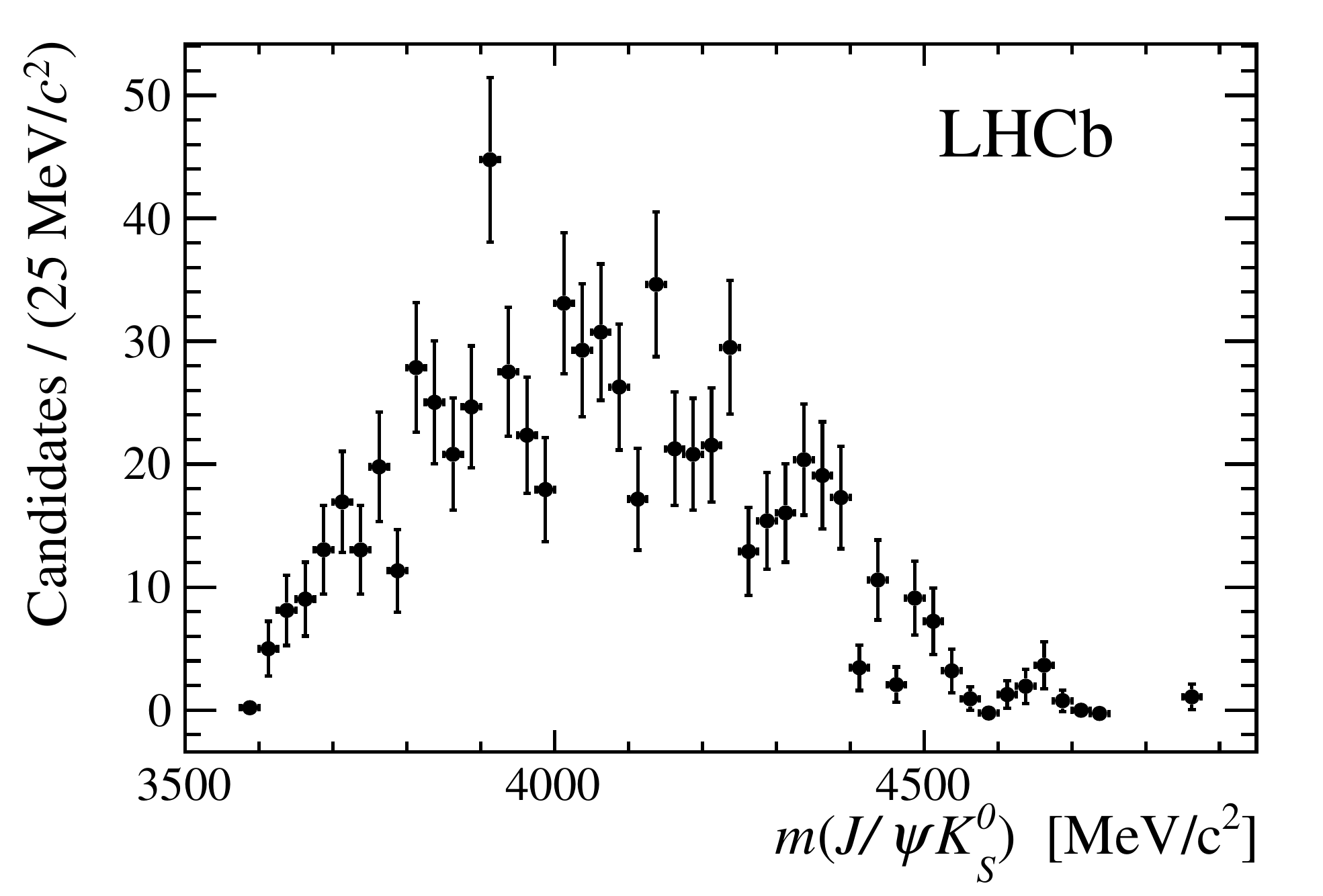}
  \includegraphics[width=0.49\textwidth]{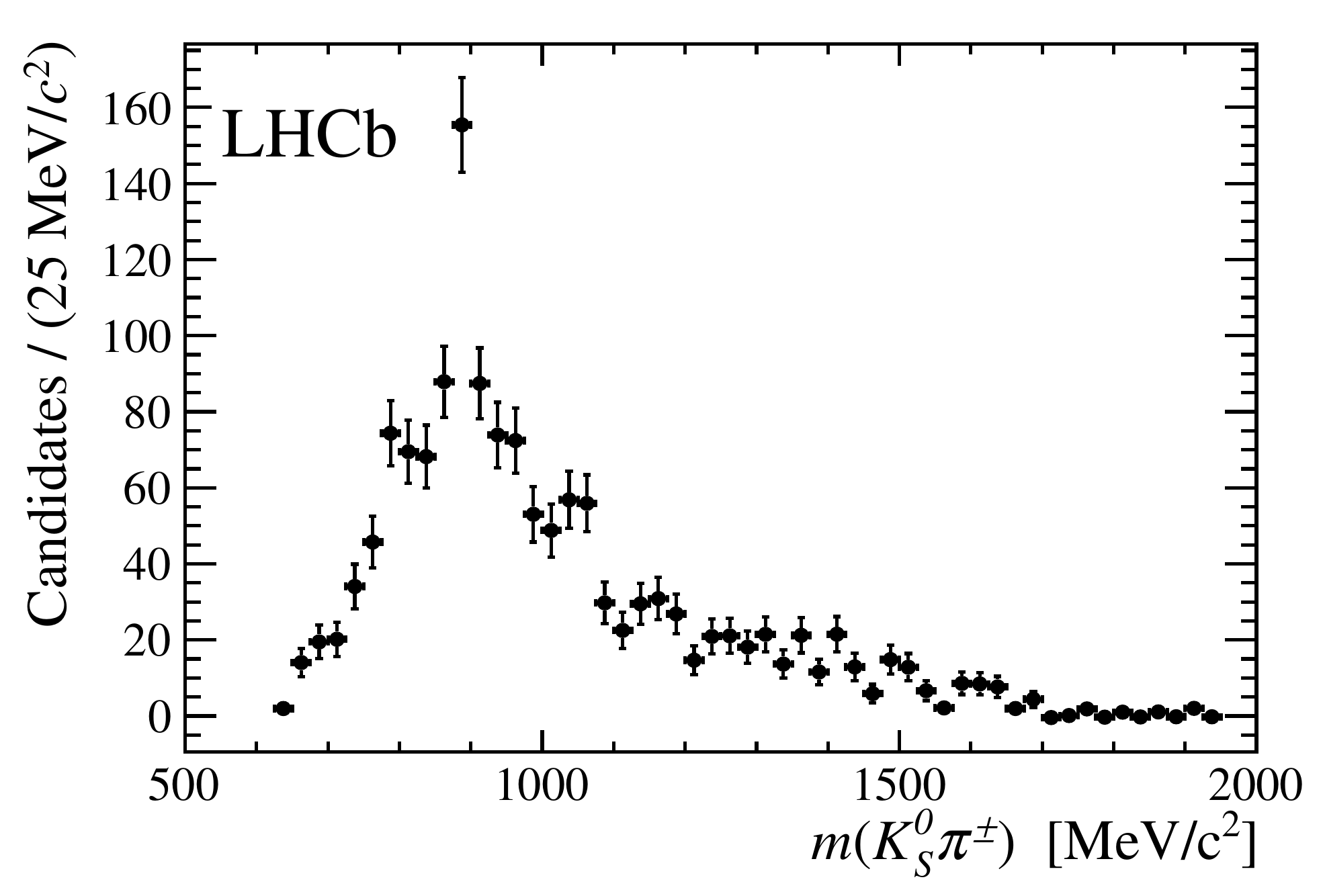}
  \includegraphics[width=0.49\textwidth]{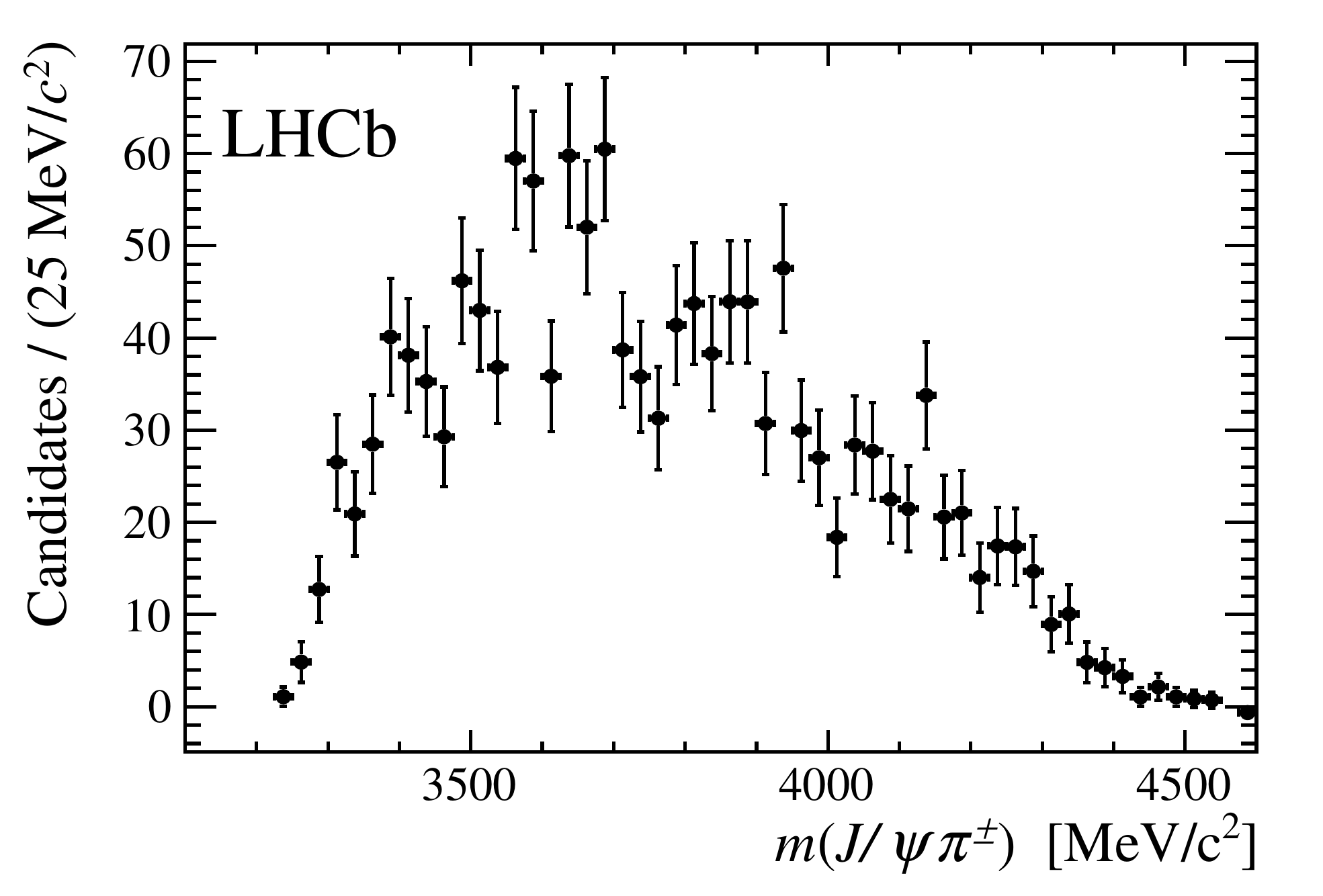}
  \caption{\small
    Background-subtracted distributions of the possible two-body invariant mass combinations in \BdToJpsiKSpipi decays.
    Contributions from the $\rho(770)^0$ and $K^*(892)^\pm$ mesons are seen in the $m(\pip\pim)$ and $m(\KS\pipm)$ distributions, respectively.
  } 
  \label{fig:twomcdataInvarMassBdpipi}
\end{figure}

\begin{figure}[!t]
  \centering
  \includegraphics[width=0.49\textwidth]{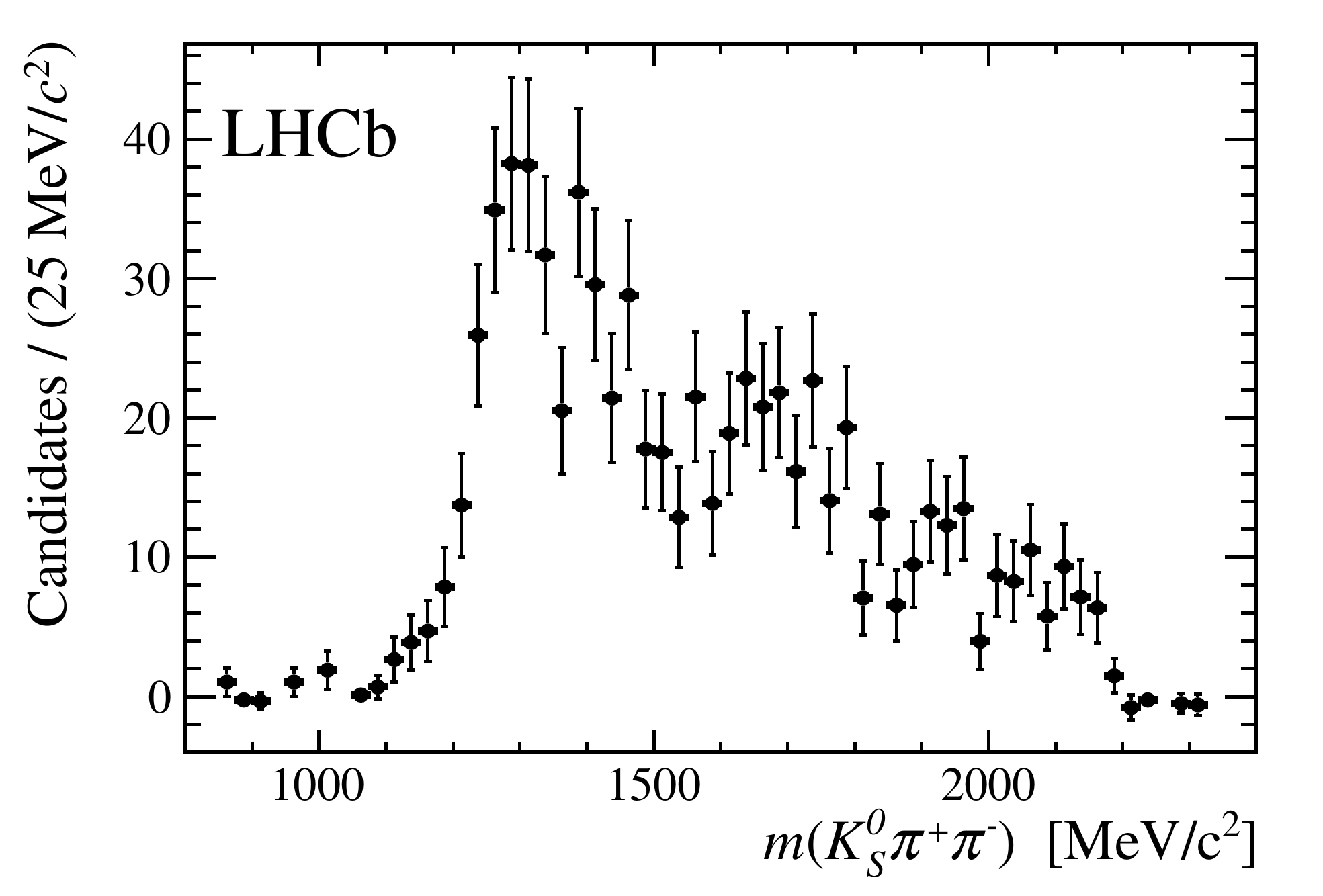}
  \includegraphics[width=0.49\textwidth]{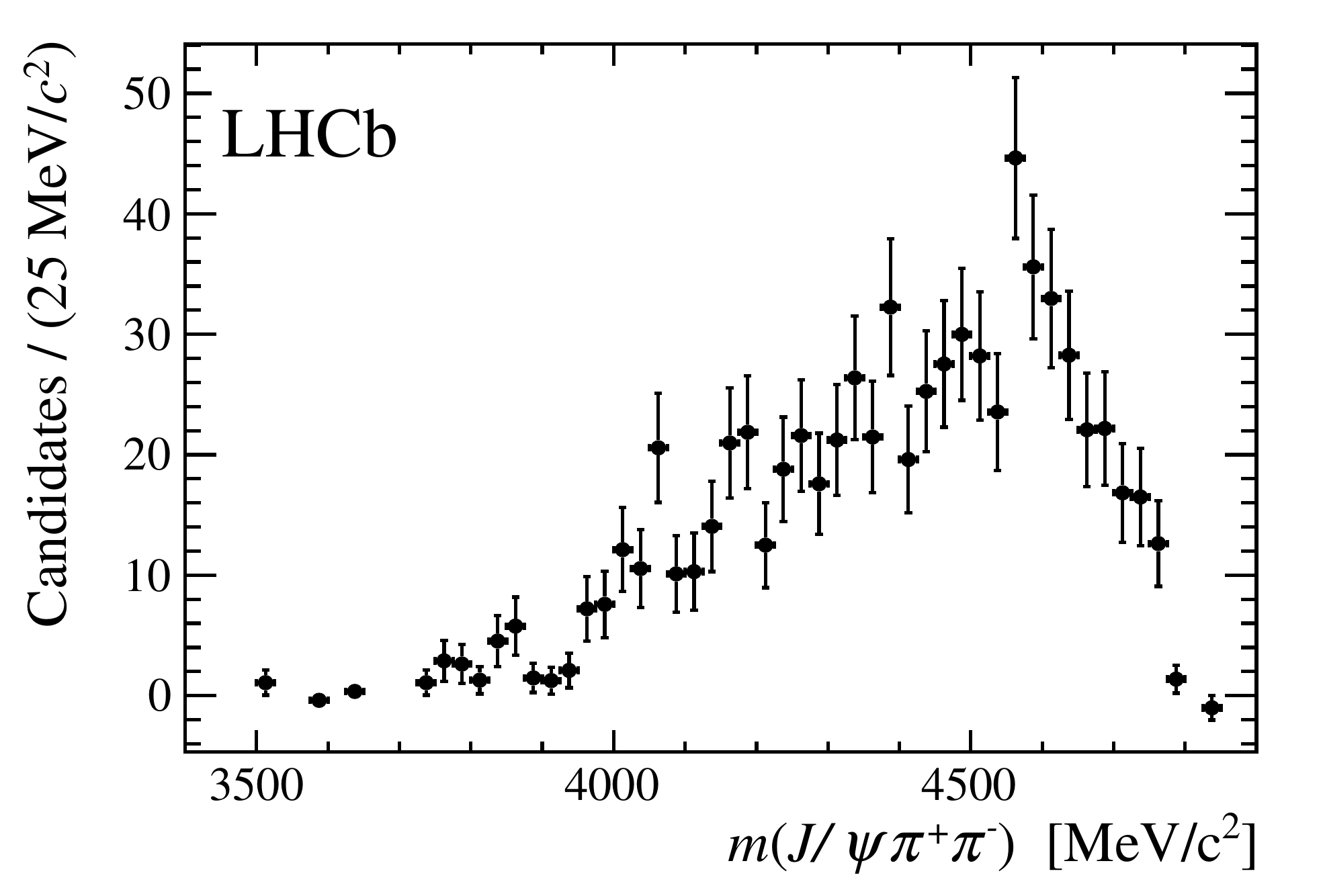}
  \includegraphics[width=0.49\textwidth]{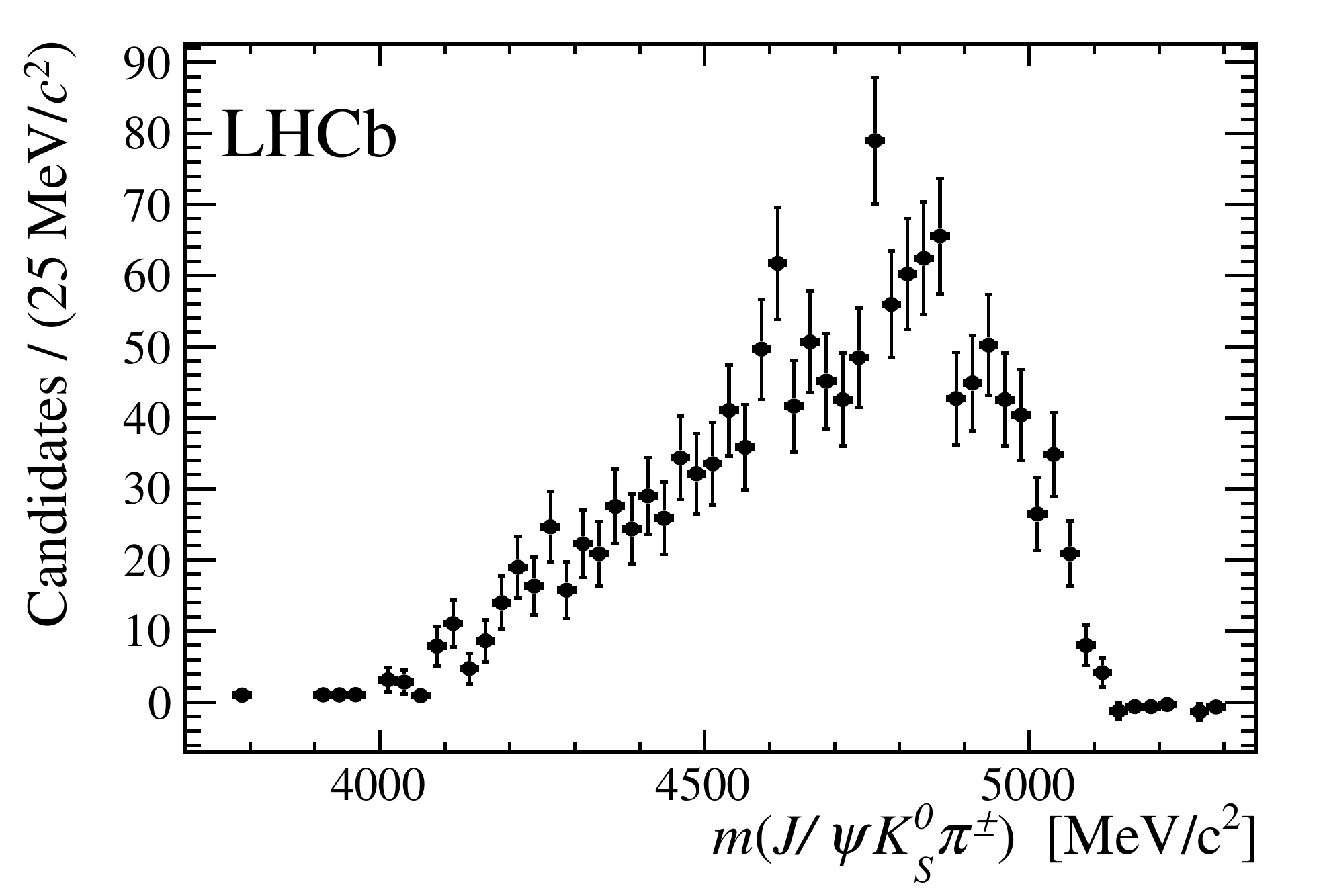}
  \caption{\small
    Background-subtracted distributions of the possible three-body invariant mass combinations in \BdToJpsiKSpipi decays.
    An enhancement from the $K_1(1400)$ state are seen in the $m(\KS\pip\pim)$ distribution.
  }
  \label{fig:threemcdataInvarMassBdpipi}
\end{figure}

\begin{figure}[!t]
  \centering
  \includegraphics[width=0.49\textwidth]{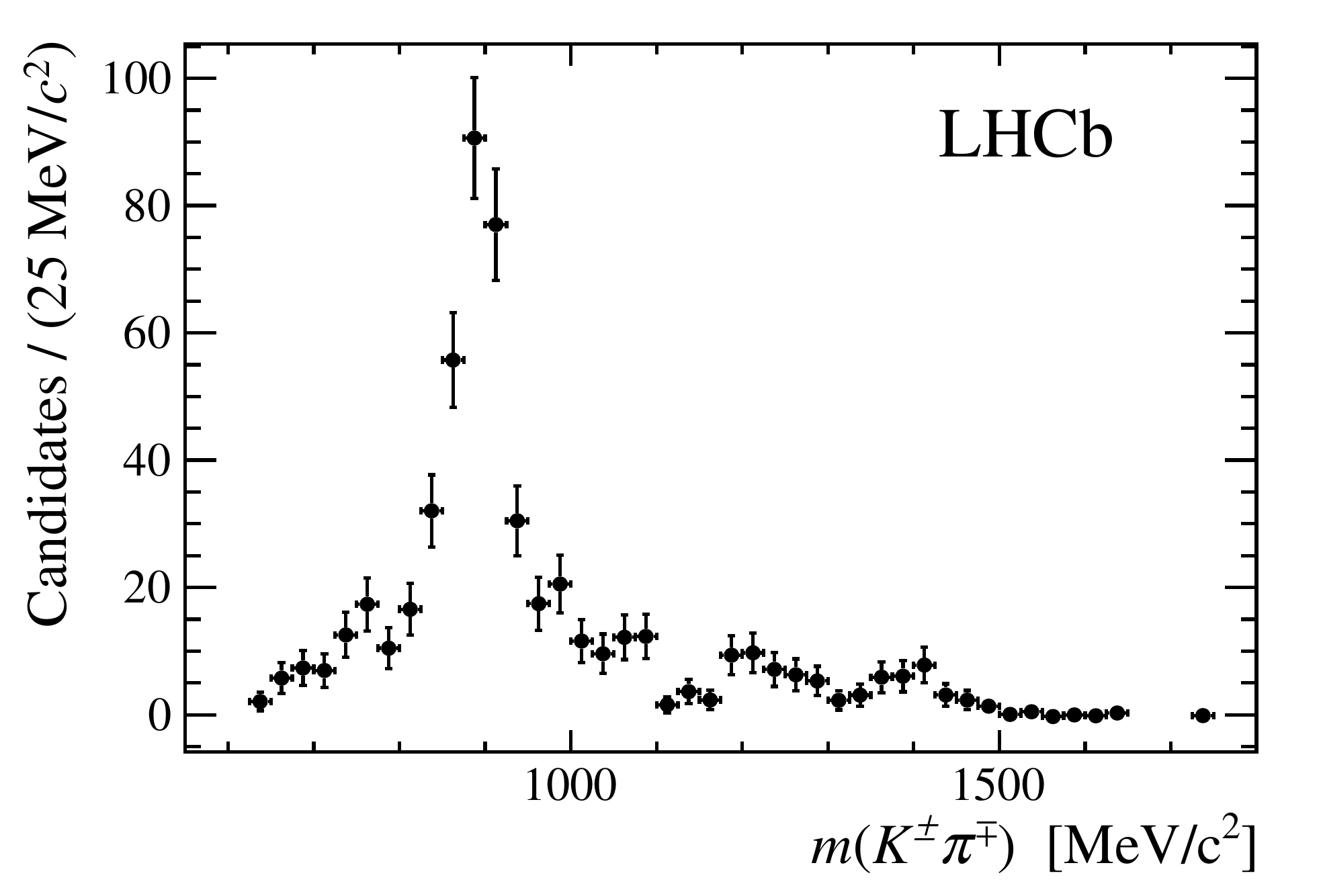}
  \includegraphics[width=0.49\textwidth]{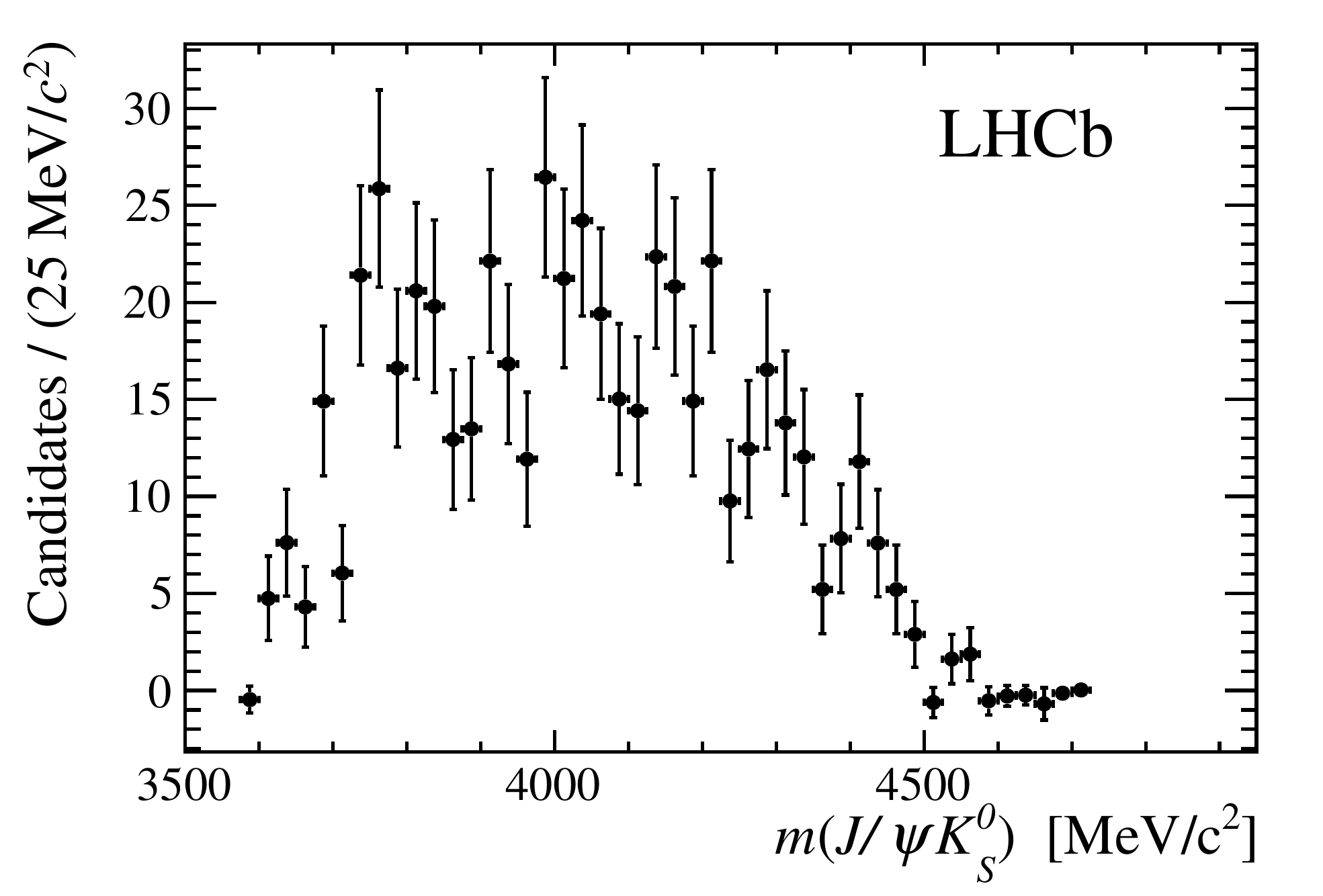}
  \includegraphics[width=0.49\textwidth]{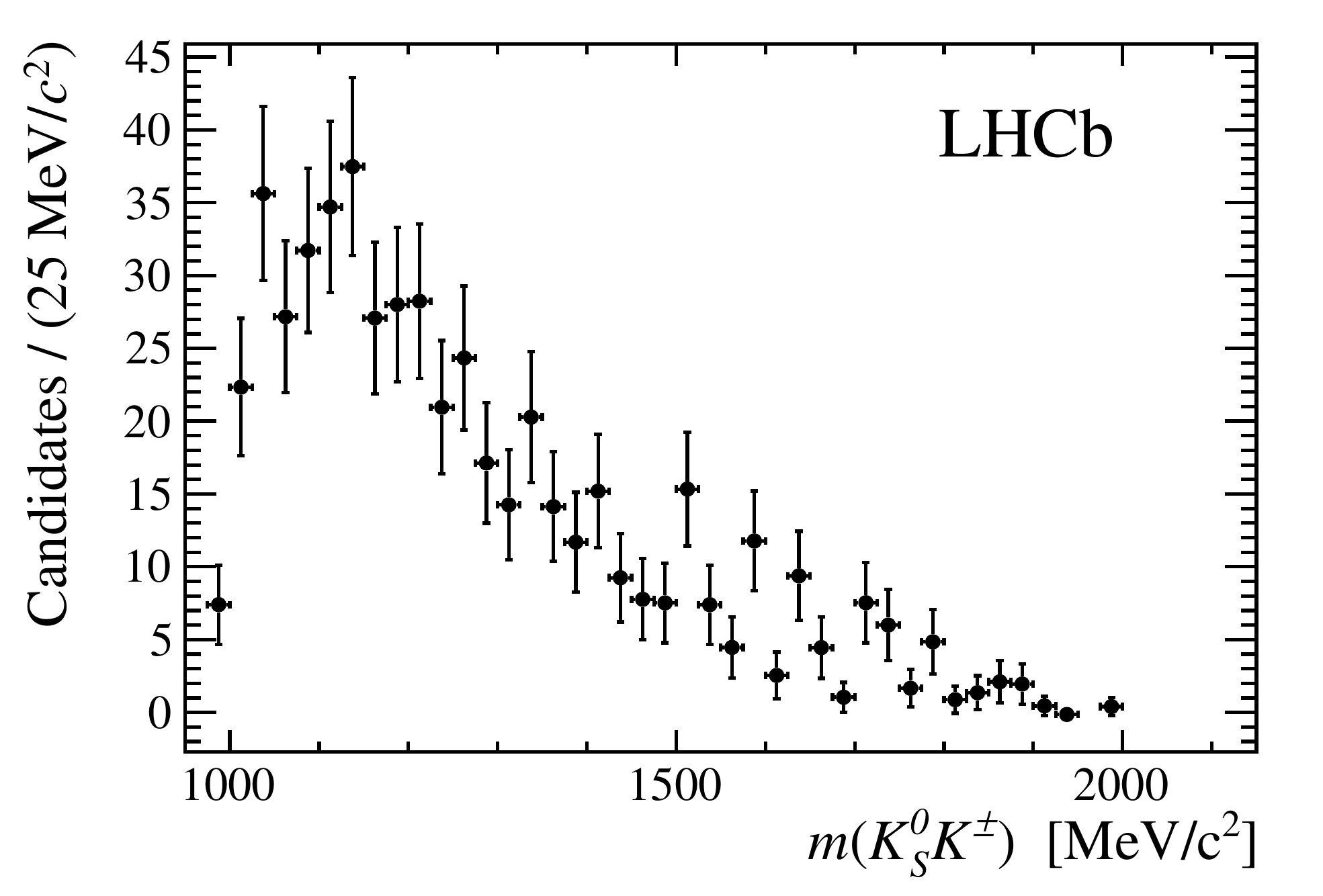}
  \includegraphics[width=0.49\textwidth]{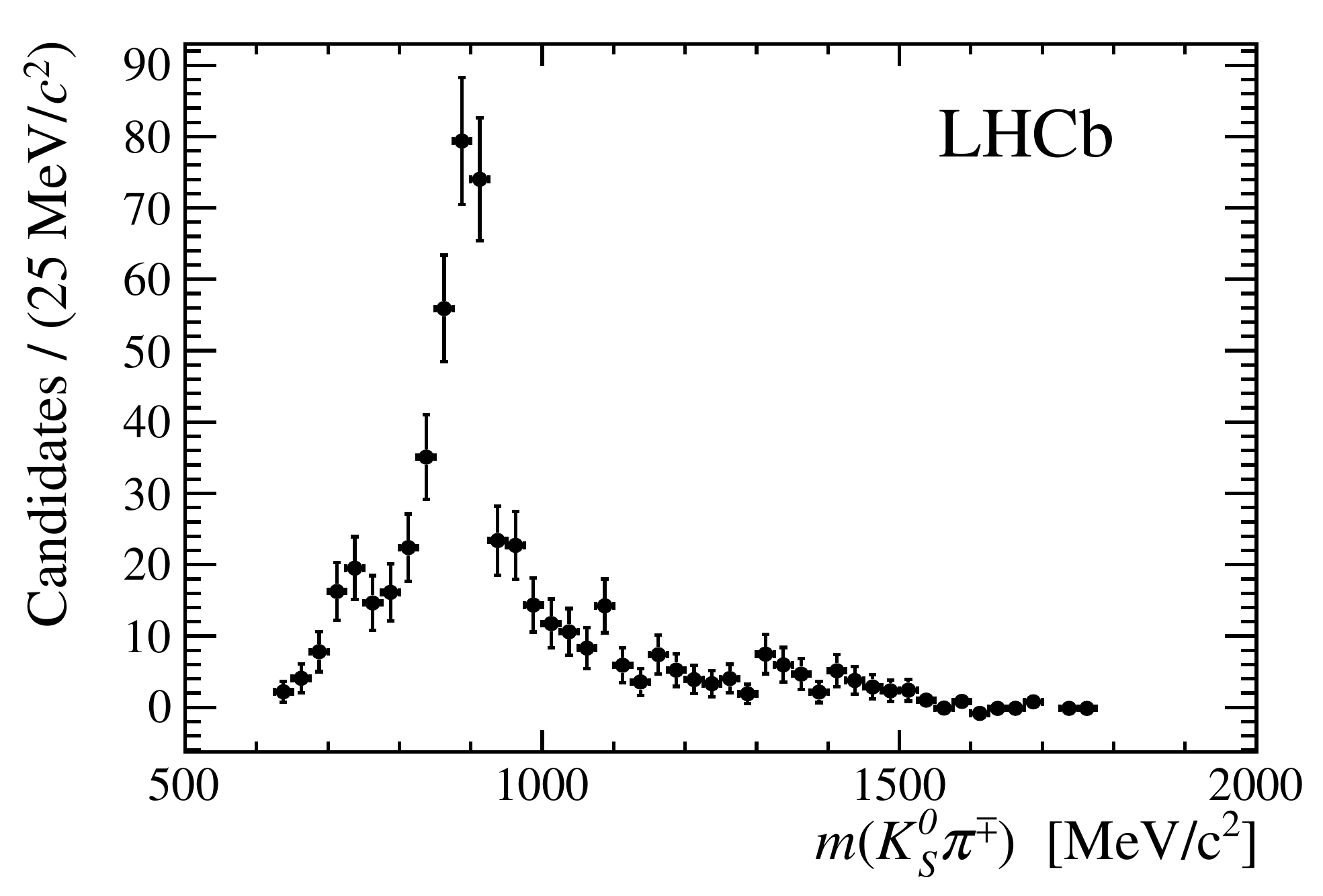}
  \includegraphics[width=0.49\textwidth]{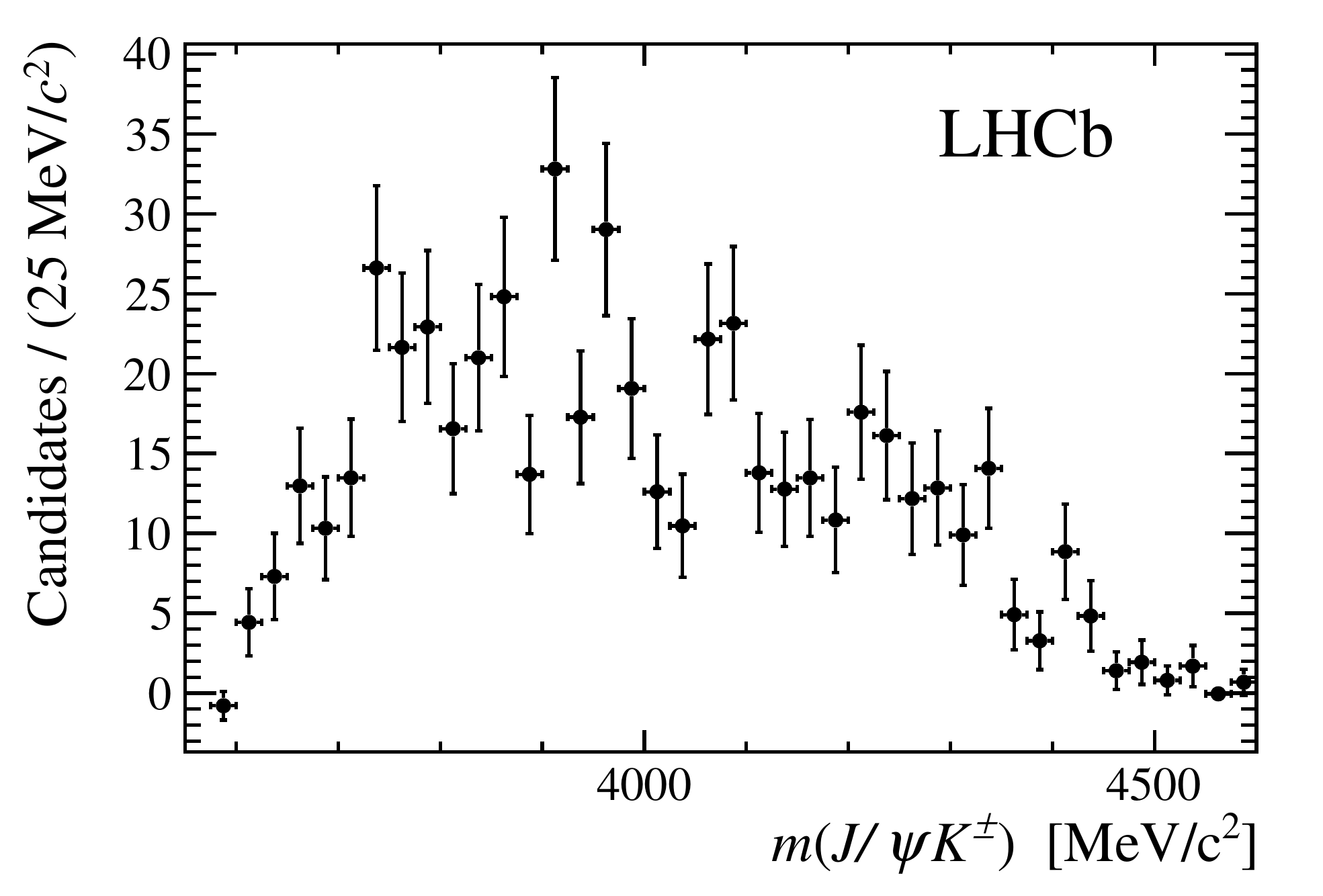}
  \includegraphics[width=0.49\textwidth]{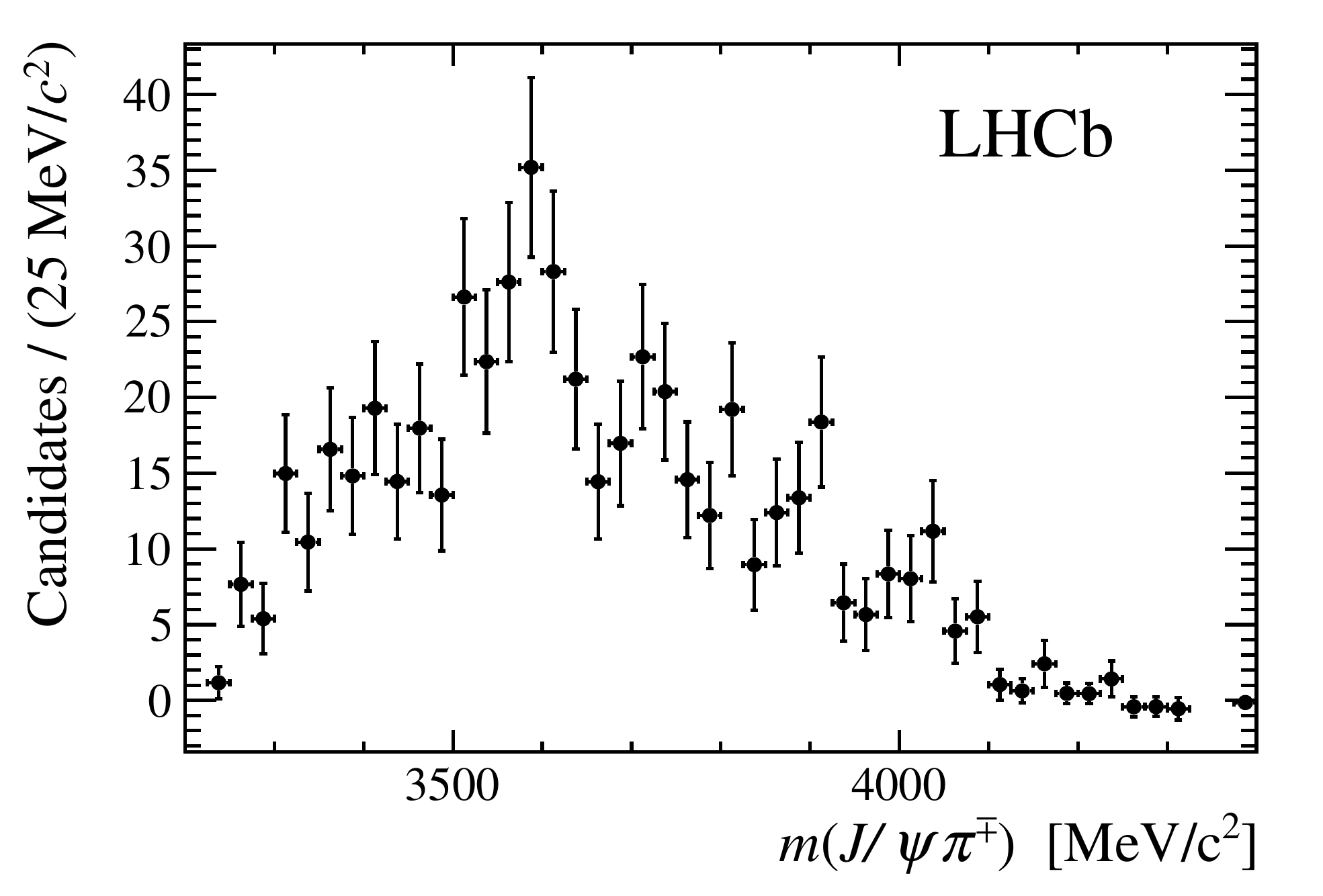}
  \caption{\small
    Background-subtracted distributions of the possible two-body invariant mass combinations in \BsToJpsiKSKpi decays.
    Contributions from the $\KorKbar^*(892)^0$ and $K^*(892)^\pm$ mesons are seen in the $m(\Kpm\pimp)$ and $m(\KS\pipm)$ distributions, respectively.
  }
  \label{fig:twomcdataInvarMassBsKpi}
\end{figure}

\begin{figure}[!t]
  \centering
  \includegraphics[width=0.49\textwidth]{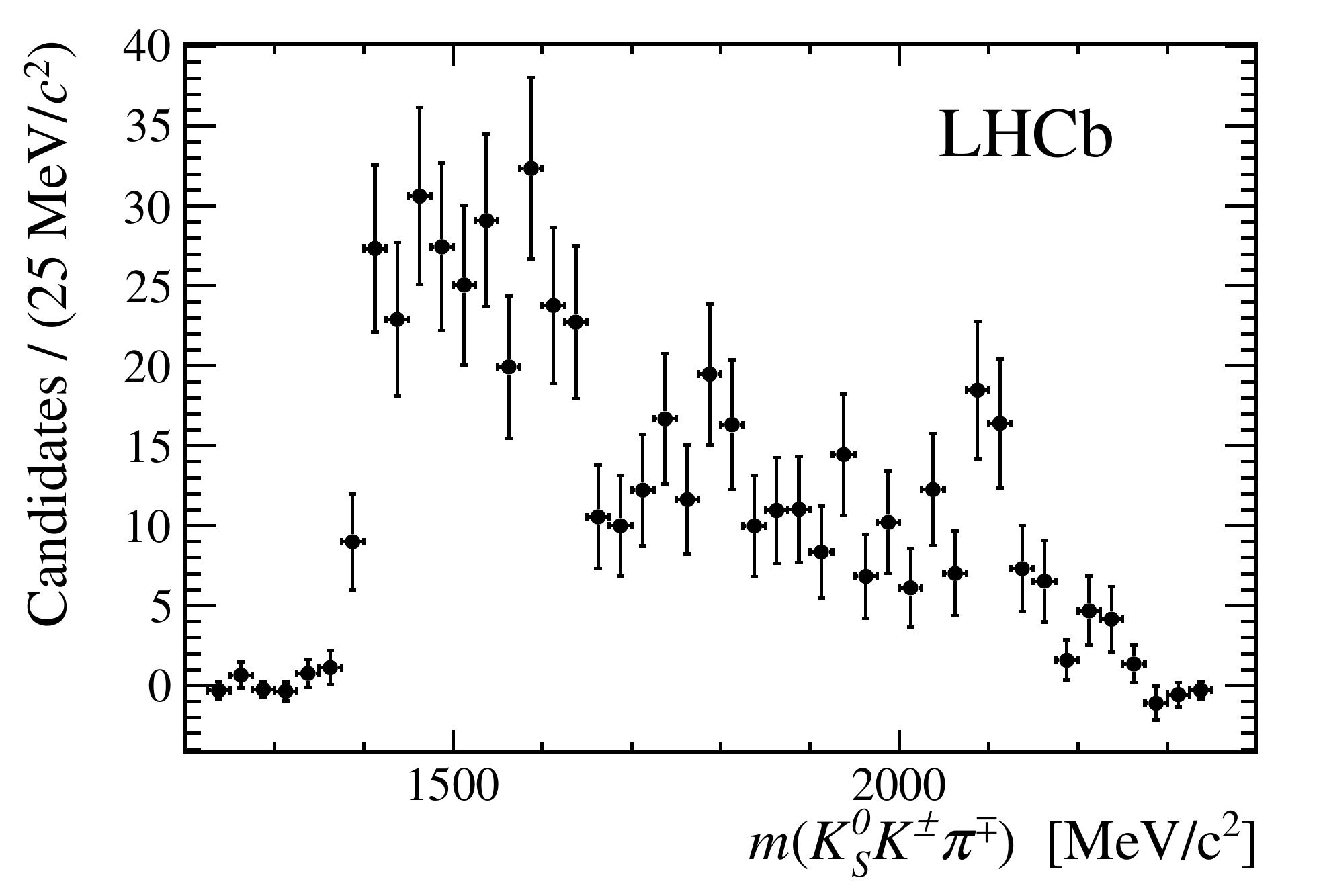}
  \includegraphics[width=0.49\textwidth]{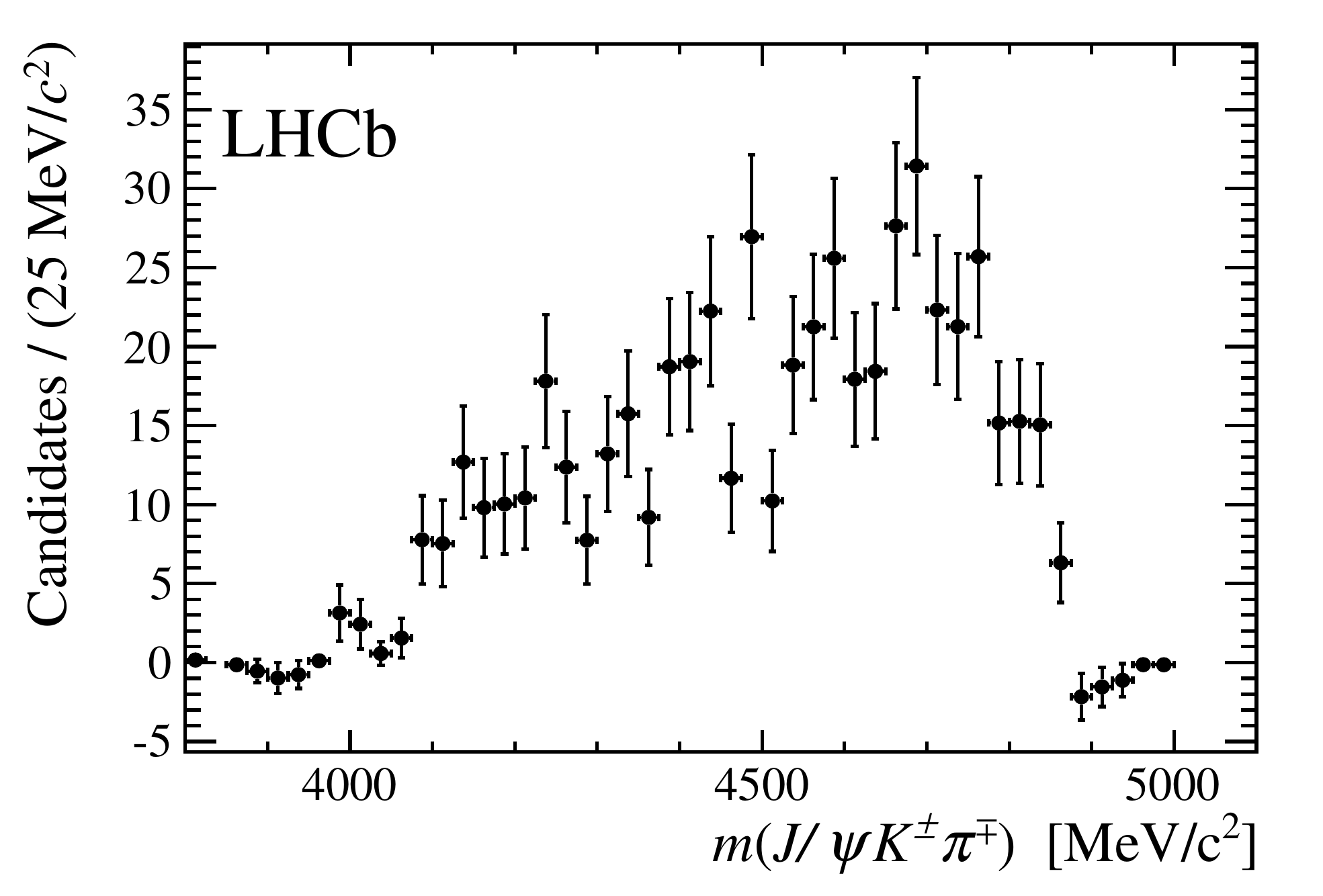}
  \includegraphics[width=0.49\textwidth]{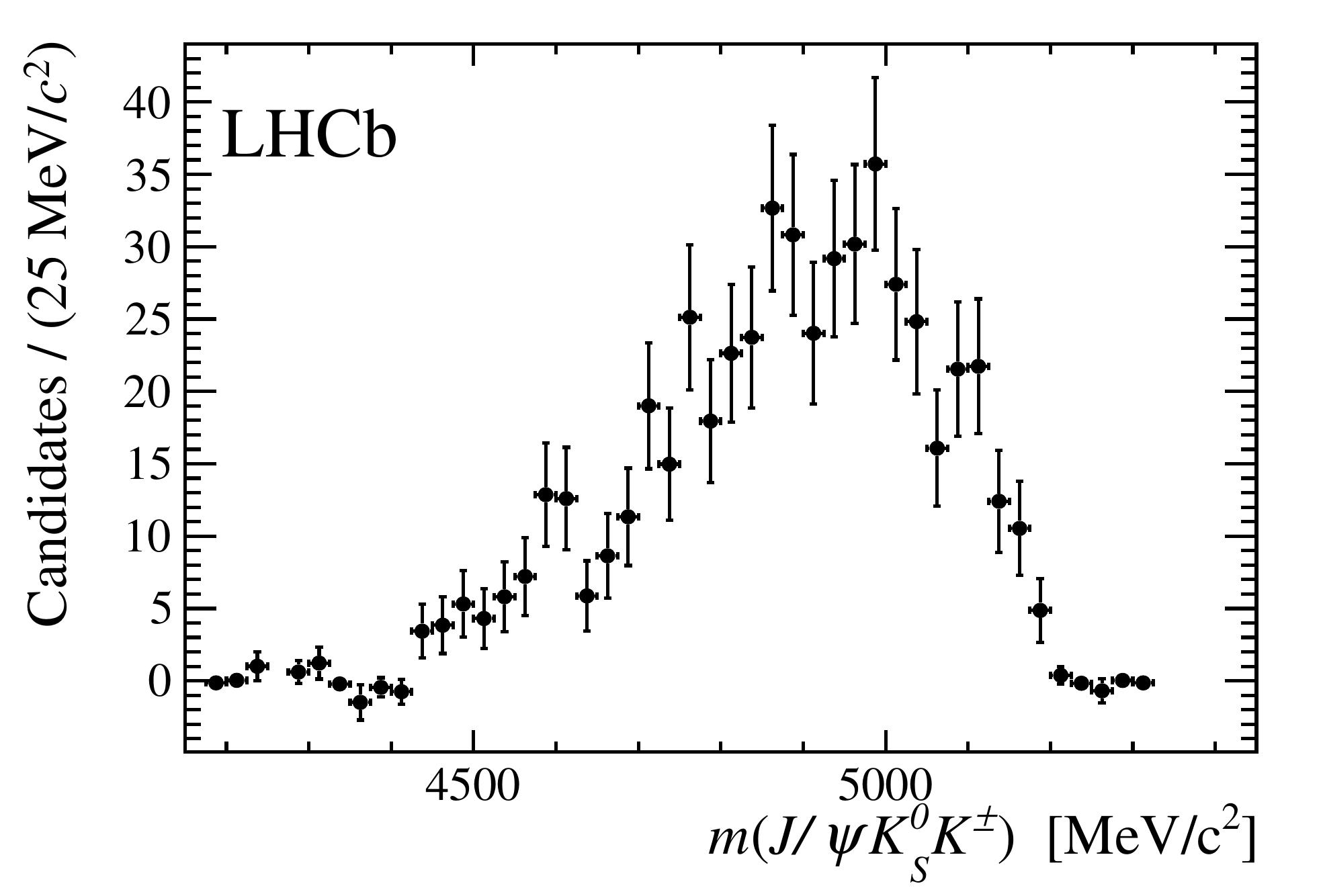}
  \includegraphics[width=0.49\textwidth]{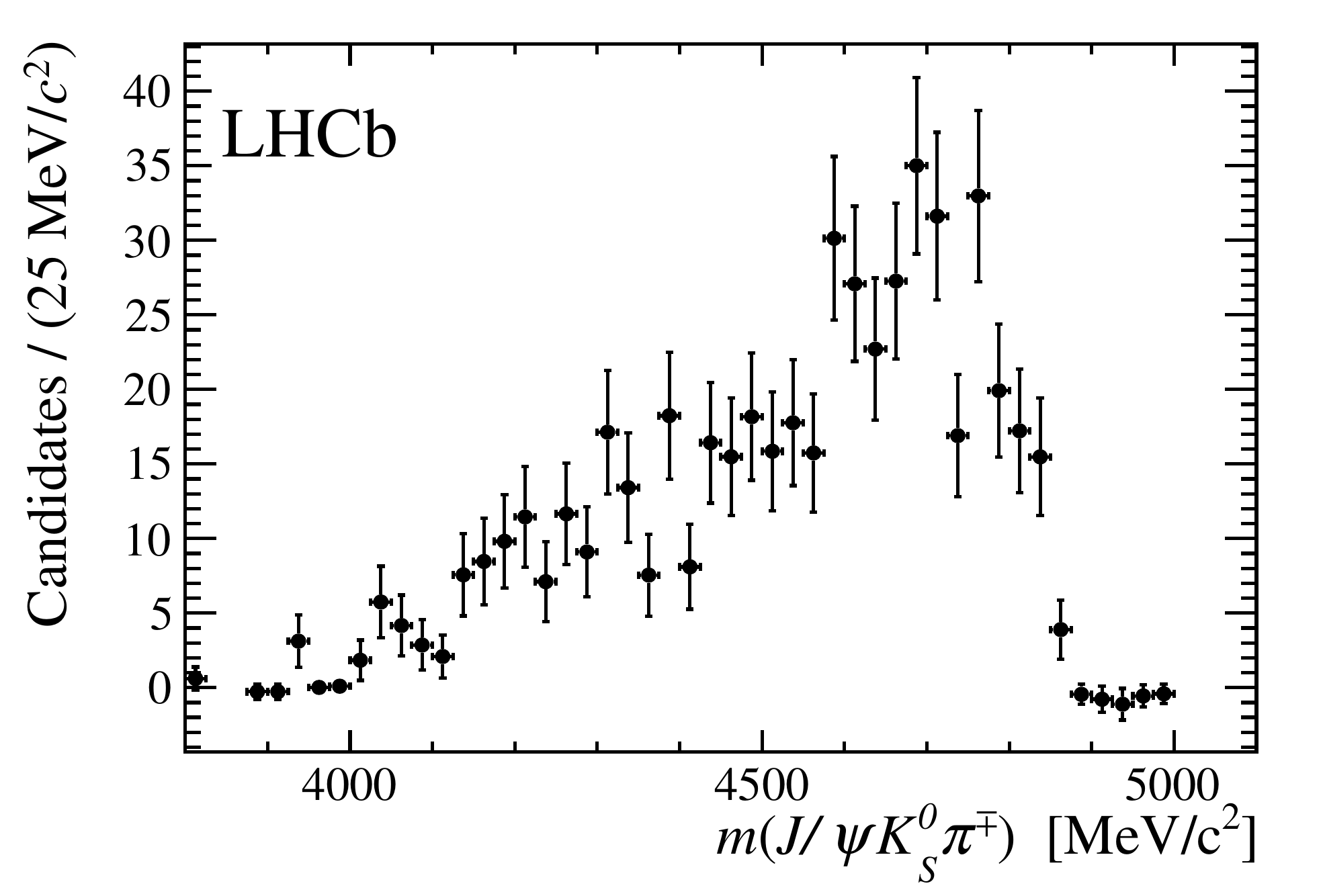}
  \caption{\small
    Background-subtracted distributions of the possible three-body invariant mass combinations in \BsToJpsiKSKpi decays.
    No clear signatures of narrow resonances are observed.
  }
  \label{fig:threemcdataInvarMassBsKpi}
\end{figure}

\begin{figure}[!t]
  \centering
  \includegraphics[width=0.49\textwidth]{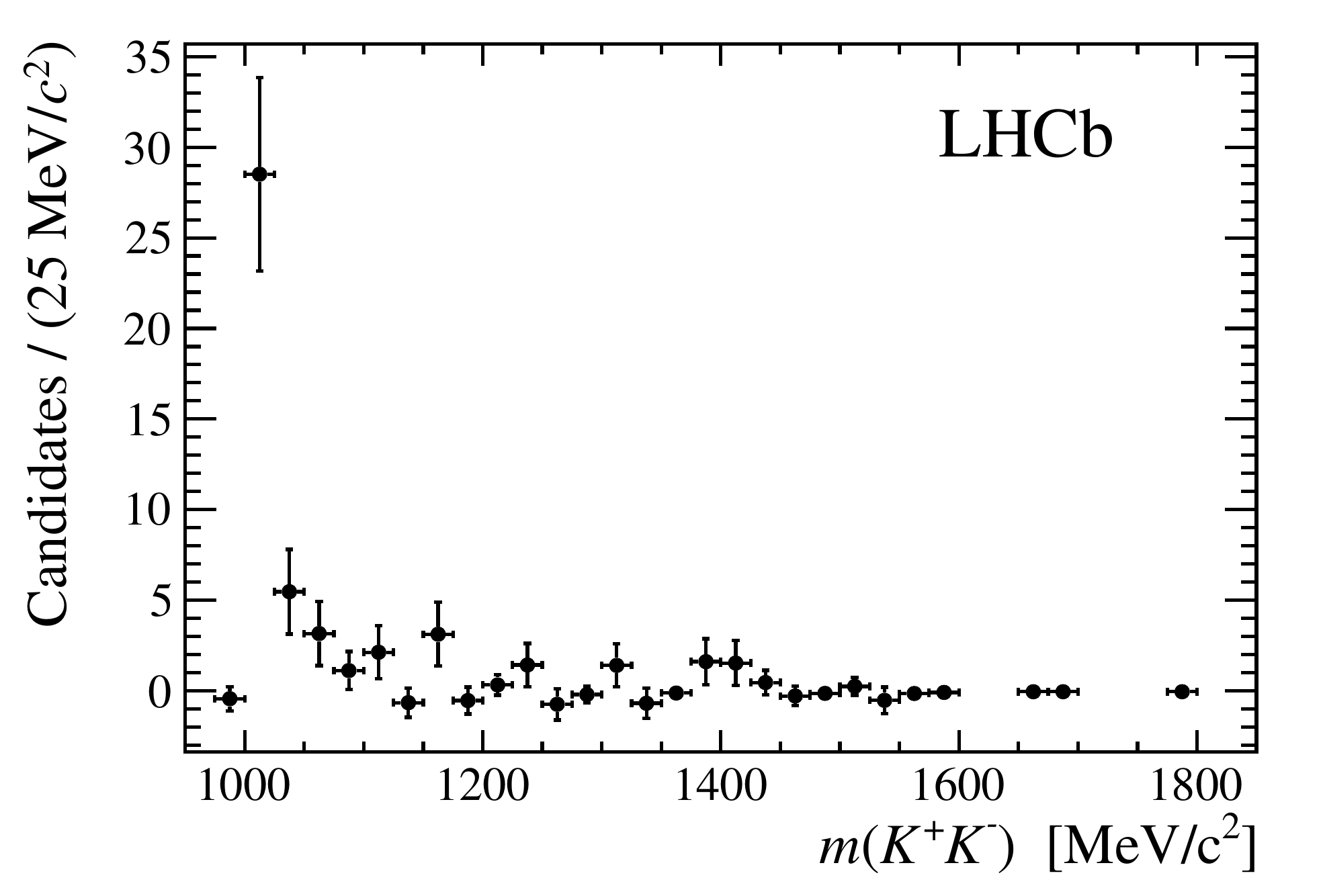}
  \includegraphics[width=0.49\textwidth]{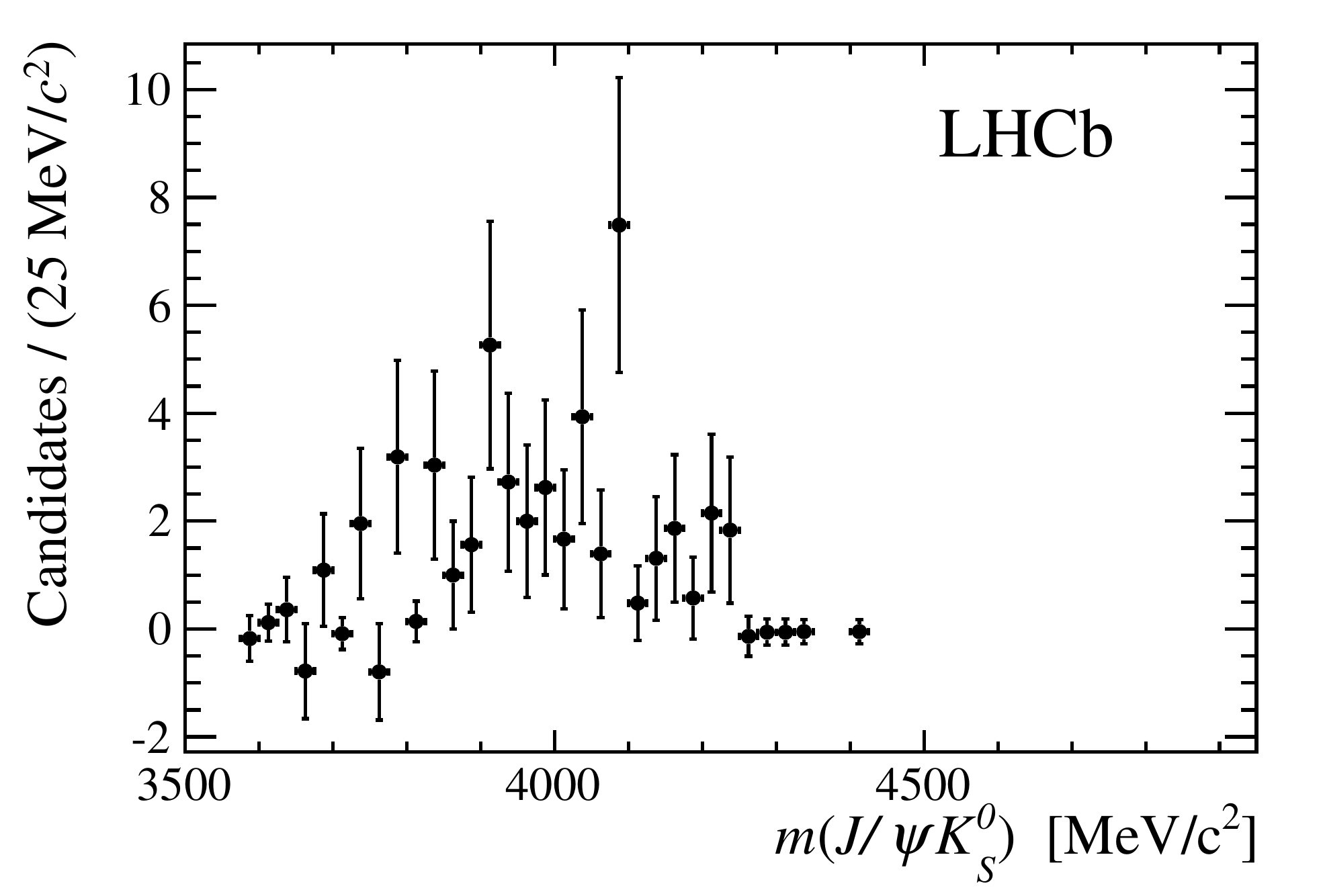}
  \includegraphics[width=0.49\textwidth]{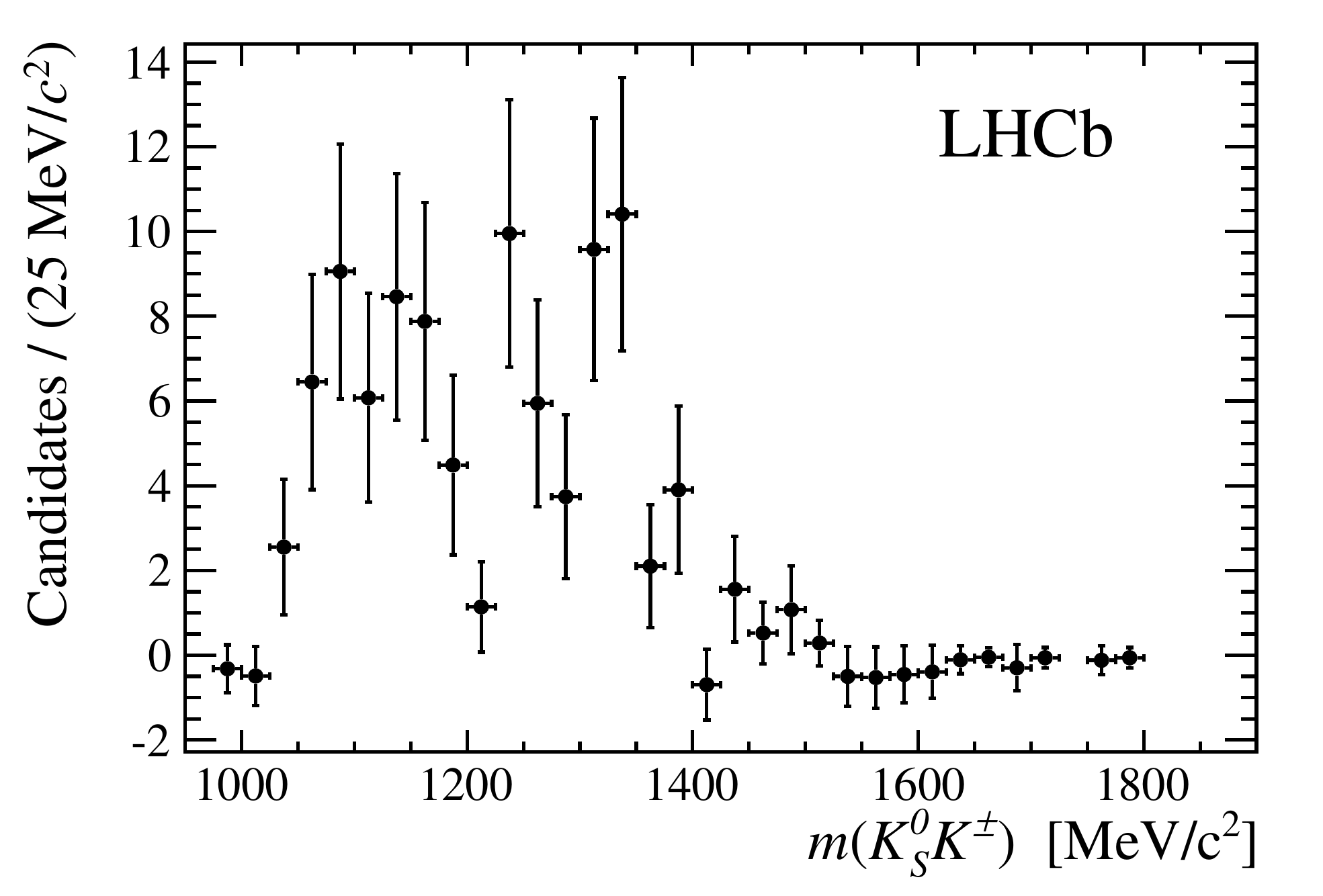}
  \includegraphics[width=0.49\textwidth]{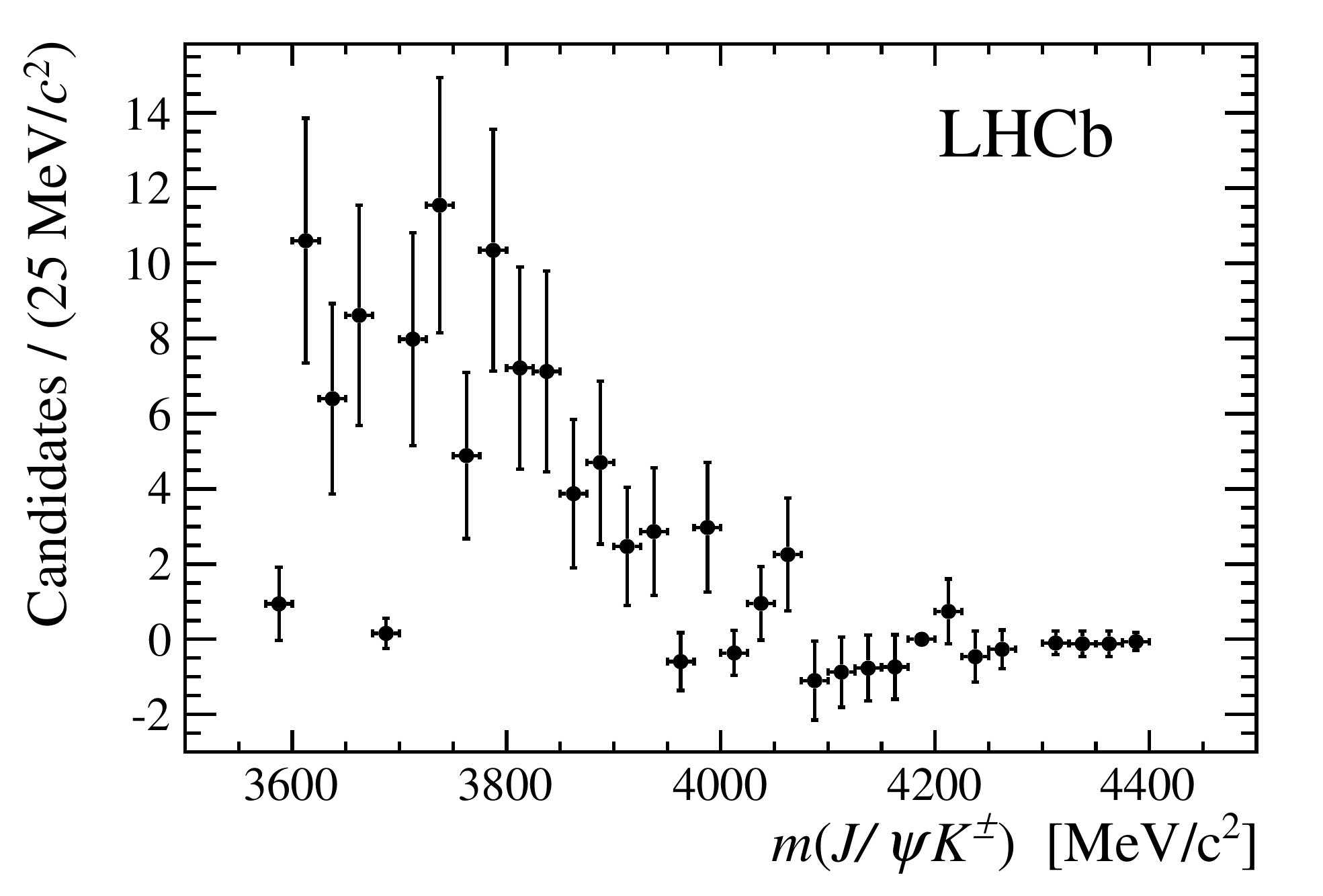}
  \caption{\small
    Background-subtracted distributions of the possible two-body invariant mass combinations in \BdToJpsiKSKK decays.
    The $\phi(1020)$ resonance is clearly seen in the $m(\Kp\Km)$ distribution.
  }
  \label{fig:twomcdataInvarMassBdKK}
\end{figure}

\begin{figure}[!t]
  \centering
  \includegraphics[width=0.49\textwidth]{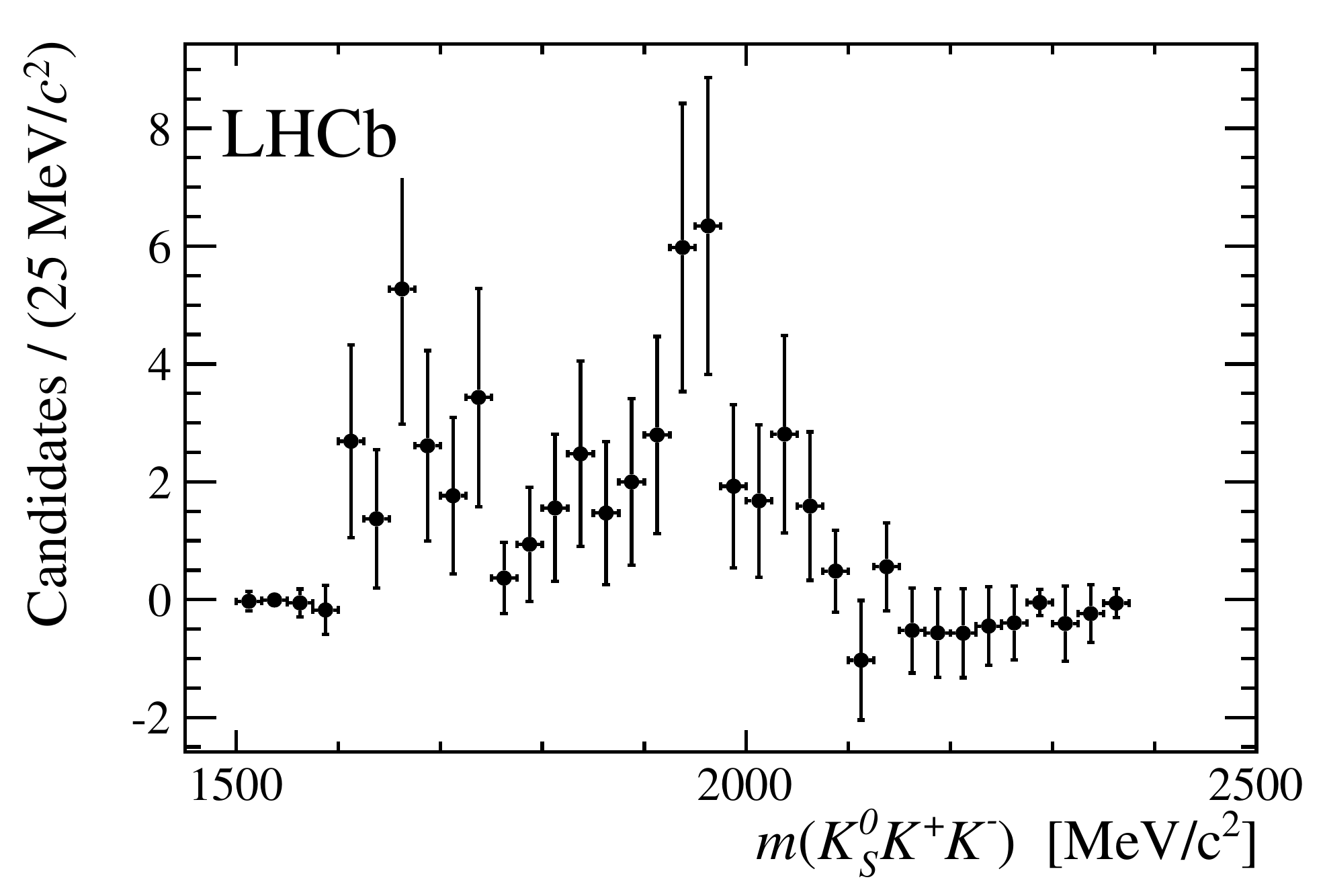}
  \includegraphics[width=0.49\textwidth]{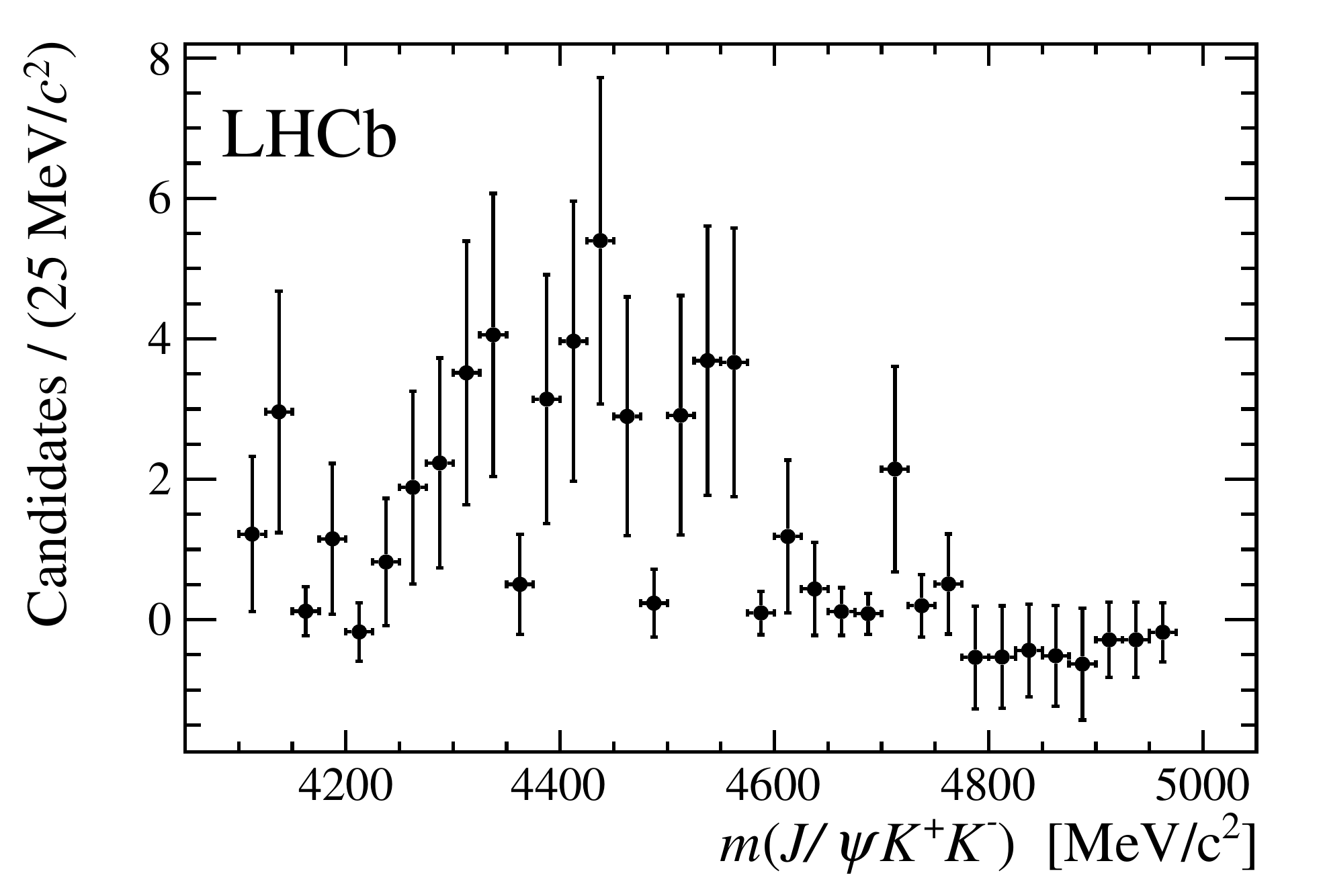}
  \includegraphics[width=0.49\textwidth]{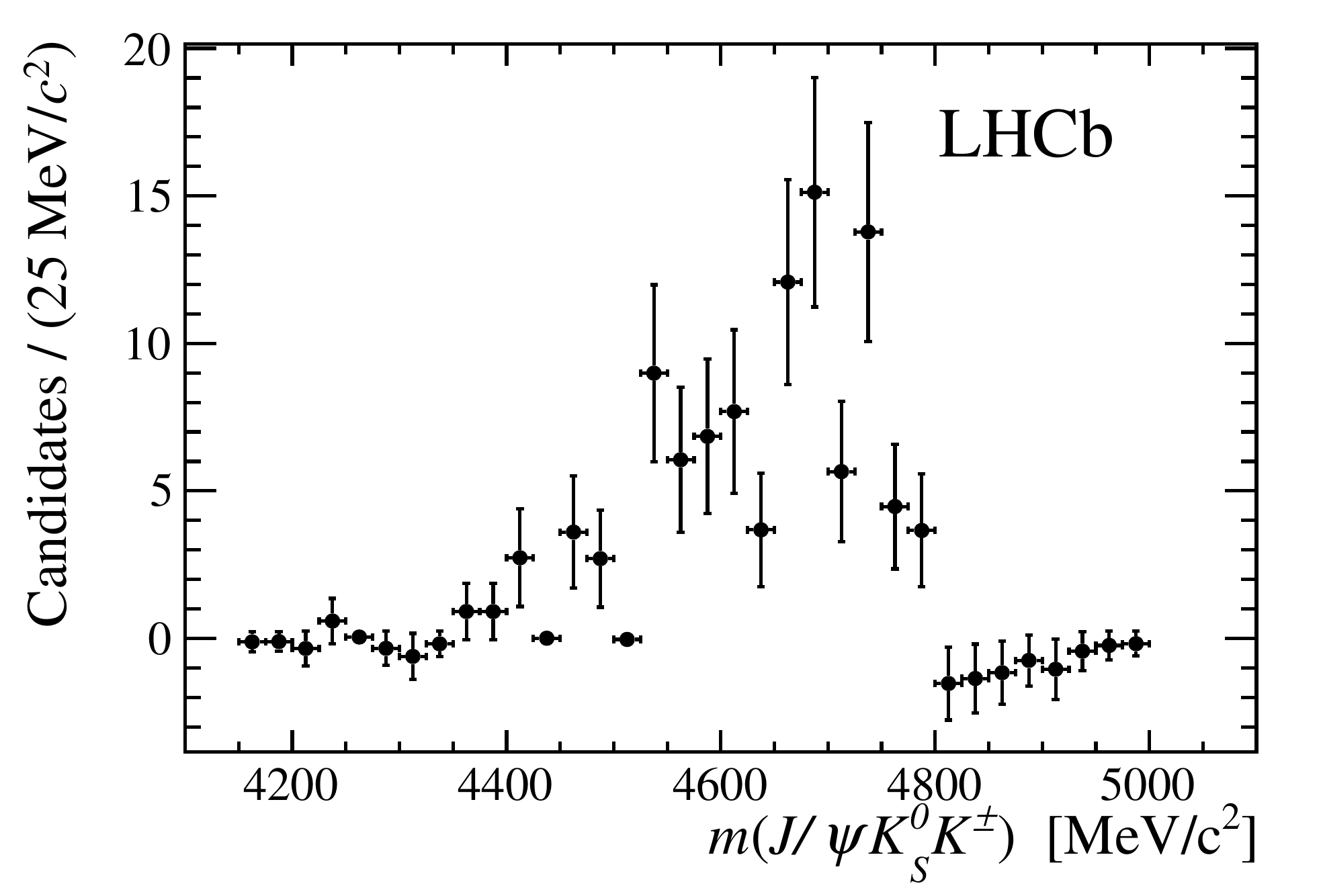}
  \caption{\small
    Background-subtracted distributions of the possible three-body invariant mass combinations in \BdToJpsiKSKK decays.
    No clear signatures of narrow resonances are observed.
  }
  \label{fig:threemcdataInvarMassBdKK}
\end{figure}

None of the channels show significant structures in any invariant mass combinations involving the \jpsi meson.
In \BdToJpsiKSpipi decays the $\psi(2S)$ contribution is vetoed and therefore does not appear in $m(\jpsi \pip\pim)$; there is also a small but not significant excess around the $X(3872)$ mass.
In the same channel, excesses from $K^*(892)$ and $\rho(770)$ mesons are seen in $m(\KS\pipm)$ and $m(\pip\pim)$ respectively, and there is an enhancement from the $K_1(1400)$ state in $m(\KS\pip\pim)$, as shown in Figs.~\ref{fig:twomcdataInvarMassBdpipi} and~\ref{fig:threemcdataInvarMassBdpipi}.
In \BsToJpsiKSKpi decays (Figs.~\ref{fig:twomcdataInvarMassBsKpi} and~\ref{fig:threemcdataInvarMassBsKpi}), excesses from $K^*(892)$ resonances are seen in $m(\KS\pipm)$ and $m(\Kpm\pimp)$, but no significant narrow structures are seen in $m(\KS\Kpm\pimp)$.
In \BdToJpsiKSKK decays (Figs.~\ref{fig:twomcdataInvarMassBdKK} and~\ref{fig:threemcdataInvarMassBdKK}), the $\phi(1020)$ state is seen in $m(\Kp\Km)$, but no other narrow structures are evident in any combination.

\section{Systematic uncertainties}
\label{sec:systematics}

Systematic uncertainties arise from possible inaccuracies in the determination of the yields, and imprecision of the knowledge of the efficiencies and fragmentation fractions that enter Eq.~(\ref{eq:standardBF1}) and Eq.~(\ref{eq:standardBF2}).
These contributions are summarised in Tables~\ref{table:Systematics1} and~\ref{table:Systematics2} for measurements with the simulation-based and data-based selection, respectively.
Total systematic uncertainties are obtained by addition in quadrature.

\begin{table}[!htbp]
  \centering
  \caption{\small
    Systematic uncertainties (\%) for the relative branching fraction measurements with \BdToJpsiKS as normalisation channel, given separately for long and downstream categories.
    The total systematic uncertainty is the sum in quadrature of all contributions.
  }
  \label{table:Systematics1}
\begin{tabular}{ccc@{\hspace{8mm}}cc}
\hline
                                & \multicolumn{2}{c}{\hspace{-10mm}Source} & Total      & Normalisation  \\
                                & Yield       & Efficiency   & systematic & sample size     \\
\hline
\multicolumn{5}{c}{long} \\ 
\hline \\ [-2.5ex]
${\cal B}(\BdToJpsiKSpipi)$     & 4.5    &  5.9         & 7.4        & 1.5 \\ [0.3ex]
${\cal B}(\BdToPsitwoSKS)$      & 3.3    &  5.5         & 6.4        & 1.5 \\ [0.3ex]
\hline
\multicolumn{5}{c}{downstream} \\
\hline \\ [-2.5ex]
${\cal B}(\BdToJpsiKSpipi)$     & 1.2    &  6.9         & 7.0        & 1.1 \\ [0.3ex]
${\cal B}(\BdToPsitwoSKS)$      & 3.3    &  7.1         & 7.8        & 1.1 \\ [0.3ex]
\hline
\end{tabular}
\end{table}

\begin{table}[!htbp]
  \centering
  \caption{\small
    Systematic uncertainties (\%) for the relative branching fraction measurements with \BdToJpsiKSpipi as normalisation channel, given separately for long and downstream categories.
    The total systematic uncertainty is the sum in quadrature of all contributions.
  }
  \label{table:Systematics2}
\resizebox{\textwidth}{!}{
\begin{tabular}{ccc@{\hspace{8mm}}ccc}
\hline
                            & \multicolumn{2}{c}{\hspace{-10mm}Source} & Total      & Fragmentation  & Normalisation  \\
                            & Yields      & Efficiencies & systematic & fractions      & sample size     \\
\hline
\multicolumn{6}{c}{long} \\
\hline \\ [-2.5ex]
${\cal B}(\BdToJpsiKSKpi)$  & 12.7 & 31.0 & 33.5 & --- & 6.6 \\ [0.3ex]
${\cal B}(\BdToJpsiKSKK)$   & \phani2.9 & \phani8.0 & \phani8.5 & --- & 6.6 \\ [0.3ex]
\hline \\ [-2.5ex]
${\cal B}(\BsToJpsiKSpipi)$ & 16.5 & 33.2 & 37.0 & 5.8 & 6.6 \\ [0.3ex]
${\cal B}(\BsToJpsiKSKpi)$  & \phani1.1 & \phani7.7 & \phani7.8 & 5.8 & 6.6 \\ [0.3ex]
${\cal B}(\BsToJpsiKSKK)$   & 39.0 & 33.2 & 51.2 & 5.8 & 6.6 \\ [0.3ex]
\hline
\multicolumn{6}{c}{downstream} \\
\hline \\ [-2.5ex]
${\cal B}(\BdToJpsiKSKpi)$  & \phani7.6 & 27.6 & 28.6 & --- & 5.0 \\ [0.3ex]
${\cal B}(\BdToJpsiKSKK)$   & \phani3.2 & \phani6.5 & \phani7.3 & --- & 5.0 \\ [0.3ex]
\hline \\ [-2.5ex]
${\cal B}(\BsToJpsiKSpipi)$ & 17.3 & 30.1 & 34.7 & 5.8 & 5.0 \\ [0.3ex]
${\cal B}(\BsToJpsiKSKpi)$  & \phani0.9 & \phani6.4 & \phani6.4 & 5.8 & 5.0 \\ [0.3ex]
${\cal B}(\BsToJpsiKSKK)$   & 18.0 & 36.7 & 40.9 & 5.8 & 5.0 \\ [0.3ex]
\hline
\end{tabular}
}
\end{table}

The systematic uncertainties on the yields are estimated by (i) varying all fixed fit parameters within their uncertainties; (ii) replacing the double Crystal Ball shape that describes the signal with a double Gaussian function; (iii) scaling the relative width of the \Bs and \Bd peaks according to the available phase-space for the decays; (iv) replacing the function that describes the combinatorial background with a second-order polynomial shape.  
The changes in the fitted yields are assigned as the corresponding uncertainties.
In addition, for channels where both signal and background yields are low, a small bias (less than 20\% of the statistical uncertainty) on the signal yield is observed in samples of pseudoexperiments.
To have a coherent treatment of all channels, each fitted yield is corrected for the bias, and the uncertainty on the bias combined in quadrature with half the correction is assigned as a systematic uncertainty.

One source of systematic uncertainty that affects the relative efficiencies arises from the particle identification requirements.  This is estimated by applying the method to determine the efficiency from control samples of $D^0 \to \Km\pip$ decays to simulated signal events, and comparing the result to the true value.  
The systematic uncertainty due to the variation of the efficiency over the phase-space is evaluated by reweighting the simulated samples for each signal decay to match the main features of the distributions seen in data (see Sec.~\ref{sec:phase-space}).
However, this method can only be applied for the channels where significant signals are observed: \BdToJpsiKSpipi, \BdToJpsiKSKpi and \BdToJpsiKSKK.
For the other decay channels the root-mean-square variation of the efficiency over the phase-space is obtained by binning the simulated events in each invariant mass combination, and this value is assigned as the associated uncertainty.  
There is also a small uncertainty arising from the limited simulation sample sizes.

The effective lifetimes of \Bs meson decays depend on the \CP-admixture of the final state~\cite{Fleischer:2011cw}.
Since the selection efficiency depends on decay time, this in principle leads to a source of uncertainty in the measurement.  
The scale of the efficiency variation is $\pm4\%$ for the extreme ranges of
possible effective lifetime distributions.  
Although knowledge of the exact composition of \CP-even and \CP-odd states requires either a detailed amplitude analysis or a measurement of the effective lifetime, all of the $\jpsi\KS h^+h^{\left(\prime\right) -}$ final states are expected to be approximately equal admixtures.
Therefore, no systematic uncertainty is assigned due to the assumption that
the effective lifetime is given by $1/\Gamma_s$, where $\Gamma_s$ is the mean
width of the two \Bs mass eigenstates~\cite{HFAG}.
The decay time distribution observed in data is consistent with that obtained
in a sample simulated with lifetime $1/\Gamma_s$, verifying that this is a
reasonable assumption.

For the relative branching fraction measurement of \BdToJpsiKSpipi to \BdToJpsiKS, there are two more tracks in the former channel than the latter.
Therefore additional small systematic uncertainties arise due to the limited knowledge of the track reconstruction and trigger efficiencies.
Uncertainty on the ratio of fragmentation fractions $f_s/f_d$ affects the measurement of the \Bs decay branching fractions relative to that of \BdToJpsiKSpipi.
Finally, for each relative branching fraction measurement the statistical uncertainty on the normalisation channel also contributes.
To allow a straightforward evaluation of the absolute branching fractions of the modes studied with the data-based selection, this source is treated separately.

\section{Results and conclusions}
\label{sec:results}

Results are obtained separately for the relative branching fractions in the long and downstream categories and then combined.
The combinations are performed using the full likelihood functions, though the uncertainties are symmetrised for presentation of the results.
Possible correlations between systematic uncertainties in the different categories, due to the fit model, particle identification efficiencies and $f_s/f_d$, are accounted for in the combinations.
All pairs of results in long and downstream categories are consistent within 2.5 standard deviations ($\sigma$).
The signal significances are obtained from the change in likelihood when the signal yields are fixed to zero.
Systematic uncertainties that affect the yield are accounted for in the calculation by smearing the likelihood with a Gaussian function of appropriate width.
The significances are summarised in Table~\ref{table:Significances}.
Since the significances of the \BdToJpsiKSKK and \BsToJpsiKSKpi signals exceed $5\,\sigma$, these results constitute the first observations of those decays.

\begin{table}[!htbp]
  \centering
  \caption{\small
    Significances ($\sigma$) for previously unobserved channels obtained from the fits to the samples with data-based selection.
    The values quoted for long and downstream categories include only statistical effects, while the combined results include systematic uncertainties.
  }
  \label{table:Significances}
  \begin{tabular}{cccc}
    \hline
    Mode            & \multicolumn{3}{c}{Significance} \\
                    & long & downstream & combined \\
    \hline \\ [-2.5ex]
    \BdToJpsiKSKpi  & $\phani0.8$ & $\phani2.5$ & $\phani2.4$ \\ [0.3ex]
    \BdToJpsiKSKK   & $\phani6.2$ & $\phani5.1$ & $\phani7.7$ \\ [0.3ex]
    \BsToJpsiKSpipi & $\phani2.3$ & $\phani1.9$ & $\phani2.7$ \\ [0.3ex]
    \BsToJpsiKSKpi  & $17.9$ & $25.8$ & $30.0$ \\ [0.3ex]
    \BsToJpsiKSKK   & $\phani0.7$ & $\phani0.6$ & $\phani0.5$ \\ [0.3ex]
    \hline
\end{tabular}
\end{table}

The results from the simulation-based selection are 
\begin{equation*}
  \frac{{\cal B}(\BdToJpsiKSpipi)}{{\cal B}(\BdToJpsiKS)} = 0.493 \pm 0.034 \stat \pm 0.027 \syst \, ,
\end{equation*} 
and
\begin{equation*}
  \frac{{\cal B}(\BdToPsitwoSKS)\times{\cal B}(\psi(2S)\to\jpsi\pip\pim)}{{\cal B}(\BdToJpsiKS)} = 0.183 \pm 0.027 \stat  \pm 0.015 \syst \, ,
\end{equation*}
where the first uncertainties are statistical and the second systematic.
The measurement of ${\cal B}(\BdToJpsiKSpipi)$ excludes the contribution from $\psitwos\to\jpsi\pip\pim$ decays.
These results are converted to absolute branching fraction measurements using known
values of the other branching fractions involved in the ratios~\cite{PDG2012}.
For consistency with the standard convention, the results for the absolute
branching fractions are multiplied by a factor of two to obtain values
corresponding to a $\Kz$, instead of \KS, meson in the final state, giving
\begin{eqnarray*}
  {\cal B}(\BdToJpsiKzpipi)   & = & \left( 43.0 \pm 3.0 \stat \pm 3.3 \syst \pm 1.6 \pdg \right) \times 10^{-5} \, , \\
  {\cal B}(\Bd\to\psitwos\Kz) & = & \left(  4.7 \pm 0.7 \stat \pm 0.4 \syst \pm 0.6 \pdg \right) \times 10^{-4} \, ,
\end{eqnarray*} 
where the last uncertainty is from knowledge of the normalisation branching fractions.
These results are consistent with previous measurements~\cite{PDG2012} and, in the case of the former, significantly more precise.

The results from the data-based selection are
\begin{eqnarray*}
  \frac{{\cal B}(\BdToJpsiKSKpi)}{{\cal B}(\BdToJpsiKSpipi)} & = & 0.026 \pm 0.012 \stat \pm 0.007 \syst \pm 0.001 \norm \, , \nonumber \\
  & < & 0.048\ {\rm at} \ 90\% \ {\rm CL} \, , \\
  & < & 0.055\ {\rm at} \ 95\% \ {\rm CL} \, , \nonumber \\
  \frac{{\cal B}(\BdToJpsiKSKK)}{{\cal B}(\BdToJpsiKSpipi)} & = & 0.047 \pm 0.010 \stat \pm 0.004 \syst \pm 0.002 \norm \, , \\
  \frac{{\cal B}(\BsToJpsiKSpipi)}{{\cal B}(\BdToJpsiKSpipi)} & = & 0.054 \pm 0.031 \stat \pm 0.020 \syst \pm 0.003 \fsfd \pm 0.004 \norm \, , \nonumber \\
  & < & 0.10\ {\rm at} \ 90\% \ {\rm CL} \, , \\
  & < & 0.12\ {\rm at} \ 95\% \ {\rm CL} \, , \nonumber \\
  \frac{{\cal B}(\BsToJpsiKSKpi)}{{\cal B}(\BdToJpsiKSpipi)} & = & 2.12 \pm 0.15 \stat \pm 0.14 \syst \pm 0.08 \fsfd \pm 0.08 \norm \, , \\ 
  \frac{{\cal B}(\BsToJpsiKSKK)}{{\cal B}(\BdToJpsiKSpipi)} & = & 0.011 \pm 0.020 \stat \pm 0.006 \syst \pm 0.001 \fsfd \pm 0.001 \norm \, , \nonumber \\
  & < & 0.027\ {\rm at} \ 90\% \ {\rm CL} \, , \\
  & < & 0.033\ {\rm at} \ 95\% \ {\rm CL} \, , \nonumber
\end{eqnarray*} 
where the uncertainties due to $f_s/f_d$ and the size of the \BdToJpsiKSpipi normalisation sample are quoted separately.
Upper limits, obtained from integrating the likelihood in the positive region, are quoted at both 90\% and 95\% confidence level (CL) for all channels with combined significance less than $3\,\sigma$.

These results are converted to absolute branching fraction measurements by multiplying by the value of the normalisation channel branching fraction determined with the simulation-based selection.
In this process, the statistical uncertainty of the \BdToJpsiKSpipi yield is taken to be 100\% correlated between the samples with simulation-based and data-based selection, since differences are small enough to be neglected.
The results are
\begin{eqnarray*}
  {\cal B}(\BdToJpsiKzKpi) & = & \left( 11 \pm 5 \stat \pm 3 \syst \pm 1 \pdg \right) \times 10^{-6} \, , \\
                           & < & 21\times 10^{-6} \ {\rm at} \ 90\% \ {\rm CL} \, , \\
                           & < & 24\times 10^{-6} \ {\rm at} \ 95\% \ {\rm CL} \, , \\
  {\cal B}(\BdToJpsiKzKK) & = & \left( 20.2 \pm 4.3 \stat \pm 1.7 \syst \pm 0.8 \pdg \right) \times 10^{-6} \, , \\
  {\cal B}(\BsToJpsiKzbpipi) & = & \left( 2.4 \pm 1.4 \stat \pm 0.8 \syst \pm 0.1 \fsfd \pm 0.1 \pdg \right) \times 10^{-5} \, , \\
                             & < & 4.4 \times 10^{-5} \ {\rm at} \ 90\% \ {\rm CL} \, , \\
                             & < & 5.0 \times 10^{-5} \ {\rm at} \ 95\% \ {\rm CL} \, , \\
  {\cal B}(\BsToJpsiKzKpi) & = & \left( 91 \pm 6 \stat \pm 6 \syst \pm 3 \fsfd \pm 3 \pdg \right) \times 10^{-5} \, , \\ 
  {\cal B}(\BsToJpsiKzbKK) & = & \left( 5 \pm 9 \stat \pm 2 \syst \pm 1 \fsfd \right) \times 10^{-6} \, , \\
                           & < & 12 \times 10^{-6} \ {\rm at} \ 90\% \ {\rm CL} \, , \\
                           & < & 14 \times 10^{-6} \ {\rm at} \ 95\% \ {\rm CL} \, , 
\end{eqnarray*} 
where the contribution from the PDG uncertainty to the last result is
negligible.
The expression ${\cal B}(\BxToJpsiKzKpi)$ denotes the sum of the branching
fractions for $\Bds\to\jpsi\Kz\Km\pip$ and $\Bds\to\jpsi\Kzb\Kp\pim$ decays.
In all results the strangeness of the produced neutral kaon is assumed to be
that which is least suppressed in the SM.

In summary, using a data sample corresponding to an integrated luminosity of $1.0 \invfb$ of $pp$ collisions at centre-of-mass energy $\sqrt{s} = 7 \tev$ recorded with the \lhcb detector at CERN, searches for the decay modes \BxToJpsiKShhp have been performed. 
The most precise measurement to date of the \BdToJpsiKSpipi branching fraction and the first observations of the \BdToJpsiKSKK and \BsToJpsiKSKpi decays are reported.
The first limits on the branching fractions of \BsToJpsiKSpipi, \BdToJpsiKSKpi and \BsToJpsiKSKK decays are set.
Inspection of the phase-space distributions of the decays with significant signals does not reveal any potentially exotic narrow structure,
nor is any significant excess from a narrow resonance seen in the $\KS\Kpm\pimp$ invariant mass distribution in \BsToJpsiKSKpi decays.
Further studies will be needed to investigate the underlying dynamics of these channels, and to understand whether they can in future be used for \CP violation studies.

\section*{Acknowledgements}

\noindent We express our gratitude to our colleagues in the CERN
accelerator departments for the excellent performance of the LHC. We
thank the technical and administrative staff at the LHCb
institutes. We acknowledge support from CERN and from the national
agencies: CAPES, CNPq, FAPERJ and FINEP (Brazil); NSFC (China);
CNRS/IN2P3 and Region Auvergne (France); BMBF, DFG, HGF and MPG
(Germany); SFI (Ireland); INFN (Italy); FOM and NWO (The Netherlands);
SCSR (Poland); MEN/IFA (Romania); MinES, Rosatom, RFBR and NRC
``Kurchatov Institute'' (Russia); MinECo, XuntaGal and GENCAT (Spain);
SNSF and SER (Switzerland); NASU (Ukraine); STFC and the Royal Society (United
Kingdom); NSF (USA). We also acknowledge the support received from EPLANET, 
Marie Curie Actions and the ERC under FP7. 
The Tier1 computing centres are supported by IN2P3 (France), KIT and BMBF (Germany),
INFN (Italy), NWO and SURF (The Netherlands), PIC (Spain), GridPP (United Kingdom).
We are indebted to the communities behind the multiple open source software packages on which we depend.
We are also thankful for the computing resources and the access to software R\&D tools provided by Yandex LLC (Russia).

\addcontentsline{toc}{section}{References}
\setboolean{inbibliography}{true}
\bibliographystyle{LHCb}
\bibliography{main,LHCb-PAPER,LHCb-CONF,LHCb-DP}

\end{document}